\documentclass[twocolumn,hyperpdf,amsmath,amssymb,aps,prd,10pt,superscriptaddress,nofootinbib,noeprint,preprintnumbers,floatfix]{revtex4-2}

\usepackage{orcidlink}
\usepackage{graphicx, color}

\usepackage[letterspace=-10]{microtype} 

\usepackage{bm, amsmath, amsfonts, amssymb,xfrac}
\usepackage{multirow, tabularx, dcolumn, soul}
\usepackage{mathtools, leftidx, braket, slashed, cancel, bigdelim}
\usepackage{blkarray}
\usepackage[figures]{rotating}

\usepackage{tikz}
\usetikzlibrary[plotmarks]
\usepackage{anyfontsize}

\usepackage[utf8]{inputenc} 
\usepackage{hyperref}
\newcommand\bef{\begin{figure}}
\newcommand\eef[1]{\label{fg:#1}\end{figure}}
\newcommand\beq{\begin{equation}}
\newcommand\eeq[1]{\label{#1}\end{equation}}
\newcommand\beqa{\begin{eqnarray}}

\newcommand\bet{\begin{table}}
\newcommand\eet[1]{\label{tb:#1}\end{table}}
 
\newcommand\eqn[1]{Eq.~(\ref{#1})}

\def\IMSc{The Institute of Mathematical Sciences, CIT Campus, Chennai, 600113, India}
\def\HBNI{Homi Bhabha National Institute, Training School Complex, Anushaktinagar, Mumbai 400094, India}
\newcommand{\imsc}{\affiliation{\IMSc}}
\newcommand{\hbni}{\affiliation{\HBNI}}

\definecolor{link_blue}{RGB}{51,102,204}

\hypersetup{%
pdftitle = {title},
pdfsubject = {},
pdfkeywords = {},
colorlinks = {true},
filecolor = {black},
linkcolor = {link_blue},
menucolor = {black},
citecolor = {link_blue},
urlcolor = {link_blue},
}{}

\begin{document}

\title{Precise study of triply charmed baryons $\Omega_{ccc}$}

\author{Navdeep Singh Dhindsa~\orcidlink{0000-0002-3133-6979}}
\email{navdeep@theory.tifr.res.in}
\imsc
\affiliation{Department of Theoretical Physics, Tata Institute of Fundamental Research, \\ Homi Bhabha Road, Mumbai 400005, India }

\author{Debsubhra Chakraborty~\orcidlink{0000-0001-5815-4182}}
\email{debsubhra.chakraborty@tifr.res.in}
\affiliation{Department of Theoretical Physics, Tata Institute of Fundamental Research, \\ Homi Bhabha Road, Mumbai 400005, India }

\author{Archana Radhakrishnan~\orcidlink{0000-0001-9357-1360}}
\email{archana.radhakrishnan@tifr.res.in}
\affiliation{Department of Theoretical Physics, Tata Institute of Fundamental Research, \\ Homi Bhabha Road, Mumbai 400005, India }

\author{Nilmani Mathur~\orcidlink{0000-0003-2422-7317}}
\email{nilmani@theory.tifr.res.in}
\affiliation{Department of Theoretical Physics, Tata Institute of Fundamental Research, \\ Homi Bhabha Road, Mumbai 400005, India }

\author{M. Padmanath~\orcidlink{0000-0001-6877-7578}}
\email{padmanath@imsc.res.in}
\imsc
\hbni

\preprint{TIFR/TH/24-22}

\begin{abstract}
We present the most precise results for the ground state mass of the triply-charmed spin-$3/2$ baryon using lattice quantum chromodynamics.
The calculations are performed on six $N_f=2+1+1$ Highly Improved Staggered Quark (HISQ) lattice ensembles generated by the MILC collaboration. Two different lattice setups are employed: in the first one, a fully dynamical calculation with HISQ action is performed, while in the second calculation, an overlap action is utilized for the valence charm quark dynamics. Following the continuum extrapolation of our results, obtained at five different lattice spacings, two different volumes, and two different actions, our prediction for the mass of the lowest triply charmed spin-3/2 baryon, $\Omega_{ccc} (3/2^{+})$, is $4793 (5) \left(^{+11}_{-8}\right)$ MeV.  This is the most precise determination to date, fully addressing the systematic uncertainties. 
 We also predict the $\Omega_{ccc} (3/2^{-})$ mass to be $5094 (12) \left(^{+19}_{-17}\right)$ MeV.
\end{abstract}

\maketitle

 \noindent The role of single-flavored baryons was paramount in identifying the color degrees of freedom and the subsequent development of quantum chromodynamics (QCD), the theory of strong interactions. The discovery of $\Omega(sss)(J^P=3/2^+)$ baryon firmly established the quark model as an effective phenomenological framework for describing subatomic particles, which further lead to many key advancements in our understanding of nuclear and particle physics.
Although the single-flavored baryons have only been discovered  with light and strange flavors ($\Delta$ and $\Omega$), the existence of their counterparts with charm and bottom flavors--namely $\Omega_{ccc}$ and  $\Omega_{bbb}$--is beyond any doubt within QCD. The $\Omega_{QQQ}$ ($Q\equiv c,b$) baryons, which are baryon analogues of heavy quarkonia, offer a simplistic system for studying quark-quark interactions and quark confinement without the complexities of valence light quark dynamics \cite{Bjorken:1985ei, Padmanath:2013zfa, Meinel:2012qz}. As pointed out by Bjorken, such baryons may provide a new window for understanding the internal parton level dynamics within baryons \cite{Bjorken:1985ei}. 
These baryons are expected to be discovered as experimental facilities reach the required center-of-momentum energy and luminosity, with $\Omega_{ccc}^{++}$ likely being the first to be observed.
 Precise theoretical predictions using first principles calculations of QCD  can guide the experiments in discovering the $\Omega_{ccc}^{++}(3/2^+)$ baryon. In this work, we present the most precise determination of the ground state mass of the $\Omega_{ccc}^{++}(3/2^+)$ baryon using the first-principles lattice QCD methodology, carefully accounting both statistical and all potential systematic uncertainties, which ensures that the final mass prediction is reliable and accurate.
 
Since the strong decay of the lightest spin-3/2 $\Omega_{ccc}$ baryon is forbidden, it should have a relatively long lifetime. Probable decay modes, where they could be searched for, are semileptonic decay processes such as $\Omega_{ccc}^{++} \rightarrow \Omega_{ccs}^{+} + \{l\} \rightarrow \Xi_{ccu}^{++} + 2\{l\}$,  $\Omega_{ccc}^{++} \rightarrow \Xi_{ccd}^{+} + \{l\}$, among others ($\{l\}$ refers to lepton pair involved in the decay). Direct detection of such long lived particles could be difficult, unless the lighter hadrons to which they decay are well understood. However, recent experimental reports on tri-$J/\psi$ production \cite{CMS:2021qsn}, and recent
proposals of inclusive search strategies \cite{Gershon:2018gda, Qin:2021zqx} and extensions of new sensitive timing based variables \cite{Bhattacherjee:2021qaa} to identify signatures for long lived hadrons augment promising prospects for the discovery of $\Omega_{ccc}^{++}$ baryons in the near future.

Considering the heavy quark constituents within, it is expected that the phenomenological models would be able to reliably predict the masses of  $\Omega_{ccc}$ baryons. Indeed, within model calculations, starting with early works of Bjorken~\cite{Bjorken:1985ei}, these baryons have been studied with
non-relativistic ~\cite{Silvestre-Brac:1996myf, Roberts:2007ni, Vijande:2015faa, Yang:2019lsg, Liu:2019vtx, Llanes-Estrada:2011gwu, PhysRevD.108.054014} and
relativistic~\cite{Martynenko:2007je, Faustov:2021qqf}  quark models, QCD sum rule~\cite{Zhang:2009re, Wang:2020avt, Aliev:2014lxa, Wang:2010vn, Wu:2021tzo}, Faddeev equation~\cite{Sanchis-Alepuz:2011xjl, Radin:2014yna, Qin:2019hgk}, diquark model~\cite{Thakkar:2016sog, Yin:2019bxe}, variational method~\cite{Jia:2006gw, Flynn:2011gf}, bag model~\cite{Bernotas:2008bu, Hasenfratz:1980ka}, hypercentral
constituent quark model~\cite{Shah:2017jkr, Patel:2008mv, doi:10.1142/S0217732321502709}, Regge theory~\cite{Wei:2015gsa}, Bethe-Salpeter equation~\cite{Migura:2006ep} and renormalization group procedure for effective particles approach~\cite{Serafin:2018aih}.
However, depending on the model parameters, these predictions are spread over 400 MeV uncertainty. 

 In this context, first principles lattice QCD calculations with controlled systematics can play a vital role in narrowing the possible mass window for these baryons. Most importantly, lattice QCD calculations can rigorously incorporate non-perturbative aspects that may be overlooked by potential models and other phenomenological approaches. 
Several lattice QCD studies of $\Omega_{ccc}$ baryons have also already been performed, starting with studies using pure $SU(3)$ gauge theory~\cite{Chiu:2005zc}, dynamical simulations with quenched charm quarks \cite{Alexandrou:2012xk, PACS-CS:2013vie, Padmanath:2013zfa, Brown:2014ena, Can:2015exa,Bahtiyar:2020uuj,Li:2022vbc, Lyu:2021qsh,Durr:2012dw, Alexandrou:2017xwd}, and those involving dynamical quarks up to charm flavor \cite{Briceno:2012wt, Basak:2012py, Alexandrou:2014sha,Chen:2017kxr,Alexandrou:2023dlu}. 
Most of these calculations were performed at a single lattice spacing, whereas addressing lattice artifacts through continuum extrapolation with several lattice spacings is essential to assess the systematic uncertainties from cutoff effects, particularly considering the presence of three heavy quarks in these baryons.
The studies in Refs.~\cite{Briceno:2012wt, Brown:2014ena, Alexandrou:2014sha, Alexandrou:2012xk, Alexandrou:2023dlu} have performed continuum extrapolation 
with up to three lattice spacings, leading to mass estimates for $\Omega_{ccc}(3/2^+)$ that are spread over a range of 100 MeV. 
There are no continuum extrapolated results for the mass of $3/2^-$ ground state, with existing predictions scattered over a 250 MeV energy interval. 

\begin{figure}[htb!]
      \includegraphics
      [width=0.45\textwidth]{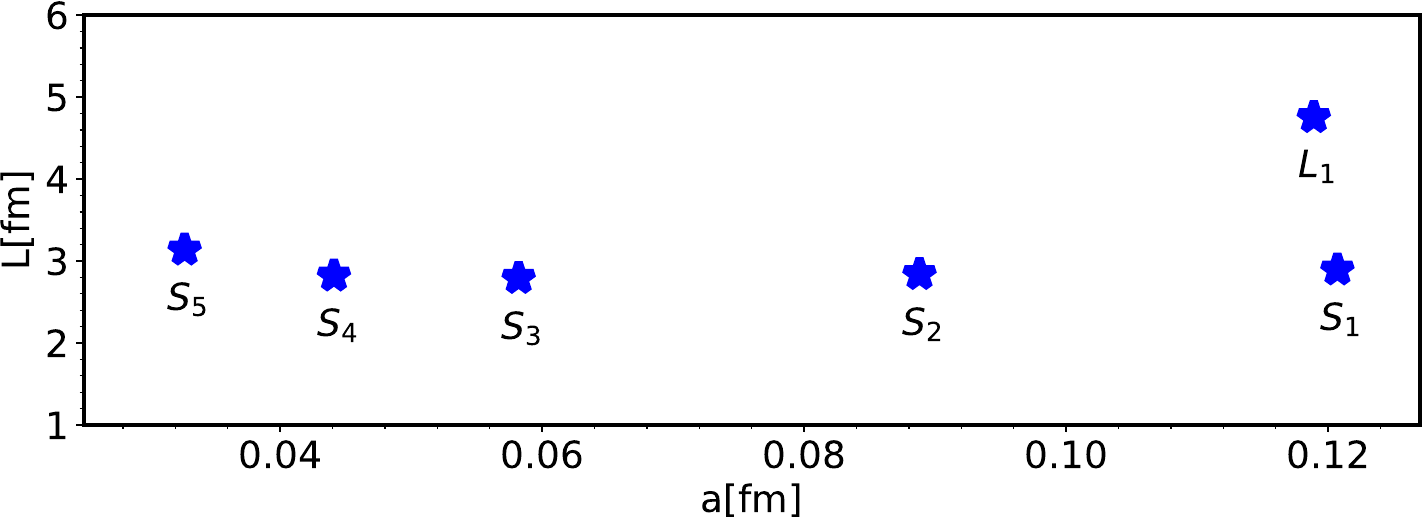}
        \caption{Six lattice QCD ensembles used in this work and are labeled in the plot as S$_5$ ($96^3 \times 288$), S$_4$ ($64^3 \times 192$), S$_3$ ($48^3 \times 144$), S$_2$ ($32^3 \times 96$), S$_1$ ($24^3 \times 64$), and L$_1$ ($40^3 \times 64$).
        }
        \label{fg:latt}
\end{figure}

It is therefore important to perform a systematic lattice QCD determination of the masses of low lying $\Omega_{ccc}^{++}$ baryons,
with an accuracy similar to the state-of-the-art lattice determination of other hadrons, such as that in low-lying charmonia and proton. In this work, we present such a study utilizing up to six lattice QCD ensembles, five different lattice spacings, two lattice volumes and $N_f=2+1+1$ Highly Improved Staggered Quark (HISQ) dynamics
\cite{Follana:2006rc}, generated by the MILC Collaboration \cite{MILC:2012znn, PhysRevD.98.074512} (see Fig. \ref{fg:latt}, and Table I in Ref. \cite{Suppl}).
In the valence sector, we employ two different approaches. In the first, a fully dynamical evaluation of the $\Omega_{ccc}^{++}$ baryon masses is made with HISQ action for the valence charm quark evolution. In the second approach, an overlap action \cite{Neuberger:1997fp,Neuberger:1998wv}, which has no 
${\cal{O}}(ma)$ error,
 is employed for the valence charm quark dynamics, similar to many of our previous publications \cite{Mathur:2018epb,Junnarkar:2018twb, Mathur:2018rwu,Padmanath:2023rdu,Radhakrishnan:2024ihu}.
With such a diverse and state-of-the-art setup of multiple lattice spacings, lattice volumes, and two different valence charm quark actions, we predict the ground state mass of   $\Omega_{ccc}^{++} (3/2^{+})$ baryon to be $4793 (5) \left(^{+11}_{-8}\right)$ MeV.
The mass for the lowest $J^{P} \equiv 3/2^{-}$ state is also predicted with per mille precision. Below, we detail the methods and procedures employed to obtain these results.

We evaluate the hadron masses from the two point correlation functions with a wall source and a point sink smearing, where the correlators are fitted with an exponential form at large source-sink separation.
Such a setup was demonstrated to be robust and reliable for extracting single-hadron masses, particularly for their ground states  
\cite{PhysRevLett.119.042001,Mathur:2018epb,Mathur:2018rwu,Lyu:2023xro, 10.1093/ptep/pts010,PhysRevLett.128.052003, PhysRevD.90.074509, Hudspith:2024kzk}.
We also observe that to be the case for the $\Omega_{ccc}$ baryon masses studied in this work. 
The charm quark mass, being heavy, leads to relatively clean and stable signals. Further, to mitigate the excited state effects from the extracted ground state hadron masses, we evaluate correlation matrices that are analyzed variationally following the solutions of a generalized eigenvalue problem \cite{Michael:1985ne, Luscher:1990ck}. 

\noindent{\it $\Omega_{ccc}$ baryons with valence overlap fermions}:-- In this setup, we utilize only spatially local interpolators, ($\mathcal{O}_b$). This is well justified considering the predicted smaller charge radius of $\Omega_{ccc}$ \cite{Can:2015exa}. The trivial flavor symmetry and color antisymmetry imply the total spin structure for the $\Omega_{ccc}$ baryon interpolator has to be symmetric. With a four component fermion spinor in the Dirac-Pauli representation of Dirac-$\gamma$ matrices \cite{sakurai1967advanced} with $\gamma_4=diag[1,1,-1,-1]$ representing the quark fields, one can have two embeddings of finite volume irrep $H$ ($\in \{\Lambda_{O_h}\}$) possessing fully symmetric spin structure and with spin-3/2 being the lowest contributing spin \cite{Basak:2005ir, Padmanath:2013zfa}. Here, $\Lambda_{O_h}$ refers to the irrep of the Octahedral group $O_h$. Correlation matrices are evaluated using the operator basis composed of these two operators and are then analyzed following variational techniques \cite{Michael:1985ne, Luscher:1990ck}. The antisymmetry along the temporal extent in baryon correlators can be utilized to average over the forward and the time-reversed backward propagations and thus doubling the statistics for each parity \cite{Datta:2012fz,Padmanath:2014arv,Montvay:1994cy}.
Further details on this are provided in Section II of the supplement \cite{Suppl}.

\noindent{\it $\Omega_{ccc}$ baryons with HISQ fermions}:--
The $\Omega_{sss}(3/2^+)$ has been studied with HISQ valence quark previously \cite{Borsanyi:2022ygn}, and following that we use the following interpolating field for $\Omega_{ccc}(3/2^+)$,
\begin{eqnarray}
    \mathcal{O}_{\Omega_{ccc}}(t) =  \epsilon_{abc} D_1c^a(\mathbf{x},t)D_2c^b(\mathbf{x},t)D_3c^c(\mathbf{x},t) \; ,\label{OmegaOp_main}
\end{eqnarray}
where $c^a(\mathbf{x},t)$ is the HISQ charm quark field with ($a$, $b$, $c$) as the color indices, and 
$D_i$s are the spatial derivative operators \cite{Golterman:1985dz}.
Note that the operator in Eq.(\ref{OmegaOp_main}) couples with multiple tastes. 
However, the splittings between finite-volume ground state energy of $\Omega$ baryon of different tastes are expected to be small as HISQ action removes all taste-exchange effectively through one-loop order \cite{Follana:2006rc}, and this has been shown in Ref. \cite{Borsanyi:2022ygn}.
To reduce the effects of oscillating components of staggered correlators, the two-point correlators are first smoothed and the procedure is detailed in the supplement \cite{Suppl}.  To determine the ground state masses, we employ two methods. The first one involves the standard large-time fit with a single exponential. We also use two-exponentials particularly for the spin-$3/2^{-}$ state where the contamination from the excited states may not be insignificant.
In the second method, we utilize the generalized eigenvalue problem (GEVP)\cite{Michael:1985ne,Luscher:1990ck}, using a matrix constructed from time-shifted correlation functions, as described in Ref. \cite{Aubin:2010jc}, allowing a robust estimation of the ground state. 

\begin{figure}[htb!]
      \includegraphics
      [width=0.45\textwidth]{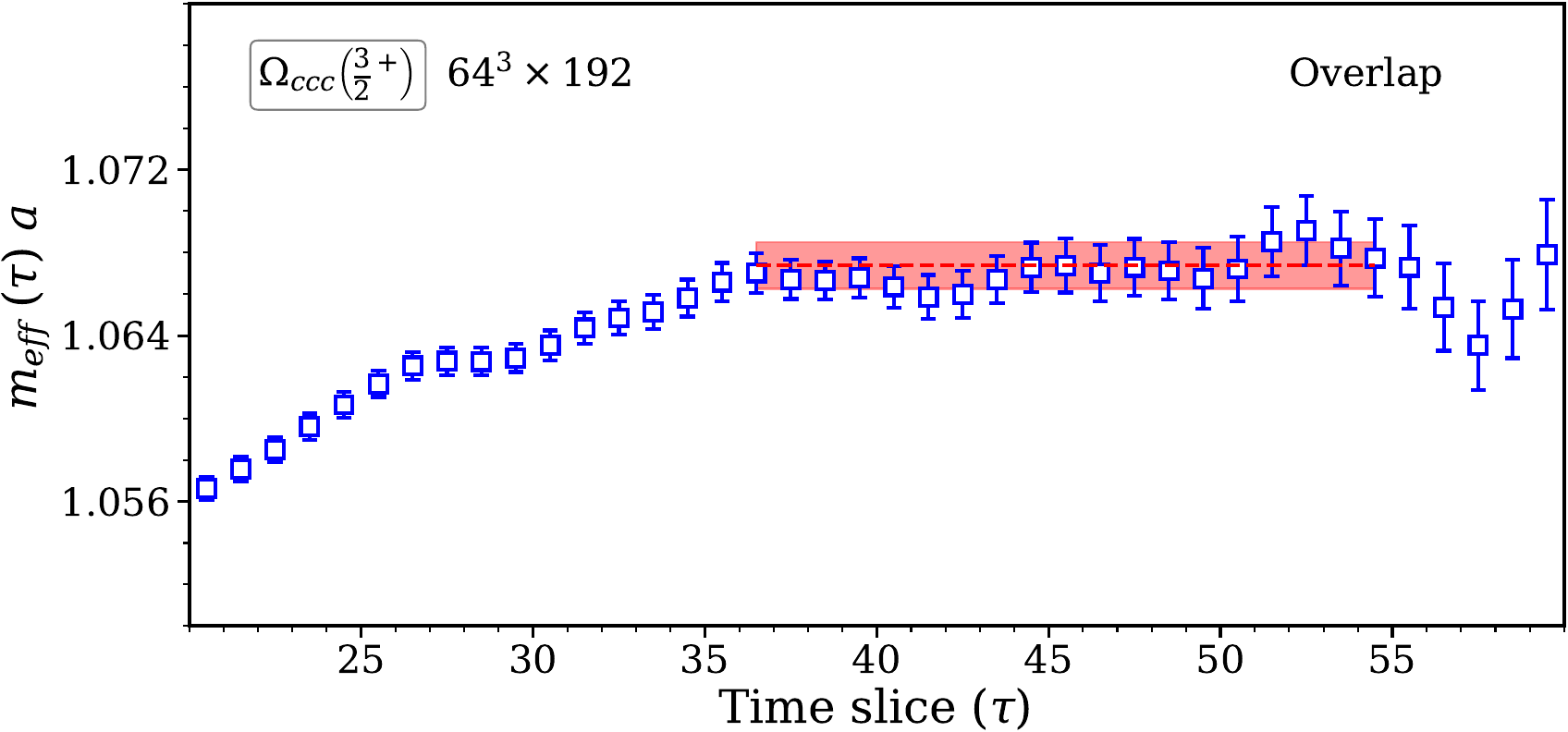} 
         \includegraphics[width=0.45\textwidth]{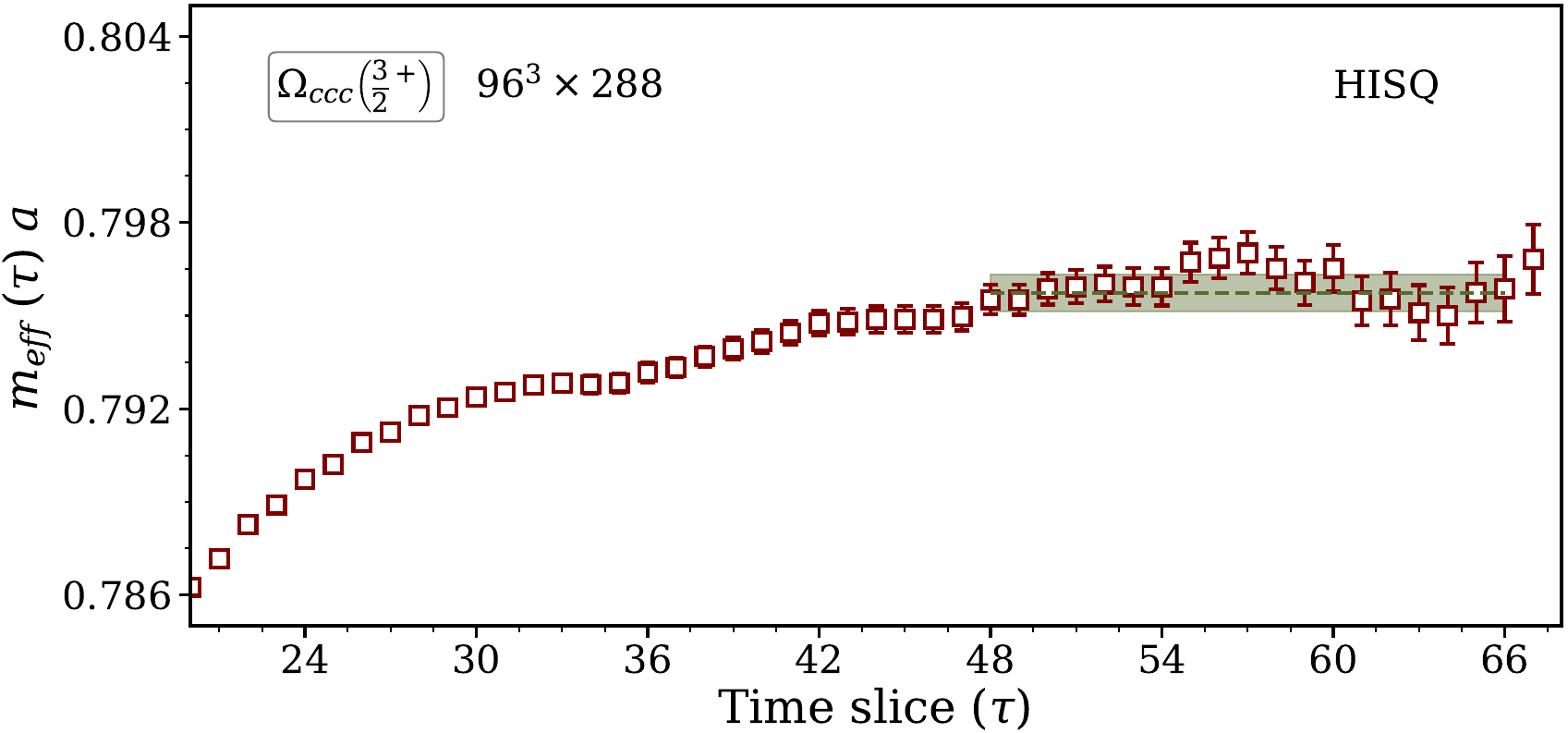}         
        \caption{Representative effective mass plots for the lowest energy state of $\Omega_{cc}(3/2^+)$ baryon. Top: overlap action, lattice size: $64^3\times 192$, Bottom: HISQ action, lattice size $96^3\times 288$.
        }
        \label{fg:eff_mass}
\end{figure}

\begin{figure}[htb!]
        \includegraphics [width=0.4\textwidth, height=0.42\textwidth]
{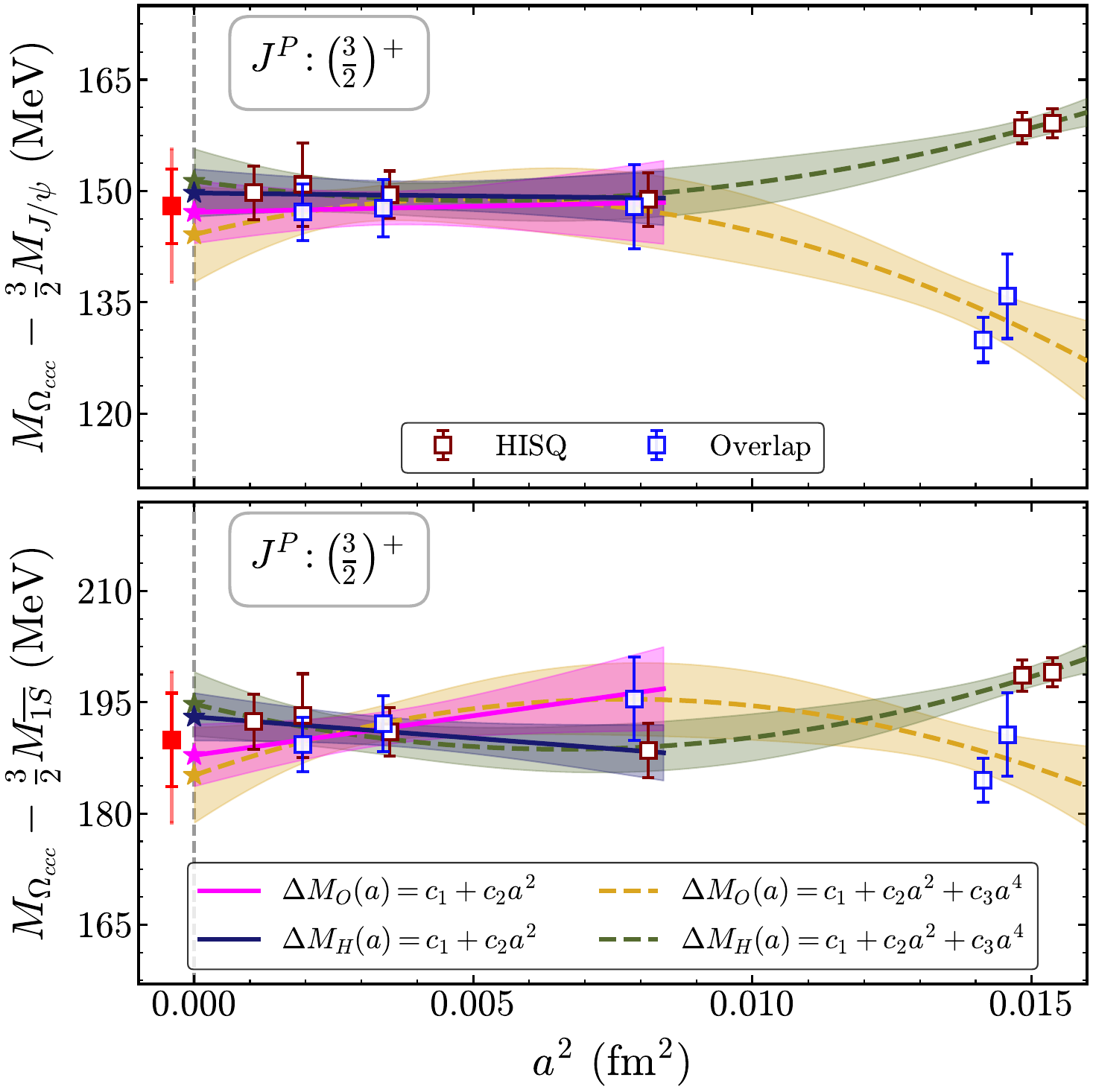}    

\includegraphics[width=0.4\textwidth, height=0.42\textwidth]
{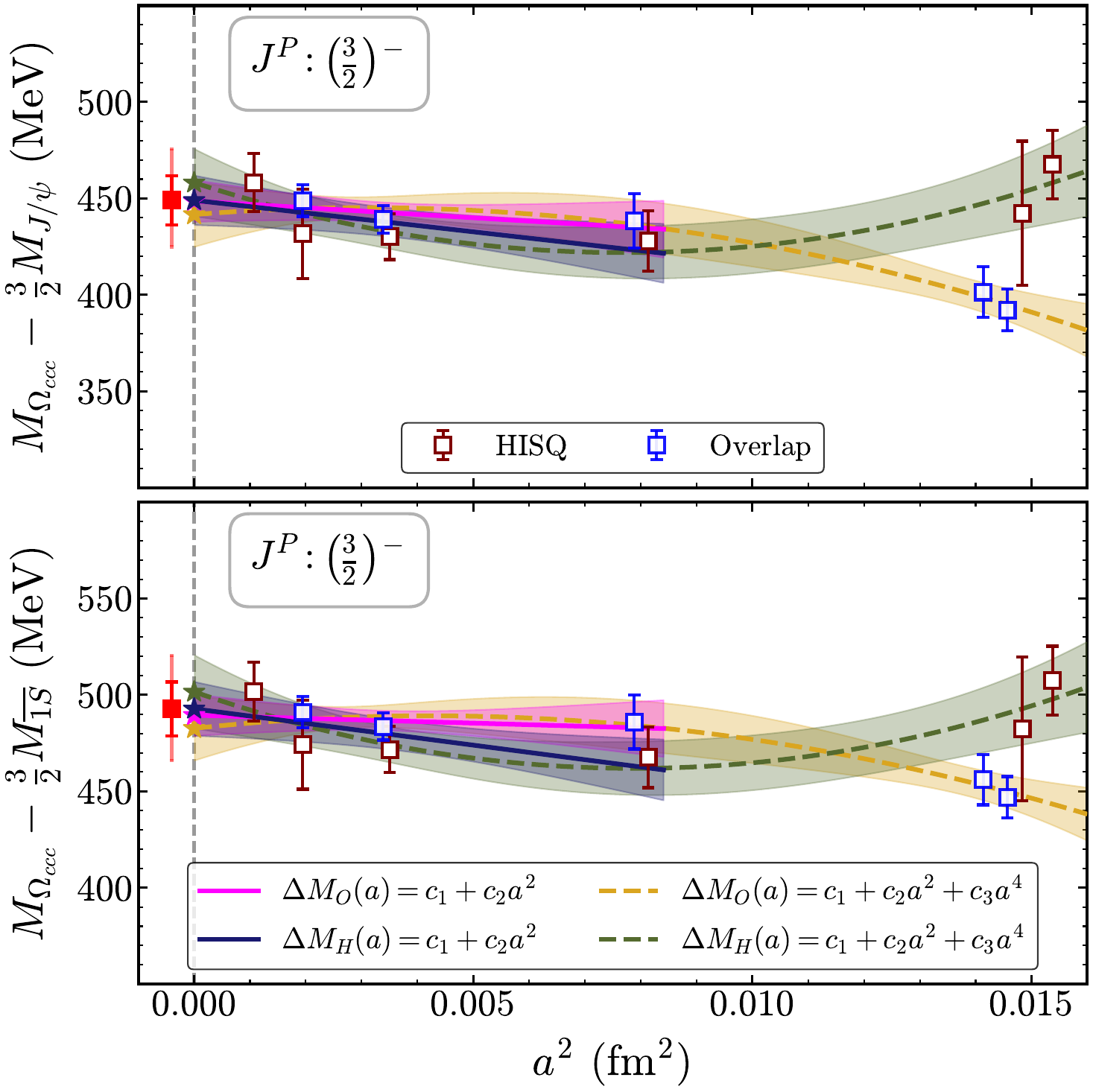}      
        \caption{Continuum extrapolations of the subtracted ground state masses $\Delta M^{\text{sub}}_{\Omega_{ccc}}(a)$ (Eq. \ref{eq:sub}). 
        Top two: $\Omega_{ccc}(3/2^+)$ and bottom two: $\Omega_{ccc}(3/2^-)$.
Subtraction of charm quark content is done with both $c\bar{c} \equiv J/\psi$ and $\overline{1S}$.
Various fit forms with their respective color schemes are shown in the legends and more details on fits are provided in the supplement \cite{Suppl}.
The shaded bands represent $1\sigma$ uncertainties, estimated via the bootstrap method. Symbol star: continuum extrapolated values. Red solid square: final value by symmetrized averaging the linear fit results of overlap and HISQ quarks, while the extended errors include other fits results.
}      
      \label{fg:omega_cont_exp_comb3}  
\end{figure}

{\it Mass extraction and results}:- In Fig.~\ref{fg:eff_mass}, we present two representative effective mass plots ($m_{eff} = \mathrm{log}[C(\tau+1)/C(\tau)]$, $C(\tau)$ being the correlation function) 
for the lowest energy level of the $\Omega_{ccc}(3/2^+)$ state on ensembles with the finest two lattice spacings. Further details are provided in the caption. The horizontal bands represent the fit estimate with $1\sigma$ error.
Further details of mass fits and results for other lattices and other hadrons studied are shown in the supplemental material \cite{Suppl}.

The large bare quark mass ($m_Q a$) introduces significant discretization effects, which can be a primary source of systematic uncertainties in heavy hadrons, especially in those containing multiple heavy quarks. The leading discretization errors in the valence quark actions that we employ (overlap and HISQ) are at $\mathcal{O}((ma)^2)$. To further mitigate these errors, instead of estimating the masses directly, we follow the practice of studying mass splittings, where the valence heavy quark content is subtracted out as in 
Refs. \cite{Mathur:2018epb,Mathur:2018rwu,Junnarkar:2018twb,Mathur:2022ovu,Padmanath:2023rdu,Radhakrishnan:2024ihu}. Here we subtract out the valence charm quark content with the following choice,
\begin{equation}
    a\Delta M_{\Omega_{ccc}} = [aM^{L}_{\Omega_{ccc}} - \frac{3}{2}aM^L_{c\bar{c}}] , \label{eq:sub}
\end{equation}
where $aM^{L}_{\Omega_{ccc}}$ and $aM^L_{c\bar{c}}$ are the finite lattice estimates for the mass of $\Omega_{ccc}$ baryon and the reference charmonia energy ($c\bar{c}$), respectively. 
Two different choices for $c\bar{c}$ are considered in our analysis: ${J/\psi}$ and the spin averaged ${\overline{1S}}$ state, $\frac{1}{4}\left({\eta_c} + 3{J/\psi}\right)$. These choices are made considering that we tune the charm quark mass using ${\overline{1S}}$, while ${J/\psi}$ masses on the lattice are found to be closer to its physical mass on each lattice. We perform the continuum extrapolation on these mass splittings, owing to the expectations of reduced cutoff effects on them. 
Here we prefer the reference energy with respect to the $J/\psi$ meson mass as the estimation of the $J/\psi$ mass on lattice is more accurate compared to the spin-averaged ${\overline{1S}}$ energy, potentially related to the dominant charm self-annihilation effects expected and also observed in $\eta_c$ meson mass \cite{ZhangRenQiang:2021gnn}. 

In Fig.~\ref{fg:omega_cont_exp_comb3} we show the continuum extrapolation of the mass splittings $\Delta M_{\Omega_{ccc}(3/2^+)}$ in physical units, with the top two panes using $c\bar{c} \equiv J/\psi$ (top figure) and  $c\bar{c} \equiv {\overline{1S}}$ (second from the top). We consider two possible cutoff dependence: (i) $\Delta M(a) = c_1 + c_2a^2$, (ii) $\Delta M(a) = c_1 + c_2a^2 + c_3a^4$. The marker and coloring convention used are explained in the figure and its caption. The bottom two figures, with the same color-coding and fit forms, represent the continuum extrapolation for the $\Omega_{ccc}(3/2^-)$ baryon.
Once the continuum extrapolated value of $\Delta M^{cont}_{\Omega_{ccc}}$ is obtained, we add that to the physical value of $\frac{3}{2}M^{exp}_{c\bar{c}}$ to obtain the physical estimates for baryon mass, as in,
\begin{equation}
    M^{phys}_{\Omega_{ccc}} = [\Delta M^{cont}_{\Omega_{ccc}} + \frac{3}{2}M^{exp}_{c\bar{c}}].
\end{equation}A clear agreement can be observed between the $\Omega_{ccc}(3/2^+)$ estimates using the overlap as well as the HISQ action across finer lattices leading to a consistent continuum extrapolated value. 
Our final result, represented by the red square at the continuum point, is a symmetrized average of the HISQ (green star) and overlap (blue star) results obtained from linear fits to the finest lattice points (starting at $a = 0.0888$ fm), with the magenta and blue bands representing the fitted results for the HISQ and overlap actions, respectively. The associated fit error, indicated by the thick red line, encompasses the fit errors for both the overlap and HISQ actions and is quoted as the statistical error. The thin red line illustrates the range of fit errors (bootstrap) arising from variations in lattice spacing dependence and differences in actions (as shown in Table III of the supplement \cite{Suppl}). The deviations of the thin red line from the thick red line are included in our systematic errors. As previously discussed, using $c\bar{c} \equiv J/\psi$ provides a more reliable basis for continuum extrapolation, and our final values are quoted accordingly. Our final estimates are 
$M_{\Omega_{ccc}} (3/2^+) = 4793 (5) \left(^{+11}_{-8}\right) $ MeV, and $M_{\Omega_{ccc}} (3/2^-) = 5094 (12) \left(^{+19}_{-17}\right)$ MeV, where the first error is statistical and the second one accounts for all other possible systematic errors, which we will discuss next.
We also estimate the mass splitting between these two baryons to be:
$\Delta_{\Omega_{ccc}}(3/2^+ - 3/2^-) = 301 (13) (14)$ MeV. 
Note that for reliable estimation of $\Omega_{ccc}(3/2^-)$ mass, one needs to consider the presence of a nearby three hadron final state $\Omega_{ccc}(3/2^+)\pi\pi$ in $P$-wave in the chiral limit \cite{Draper:2023xvu,Romero-Lopez:2021zdo,Hansen:2019nir,Mai:2017bge,Hansen:2021ofl,Blanton:2021llb,Draper:2023boj,Yan:2024gwp}. However, investigating such contamination is beyond the scope of this work. If the three quark Fock component of the negative parity state were dominant and its coupling with the three hadron state be negligible, we anticipate our estimate would be reliable.

A thorough estimation of systematic uncertainties is crucial in any precision investigation. Below, we provide a detailed account on systematics involved in this investigation covering several possible sources of errors. 

\noindent{\it Statistical--}
In this, we include bootstrap fit error, including various possible choices of fit-windows, and the final value includes the fit-error covering both overlap and HISQ results. 

\noindent{\it Discretization--}
Here, we include the difference in fit values obtained using different fit forms with two different discretized lattice actions.

\noindent{\it Scale setting--}
Measurements of
scale with Wilson ﬂow and $r_1$ parameter are found to be consistent for these lattice ensembles.   Including the scale setting errors we find the uncertainty in mass difference (Eq. \ref{eq:sub}) for $\Omega_{}$ to be $\sim$
3 MeV.

\noindent{\it Charm quark mass tuning:} 
After tuning the charm quark mass on each lattice (as detailed in the supplement \cite{Suppl}), we calculate the mass splittings (Eq. (\ref{eq:sub}), at its central value and at its variations. The effect on mass splittings was found to be, at most, 2 MeV.

\noindent{\it Unphysical sea quarks:}
The light sea-quark masses are heavier than their physical values in the set of ensembles that we utilize in this work (see Table 1 of the supplement \cite{Suppl}). 
Although it is naively expected to have only small effects from unphysical light sea quarks to the mass of an all charmed baryon \cite{McNeile:2012qf, Dowdall:2012ab, Chakraborty:2014aca}, it is better to have an estimate on it. In Sec. V of supplement \cite{Suppl} we provide a detailed account of the associated systematics estimation and we find this uncertainty to be about 5 MeV.

\noindent{\it Taste splitting (HISQ)--} For HISQ valence quarks this effect leads to a
difference in masses at finite lattice spacing with a  $\mathcal{O}(\alpha_s(1/a)(m_ca)^2)$ mixing between two tastes of $\Omega_{ccc}$ baryons.
We account for that in the continuum extrapolation, which leads to an error of 2 MeV.

\noindent{\it Mixed action effect (Overlap)--} 
In the overlap setup, unitarity violations arise for two reasons \cite{Bar:2005tu,Chen:2007ug, PhysRevD.86.014501}. One is related to the difference in action used in the sea and valence sectors, which is expected to die out in the continuum limit \cite{Basak:2014kma}. In the same lattice setup (with $a = 0.0582$ fm), we
found this effect to be small \cite{Basak:2014kma}. The second one is related to the sea and valence charm quark mass difference, which could survive in the continuum \cite{PhysRevD.86.014501}. The difference in estimates from the HISQ valence action and the overlap valence action already reflects these systematics. Based on the disagreement between the numbers from these two scenarios in the common finest lattice, we assign a conservative estimate of 2 MeV to account for this systematic. 

\noindent{\it Finite volume--}
The elastic strong decay threshold in the $\Omega_{ccc}(3/2^{\pm})$ correspond to the three particle decay mode $\Omega_{ccc}\pi\pi$. Assuming a nonrelavistic scenario, the leading finite-volume correction in binding energy estimates was found to scale $\sim$ $e^{-\kappa L}/L$, where $\kappa=\sqrt{MB_{\infty}}$ is the binding momentum \cite{Beane:2003da,Briceno:2013bda}. Choosing $M$ to be mass of the pion (lightest hadron involved in the $\Omega_{ccc}\pi\pi$ final state) and the binding $B_{\infty}$ to be same order of magnitude as the binding $B_{FV}$ determined from the largest volume utilized, one observes the corrections on the $\Omega_{ccc}^{++}(3/2^+)$ mass to be less than 1 MeV. A similar na\"ive estimate on the binding energy of $\Omega_{ccc}^{++}(3/2^-)$ mass turns out to be of the order of 15 MeV, which is included in the associated systematics. We iterate that the na\"ive finite volume systematics quoted above for the positive parity is justified as it is significantly below the elastic strong decay threshold; however the uncertainty in the negative parity state should be addressed with more involved finite-volume analysis in future. 

\noindent{\it Electromagnetism--}
The presence of two units of charge is expected to influence the mass of a $\Omega_{ccc}^{++}$ baryon. 
A leading perturbative estimate, as shown in the supplement \cite{Suppl}, yields a correction term $<$ 5 MeV.
In Ref. \cite{Lyu:2021qsh} the Coulomb repulsion in a dibaryon, made with six charm quarks and having four units of charge, was found to be about 5 MeV. 
Given the constituent composition of $\Omega_{ccc}^{++}$ relative to that system, we anticipate that the mass correction from Coulomb effects could be comparatively smaller. Additionally, the magnetic moment of  $\Omega_{ccc}^{++}$ has been predicted to be about 1 $\mu_{N}$ \cite{PhysRevD.106.113007} suggesting no large effect on its mass. 
Since our estimation of $\Omega_{ccc}^{++}$ mass is based on the subtracted mass, $M_{\Omega_{ccc}^{++}} - 3/2M_{J/\psi}$, we also estimate the electromagnetic corrections to $M_{J/\psi}$, and find it to be 2.4 MeV (details are in supplement \cite{Suppl}). After accounting the electromagnetic mass corrections to both  $\Omega_{ccc}^{++}$ and $J/\psi$, we assign a uncertainty of $\left(^{+7.8}_{+0.0}\right)$ MeV to the mass of $\Omega_{ccc}^{++}$. However, a direct calculation with Coulomb term within the gauge action would be necessary for a more rigorous assessment of these effects, but that lies beyond the scope of the present work.

Based on the above estimation, we summarize the error budget in Table \ref{error_table} below.

\bet[h]
\centering
\renewcommand\arraystretch{1.3} 
\begin{tabular}{l|c }
$Source$ & Error (MeV)\\
\hline
Statistical & 5  \\
 Discretization & 4  \\
  Scale setting& 3   \\
  $m_c$ tuning & 2  \\
  Unphysical sea-quark & 5\\
  Taste-splitting (HISQ)& 2  \\ 
  Mixed action (Overlap)& 2  \\ 
  Finite size & 1 \\
  Electromagnetism & $^{+7.8}_{+0.0}$\\
  \hline
  Total & 5 (stat) \&  $^{+11}_{-8}$ (syst)\\
  \hline
\end{tabular}
\caption{Error budget in the calculation of the ground state mass of $\Omega_{ccc}^{++}(3/2^{+})$ baryon.\label{error_table}}
\eet{error_table}

\begin{figure*}[htb!]
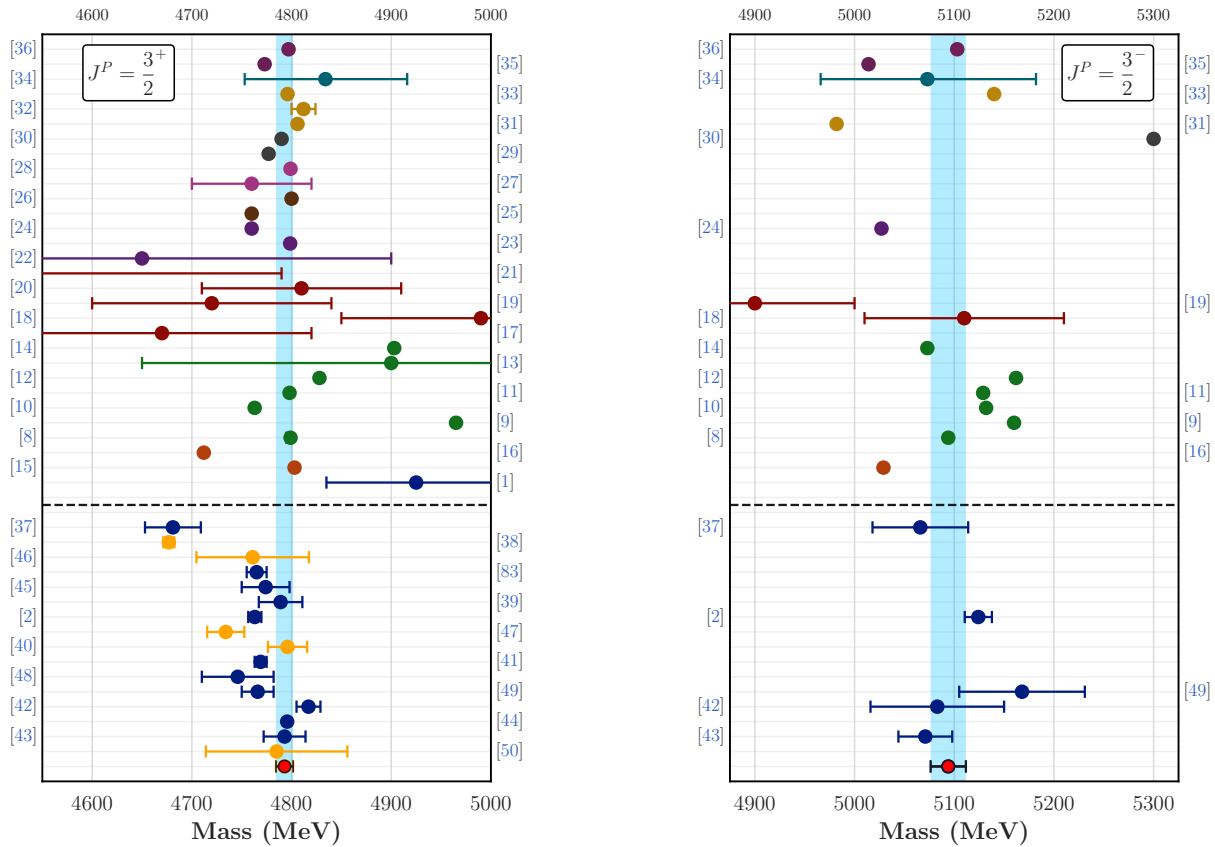
 
    \centering
    \begin{minipage}[b]{0.45\textwidth}
        \centering
        \resizebox{\linewidth}{!}{\input{Fig4a.pgf}}
    \end{minipage}
    \begin{minipage}[b]{0.45\textwidth} 
        \centering
        \resizebox{\linewidth}{!}{\input{Fig4b.pgf}} 
    \end{minipage}
    \caption{Summary of results on the ground state mass of $\Omega_{ccc}(3/2^+)$ (left) and  $\Omega_{ccc}(3/2^-)$ (right) baryons as obtained in various calculations (references on the side panels can be tracked through the faint horizontal lines). The results below the thick horizontal dashed line are from lattice calculations and above that are obtained using various models. Among the lattice results, those with orange circles are obtained with continuum extrapolations, while others (blue circles) are obtained at a single lattice spacing. The deviation of the results obtained in Refs. \cite{Alexandrou:2012xk} and \cite{Alexandrou:2014sha} 
     are most possibly due to their errors in the charm quark mass tuning. Our results at the physical point are shown by the red circles (at the bottom). }
    \label{fg:3o2}
\end{figure*}

In Fig.~\ref{fg:3o2}, we summarize the results on the ground state mass of $\Omega_{ccc}(3/2)$ baryons
showing our results (red circles at the bottom / vertical blue band) and compare it with previous results, using the blue band as reference.
A horizontal dashed line is placed to separate lattice and non-lattice results (see caption). Lattice results are also color-coded to distinguish those that include continuum extrapolation from those that do not 
({\it c.f.} see Table X in supplement \cite{Suppl}
for more details). It is evident this work clearly narrows down the broad range of estimates from previous calculations, providing the most precise lattice results to date while properly addressing the associated systematic uncertainties. 
Note a few of the earlier calculations deviates from the most recent ones. In particular, 
estimations in  Refs. \cite{Alexandrou:2012xk, Alexandrou:2014sha} differ, despite continuum and chiral extrapolations. 
We believe this is due to charm quark mistuning, which was also reflected in their estimation of $\Xi_{cc}$ mass, when compared with other lattice results and the experimental number from LHCb \cite{LHCb:2017iph}. Deviation in  
Ref. \cite{Padmanath:2013zfa} is also 
potentially due to charm quark mistuning and uncontrolled discretization errors. 
It is also worth noting that some model results align closely with the vertical line, suggesting it may be interesting to explore possible commonalities in their interaction potentials that lead to results near our lattice estimation.

\acknowledgements
This work is supported by the Department of Atomic Energy, Government of India, under Project Identification Number RTI 4002. Computations were carried out on the Cray-XC30 of ILGTI, TIFR (which has recently been closed), and the computing clusters at the Department of Theoretical Physics, TIFR, Mumbai, and IMSc Chennai. We are thankful to the MILC collaboration and, in particular, to S. Gottlieb for providing us with the HISQ lattice ensembles. We would also like to thank  Ajay Salve, Kapil Ghadiali, and T. Chandramohan for computational support. NM and DC also thank Amol Dighe for discussions.  MP gratefully acknowledges
support from the Department of Science and Technology, India, SERB Start-up Research Grant No. SRG/2023/001235. 

NSD, DC, and AR 
have contributed equally and are credited as co-first authors.

\bibliography{Omega_ccc}

\onecolumngrid
\clearpage
\onecolumngrid

\begin{center}
	{\Large \bf Supplemental material}
\end{center}

\makeatletter
\c@secnumdepth=4
\makeatother

\newif\ifsepsupp
\sepsuppfalse
\vspace{0.1in}

This supplemental material provides further information pertaining to the lattice QCD ensembles and mass extraction procedures in this study of triply charmed baryons ($\Omega_{ccc})$, along with a compilation of our results on mass splittings and continuum extrapolation. Towards the end, we also provide a consolidated table of various lattice QCD results in this regard along with various technical details involved. 

\section{Lattice Details}

We utilize six lattice QCD ensembles with dynamical $N_f = 2+1+1$ HISQ flavors ($u =d,s,c$; listed in Table \ref{Lattice_details_ov_hisq}), generated by MILC collaboration \cite{MILC:2012znn, PhysRevD.98.074512}. The ensembles $S_1-S_5$ have a spatial volume of about $(3 \,\mathrm{fm})^3$ whereas the ensemble $L_1$ has a spatial volume of approximately $(4.8\, \mathrm{fm})^3$. The number of measurements ($n_{meas}$) for the Overlap and HISQ setups are shown in the last two columns respectively. The lattice spacings were measured initially with $r_1$ parameter \cite{MILC:2012znn},  which were found to be consistent with scales obtained through Wilson flow \cite{MILC:2015tqx}. In our previous works using overlap valence action, we also have determined lattice spacings independently using $\Omega_{sss}$ mass \cite{Basak:2013oya,Mathur:2018epb} and found those to be consistent with lattice spacings shown in the column for the study based on overlap valence action. The bare charm quark mass with overlap action was tuned by equating the kinetic mass of the spin average of the 1S charmonia to its physical value following the lattice spacing determination in Refs. \cite{MILC:2012znn,MILC:2015tqx,PhysRevD.98.074512}.  As for the study using HISQ valence quarks, we utilize the valence charm quark mass as listed in Ref.~\cite{HPQCD_2020}, except for the ensemble $S_5$, for which we determine it independently by equating the $J/\psi$ mass with its physical value. Naturally we consider the lattice spacings as determined in Ref.~\cite{HPQCD_2020} for the study with HISQ valence quarks. The resulting uncertainties in mass splitting due to variations in lattice spacings between the different approaches were found to be significantly smaller than the statistical errors, and they remain within a few MeV. We have accounted for these differences in the scale setting uncertainty. 

\begin{table}[h!]
    \centering
    \renewcommand\arraystretch{1.2}  
    \addtolength{\tabcolsep}{3pt}    
    \begin{tabular}{|c|c|c|c|c|c|c|}
     \hline\hline 
      \multirow{2}{*}{Ensemble} & Dimension & \multicolumn{2}{c|}{Lattice Spacing ($a$ fm)} & \multirow{2}{*}{$M_{\pi}^{sea}$ (MeV)\cite{PhysRevD.98.074512}} & \multicolumn{2}{c|}{\(n_{\text{meas}}\)} \\ \cline{3-4} \cline{6-7}
      & $N_s^3\times N_t$ & Overlap \cite{MILC:2012znn,MILC:2015tqx,PhysRevD.98.074512} & HISQ \cite{HPQCD_2020} & & Overlap & HISQ   \\ 
     \hline    
      $S_1$ & $24^3 \times 64$ & $0.1207 (11)$ & $0.12404(67)$ & $305$ & $294$ & $1400$ \\
      $S_2$ & $32^3 \times 96$ & $0.0888 (8)$ & $0.09023(48)$ & $316$ & $188$ & $396$ \\
      $S_3$ & $48^3 \times 144$ & $0.0582 (4)$ & $0.05926(33)$ & $329$ & $186$ & $386$  \\
      $S_4$ & $64^3 \times 192$ & $0.0441 (2)$ & $0.04406(27)$ & $315$ & $142$ & $400$  \\
      $S_5$ & $96^3 \times 288$ & $-$ & $0.03271(20)$ & $309$ & $-$ & $451$ \\ 
      $L_1$ & $40^3 \times 64$  & $0.1189 (10)$ & $0.12225(64)$ & $216$ & $100$ & $200$  \\
     \hline\hline
    \end{tabular}
    \caption{Details of the lattice QCD ensembles utilized for calculations with Overlap and HISQ valence quarks.}
    \label{Lattice_details_ov_hisq}
\end{table}

\vspace*{-0.1in}

\section{Mass extraction with Overlap fermions}

We utilize the operator structure proposed in Ref.~\cite{Basak:2005ir} for the $\Omega_{ccc}^{++}$ baryon in building the correlation matrices. Owing to the inherent color antisymmetric and trivially flavor symmetric structure of the $\Omega_{ccc}^{++}$ baryon, only symmetric total spin ($J$) structures are allowed by the total antisymmetric wavefunction for local $\Omega_{ccc}^{++}$ baryon operators. This constrains the allowed total spin-parity combination to the symmetric spin state is $J^P = 3/2^+$, which reduces to the $H^+$ irreducible representation (irrep) of the cubic group on the lattice \cite{Johnson:1982yq}. With the full four component Dirac spinor, there are two possible embeddings of the $H^+$ irrep, as listed in Table \ref{tab:1H_irrep} in the Dirac-Pauli representation of $\gamma$-matrices. This setup ensures that the symmetries of the baryon are faithfully represented in our lattice calculations.

\begin{table}[htbp]
    \centering
    \renewcommand\arraystretch{1.2}  
    \addtolength{\tabcolsep}{3pt}    
    \begin{tabular}{c||c|c||c|p{0.23\textwidth}}
        $S_z$ & Op. [N] & Spin & Op. [R] & Spin \\
        \hline
        +3/2 & $^1H_{+3/2}$ & \{111\}$_S$ & $^2H_{+3/2}$ & \{133\}$_S$ \\
        +1/2 & $^1H_{+1/2}$ & \{112\}$_S$ & $^2H_{+1/2}$ & \{233\}$_S$ + \{134\}$_S$ + \{143\}$_S$ \\
        $-$1/2 & $^1H_{-1/2}$ & \{122\}$_S$ & $^2H_{-1/2}$ & \{144\}$_S$ + \{234\}$_S$ + \{243\}$_S$ \\
        $-$3/2 & $^1H_{-3/2}$ & \{222\}$_S$ & $^2H_{-3/2}$ & \{244\}$_S$ \\
    \end{tabular}
    \caption{Allowed spin structure of local single flavored baryon operator with the full four component Dirac spinor for quarks. $S_z$ refers to the third component of the baryon spin in the infinite volume continuum. $^1H^+$ refers to the nonrelativistic [N] and $^2H^+$ to the relativistic [R] embedding, with \{xyz\}$_S=xyz+yzx+zxy$ representing symmetrized spin structure of the baryons, where $x,~y,~z$ refer to the Dirac spinor components of the quark.}
    \label{tab:1H_irrep}
\end{table}

In the Dirac-Pauli representation of the $\gamma$ matrices \cite{sakurai1967advanced}, the solutions of the Dirac equation suggest the lower two components (3,4) to vanish in the nonrelativistic limit, where the velocity goes to zero. In this sense, the upper two components that survive the nonrelativistic limit are referred to as nonrelativistic, and the lower two as relativistic. Within this $\gamma$ matrix representation, one of the embedding denoted by $^1H^+$ is purely built out of the upper two components and hence is referred to as the nonrelativistic embedding. The second embedding denoted by $^2H^+$ carries lower components of the quark spinor reflecting its relativistic nature and hence is referred to as relativistic. 

\begin{figure*}[htb!]       
\includegraphics[width=0.45\textwidth]{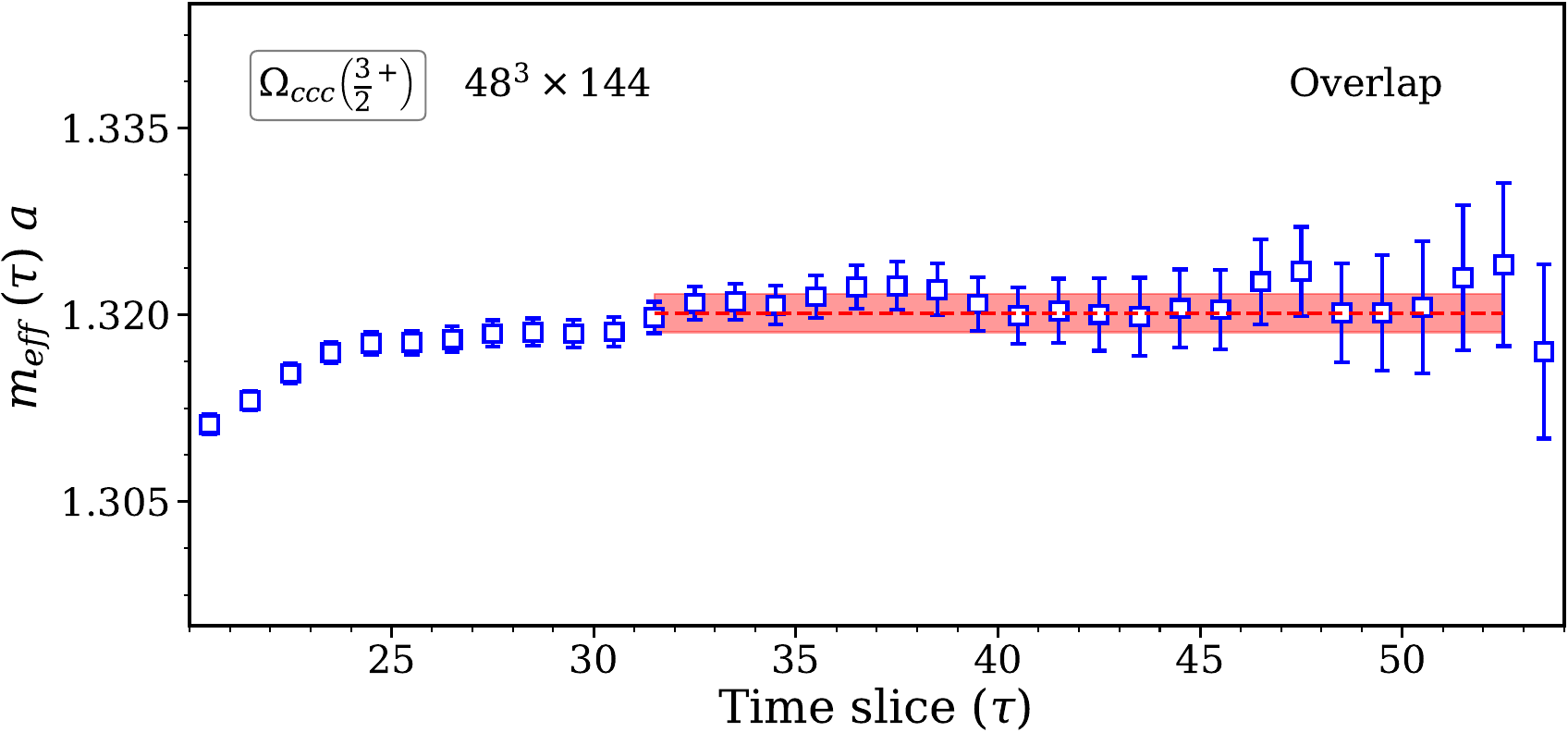} 
\includegraphics[width=0.45\textwidth]{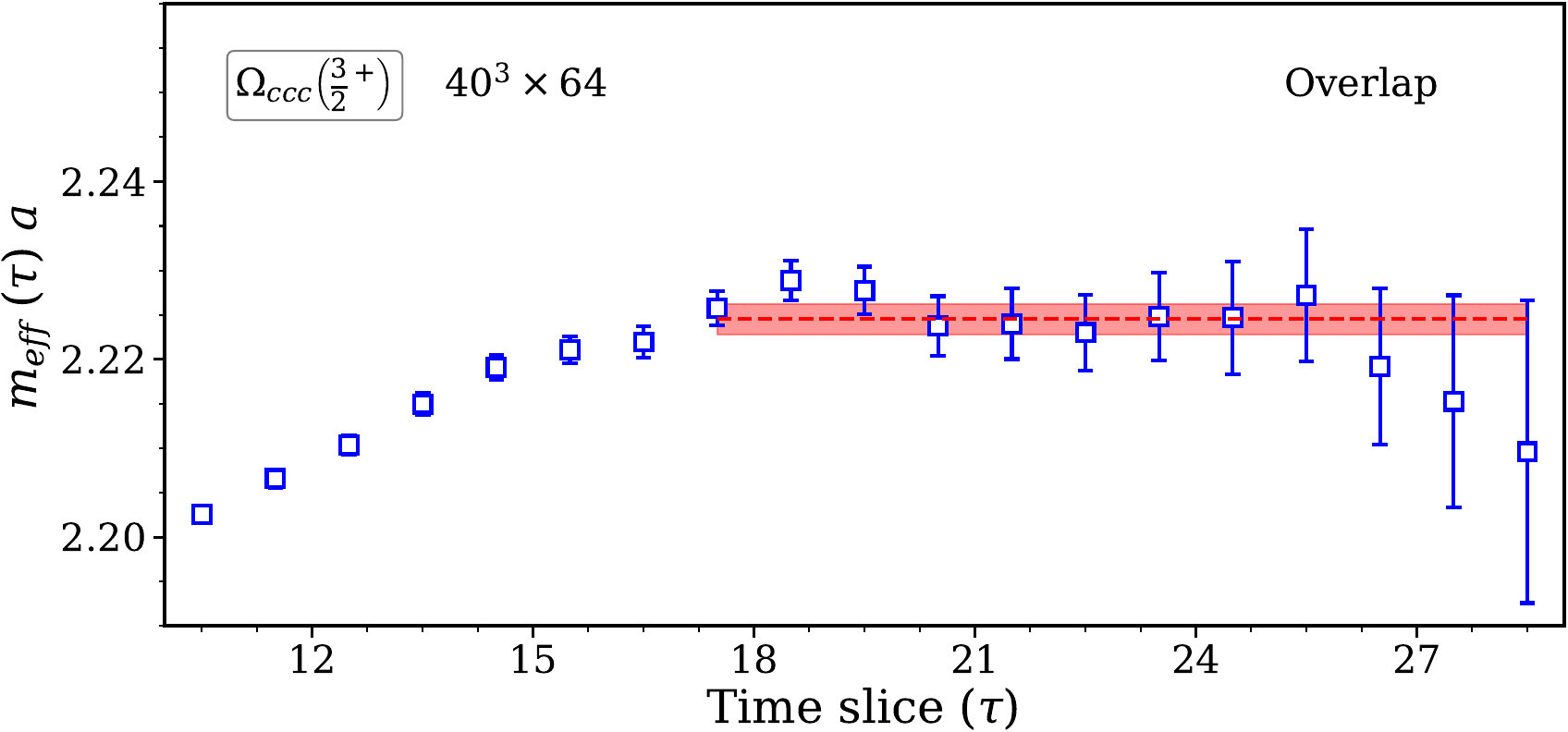} 
\includegraphics[width=0.45\textwidth]{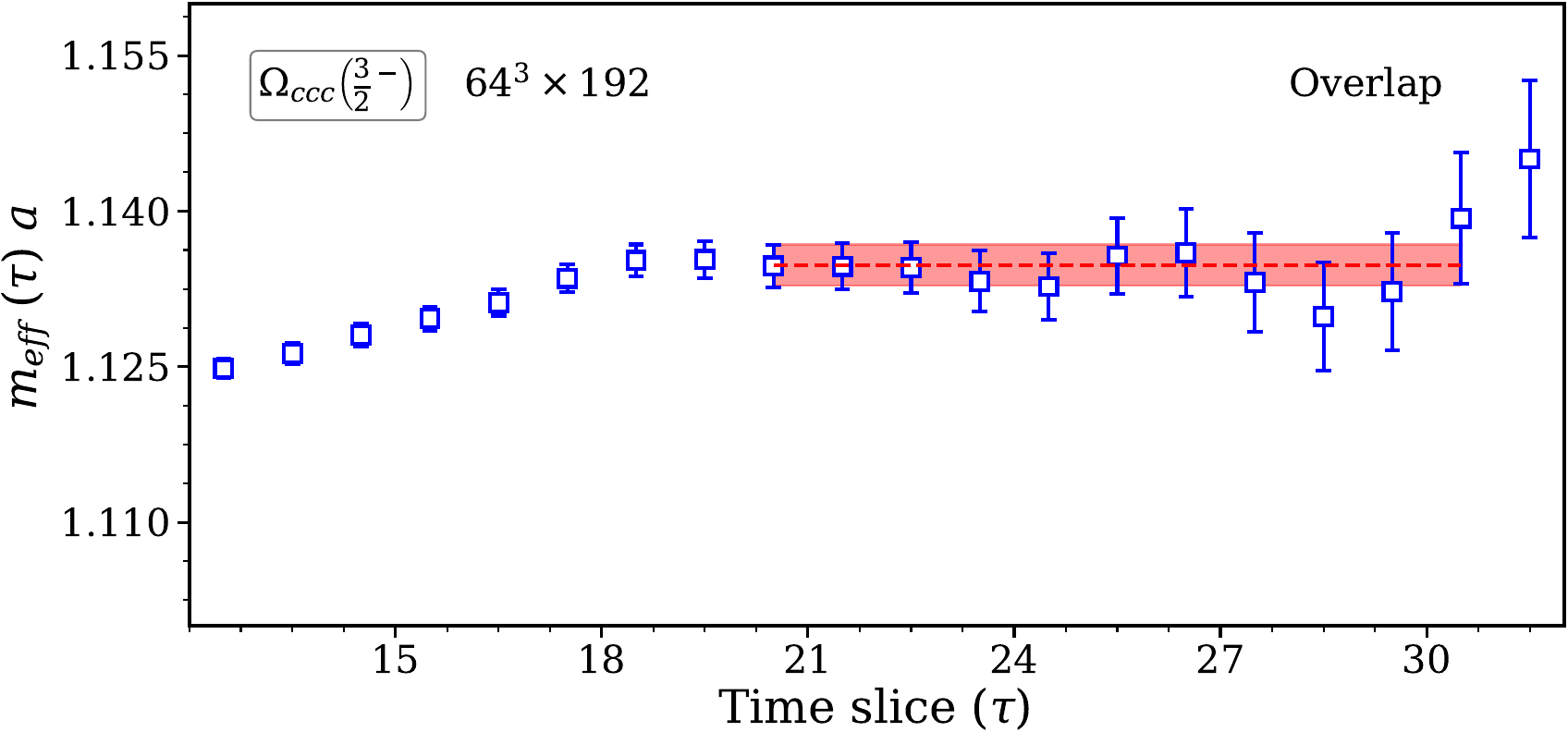} 
\includegraphics[width=0.45\textwidth]{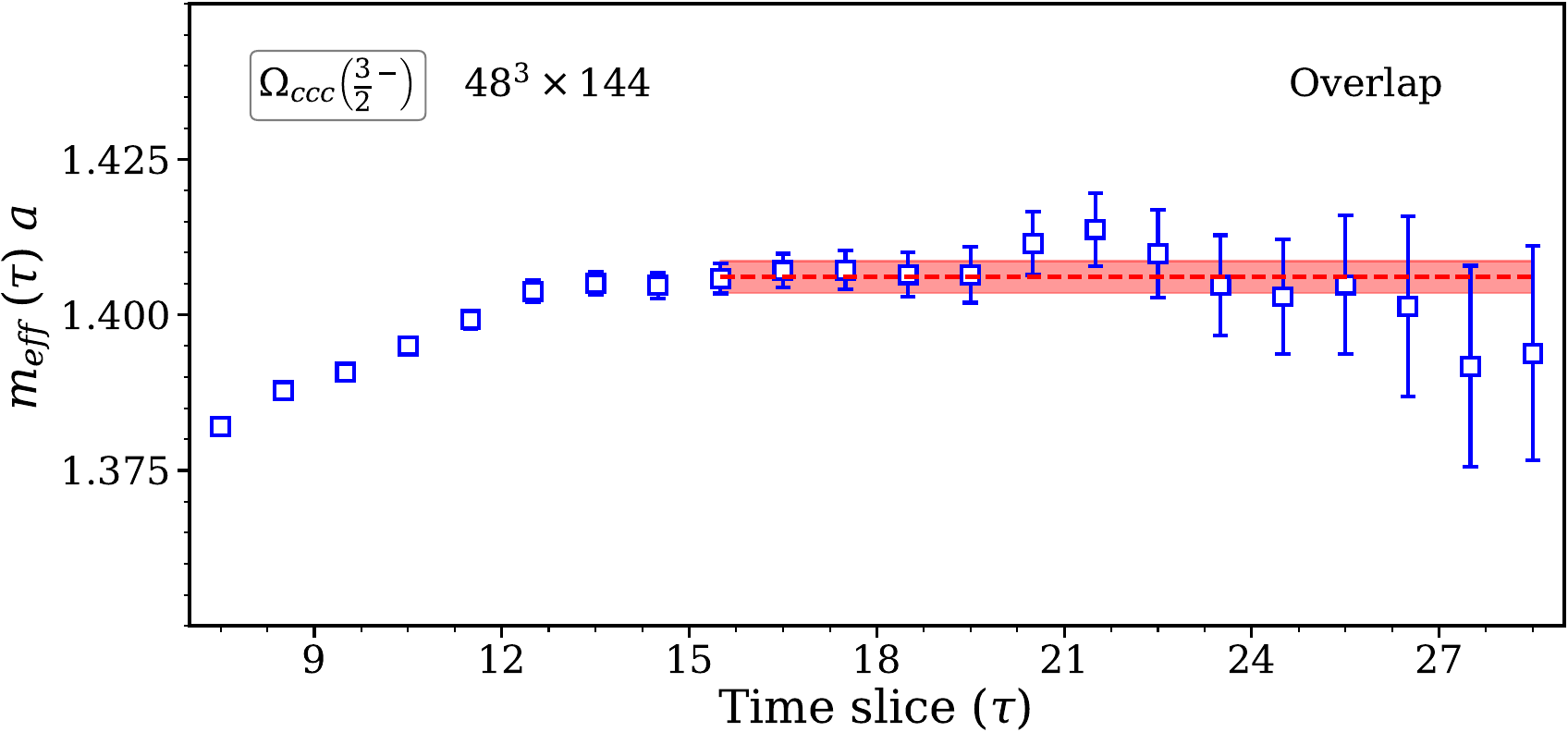} 
        \caption{Effective mass plots corresponding to the positive (top) and negative (bottom) parity lowest energy levels of $\Omega_{ccc}(3/2)$ baryons for different lattice ensembles. The dashed lines are the fit results with one exponential fits, and the bands show the fit ranges with the corresponding $1\sigma$ errors.}
    \label{fg:omega_eff_ov}
\end{figure*}

We utilize operators from both these embeddings to form an operator basis and construct $2 \times 2$ correlation matrices for baryons. The correlation matrices are built using quark propagators evaluated using a wall smearing at the source timeslice with and point smearing at the sink timeslice. We analyze these correlation matrices variationally following the solution of a generalized eigenvalue problem (GEVP) and subsequent numerical fits with the expected asymptotic forms for the eigenvalue correlators given by 
\begin{equation}
C_{ij}(t) v_{jn} = \lambda_n(t, t_0) C_{ij}(t_0) v_{jn} \mbox{      and  } \lim_{t-t_0\rightarrow \infty}\lambda_n(t,t_0) = A~e^{-m_n(t - t_0)}.
\end{equation}
The eigenvalue correlators display a few timeslices early saturation of ground state domination when compared to the behavior in a single correlation function and hence aid in a more reliable energy extraction of the ground state mass. $t_0$ is chosen as a timeslice where the ground state energy plateauing starts as one increases the time interval. We present the effective masses and energy fit estimates for the positive and negative parity $\Omega_{ccc}$ baryon ground state in different ensembles in Fig. \ref{fg:omega_eff_ov} demonstrating the signal quality. The wall-source to point-sink setup leads to an asymmetric correlation function that can have nonpositive definite spectral weights from different excited levels in the spectrum. This can result in a rising-from-below behaviour in the effective masses, as can be observed in Fig. \ref{fg:omega_eff_ov}, instead of a falling-from-above expected in a symmetric correlation function that has all positive definite spectral weights.

\begin{figure*}[htb!]        
\includegraphics[scale=0.28]{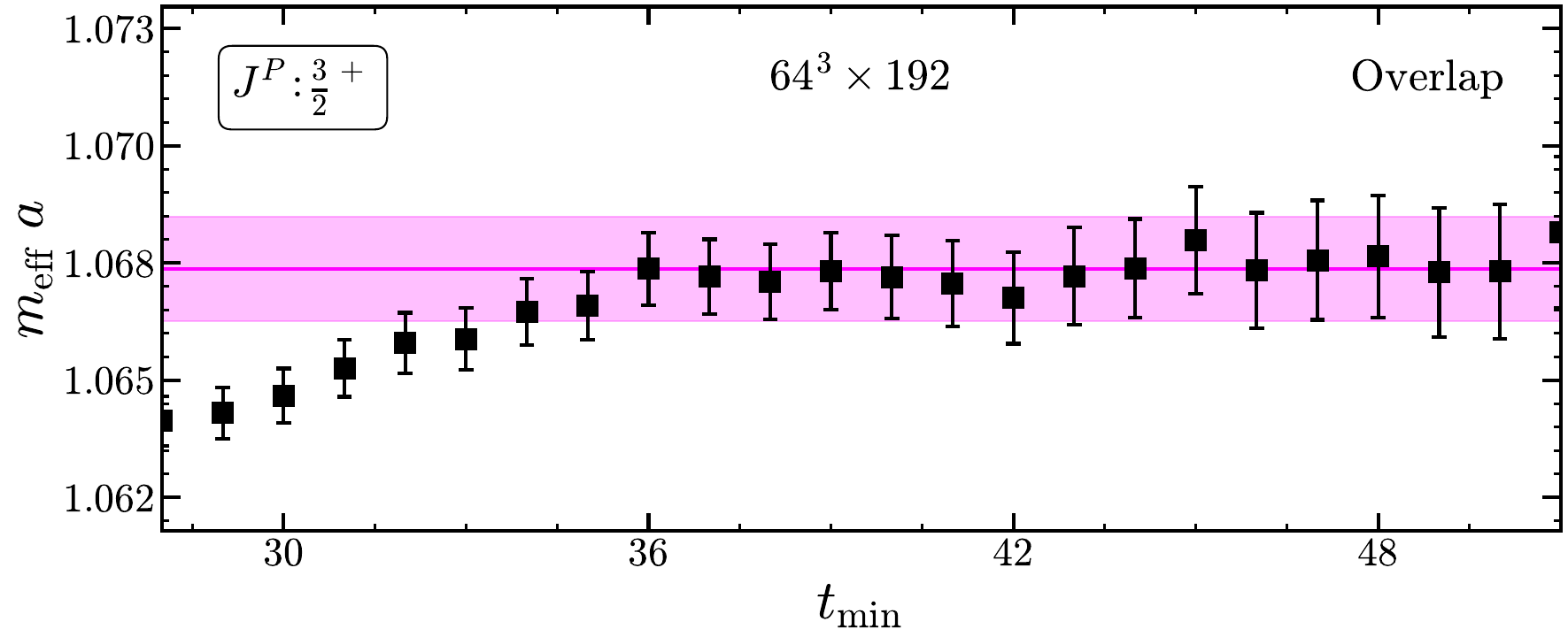} 
\includegraphics[scale=0.28]{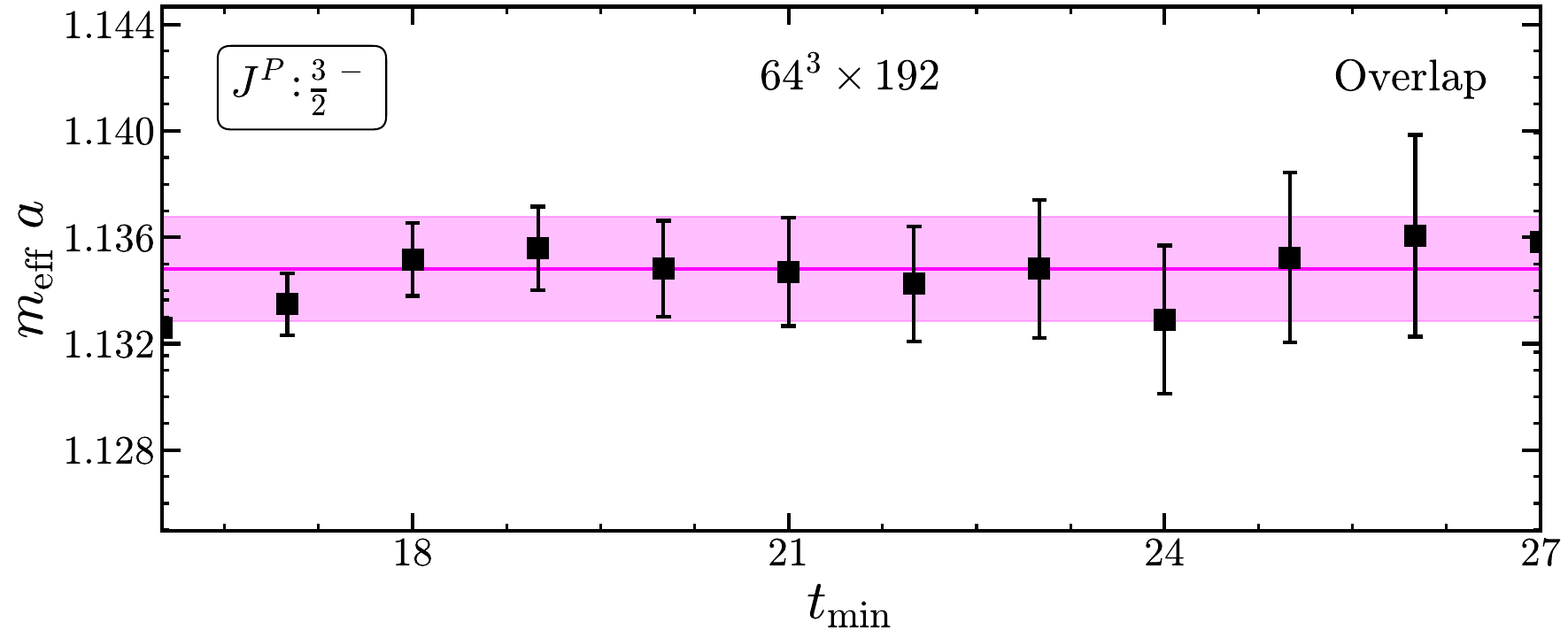}

\caption{Fit results for the lowest energy levels of $\Omega_{ccc} (\frac{3}{2})$ baryons in both positive (left) and negative (right) parity channels, obtained using fit windows with varying minimum time $t_{\text{min}}$. The black points with error bars represent the central values and $1\sigma$ uncertainties of the fitted ground-state energy across different fit windows. The light magenta band represents the quoted fit result with $1\sigma$ uncertainty, incorporating the uncertainty from the fitting window selection.}

\label{fg:omega_eff_tmin_overlap}
\end{figure*}

\begin{figure*}[htb!]        
\includegraphics[width=8.8cm, height=3.5cm]{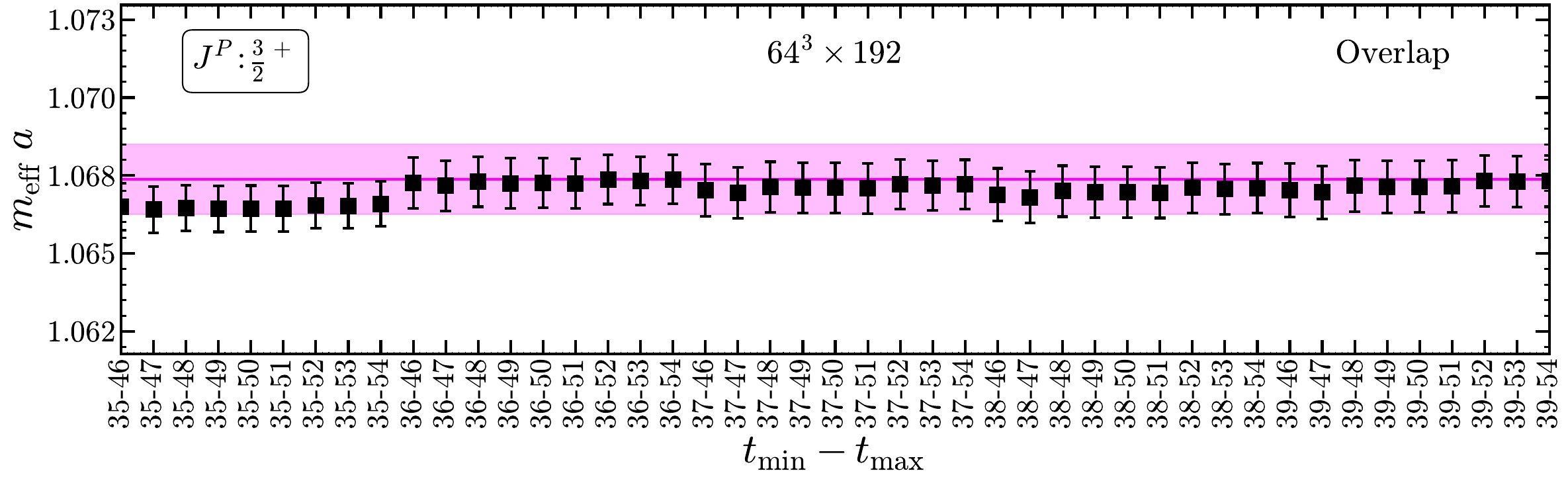} 
\includegraphics[width=8.8cm, height=3.5cm]
{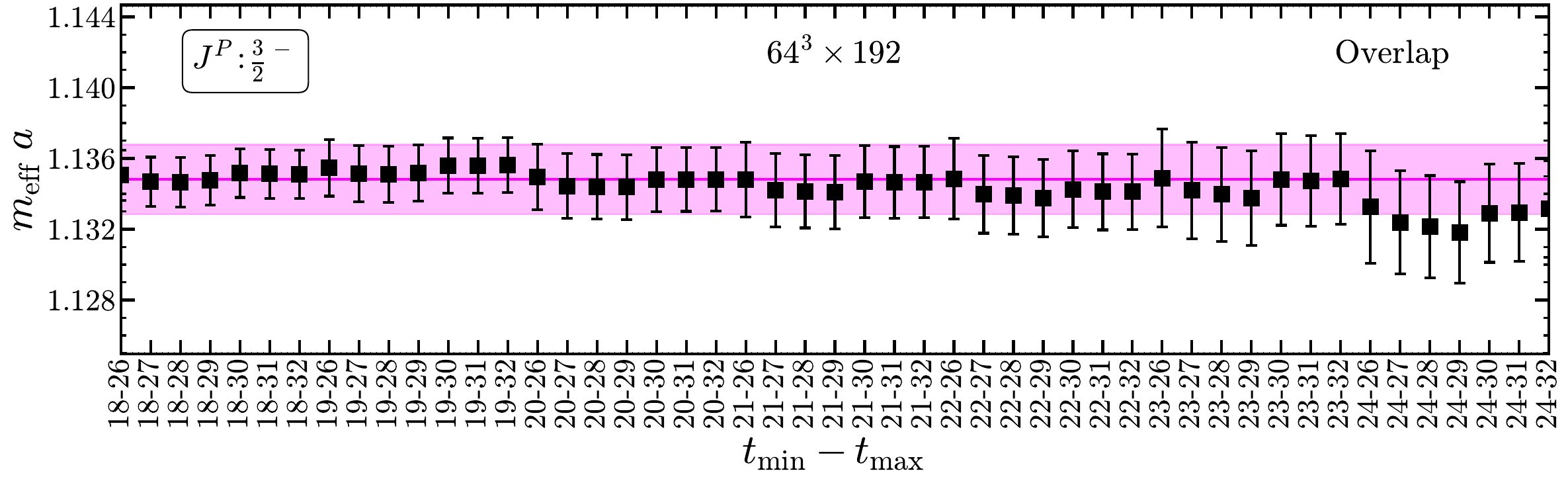}

\caption{Fit results for the lowest energy levels of $\Omega_{ccc} (\frac{3}{2})$ baryons in both positive and negative parity channels, obtained using fit windows with varying minimum and maximum time i.e. $t_{\text{min}}$ and $t_{\text{max}}$. The black points with error bars represent the central values and $1\sigma$ uncertainties of the fitted ground-state energy across different fit windows. The light magenta band represents the quoted fit result with $1\sigma$ uncertainty, incorporating the uncertainty from the fitting window selection.}

\label{fg:omega_eff_tmin_tmax_overlap}
\end{figure*}

One of the consequences of the use of asymmetric setup is that one loses control on excited state energies, although the asymptotic value representing the ground state energy is expected to remain unaffected. We perform a careful estimation of the true ground state energy. We begin by selecting $t_{\max}$ to be as large as possible while maintaining a good signal quality. Then, we vary $t_{\min}$ across a suitable range to obtain fit estimates for different $t_{\min}$ choices. The optimal $t_{\min}$ is determined based on the quality of the fit, evaluated using the goodness-of-fit ($\chi^2$) and the stability of the extracted mass. Specifically, we choose $t_{\min}$ where the effective mass shows minimal variation beyond that point, indicating a well-defined plateau. We consider neighboring fit values and incorporate the observed variations as a systematic error to account for systematic uncertainties arising from the fitting window selection. In Figs. \ref{fg:omega_eff_tmin_overlap} and \ref{fg:omega_eff_tmin_tmax_overlap}, we present representative plots showing the dependence of the fit results on $t_{\min}$ alone and on both $t_{\min}$ and $t_{\max}$, along with the final estimate, incorporating the uncertainty determined from a correlated average of the fluctuations in energy estimates around the chosen $t_{min}$ (indicated by the magenta bands). Additionally, we also ensure the consistency in the plateau saturation across different sink smearing setups as proposed in Ref. \cite{Hudspith:2020tdf} that approaches symmetric setup and thus assuring the robustness in the identified ground state energy plateaus. These results confirm the consistency and robustness of our final choices. 

\begin{figure*}[htb!]       
\includegraphics[width=0.45\textwidth]{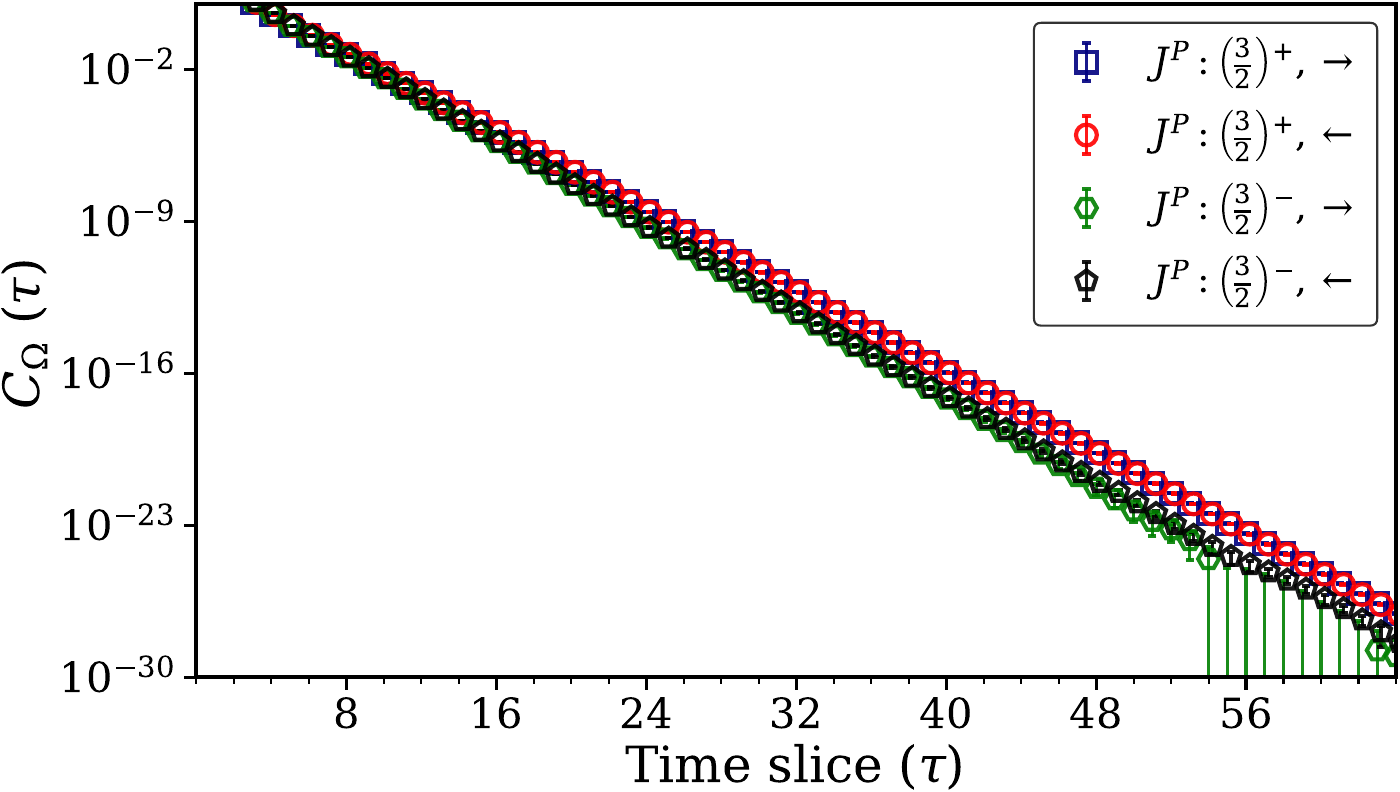} 
\includegraphics[width=0.45\textwidth]{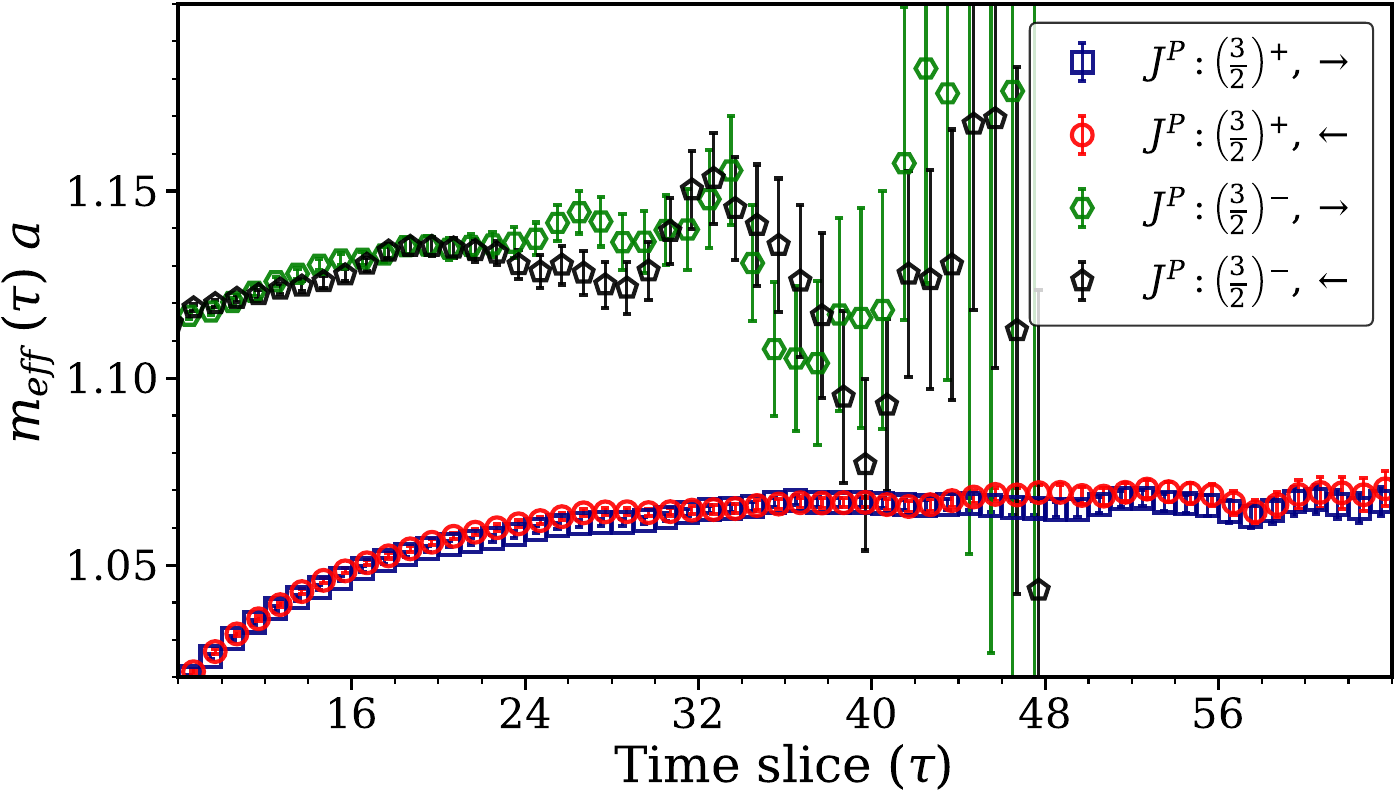} 
        \caption{Correlators and corresponding effective masses for $\Omega_{ccc}$ baryons on the $48^3 \times 144$ lattice ensembles. In the legend, ($\rightarrow$) indicates forward propagation, while ($\leftarrow$) represents backward propagation (time inversed to follow a forward propagation for the sake of comparison). }
    \label{fg:operator_comp}
\end{figure*}

Baryon correlators are typically noisier compared to that for mesons. One way to improvise the signal quality in baryon correlators is by utilizing their antisymmetric nature (for fermions) along the temporal direction. As a result of this feature, baryon correlators will carry signals for positive parity states in the forward direction from the source timeslice, whereas in the backward direction from the source timeslice, signals for negative parity baryon states with the same total spin dominate \cite{Montvay:1994cy, Datta:2012fz, Padmanath:2014arv}. We take advantage of this property to investigate the lightest spin 3/2 $\Omega_{ccc}$ baryons with both parities and also to increase the statistics by investigating forward/backward propagations of correlation matrices in the $H^+$ and $H^-$ irreps. The interpolators in the $H^-$ irrep can be built from a parity transformation to the $H^+$ irrep interpolators, which amounts to an interchange of Dirac spinor indices $(1\leftrightarrow3)$ and $(2\leftrightarrow4)$ in the Dirac-Pauli representation of $\gamma$ matrices. The large time behaviour of forward propagating correlators in the $H^+$ and $H^-$ irreps can be expressed as \cite{Padmanath:2014arv}:
\begin{eqnarray}
C_{H^+}(\tau) &\propto c_+ e^{-m_{B+} \tau} + c_- e^{-m_{B-} (N\tau - \tau)},\nonumber\\
C_{H^+}(\tau) &\propto c_- e^{-m_{B-} \tau} + c_+ e^{-m_{B+} (N\tau - \tau)}.
\label{eqn:corr_op}
\end{eqnarray}
Here, $c_+$ and $c_-$ are normalization factors in the correlator, while $m_{B+}$ and $m_{B-}$ are the masses associated with the positive and negative parity baryon states. In Fig. \ref{fg:operator_comp}, we present the forward and backward (time inversed to follow a forward propagation for the sake of comparison) propagating correlation functions (left) in the $H^+$ and $H^-$ irreps and the respective effective masses (right). It can be seen that the forward propagating correlator in the $H^+$ irrep matches with the backward propagating correlator in the $H^-$ irrep, whereas the forward propagating correlator in the $H^-$ irrep matches with the backward propagating correlator in the $H^+$ irrep. These two cases represent the positive and negative parity $\Omega_{ccc}$ spin 3/2 baryon respectively. This consistency is transparent in the effective masses presented in the figure. As one would naively expect, it can also be seen that the negative parity state has a relatively heavier effective mass, which can also be inferred from the steeper slope in the correlation function plotted on the left. 

\vspace*{0.1in}

\section{Mass extraction with HISQ fermions}
We utilize six $N_f = 2+1+1$ dynamical MILC lattice ensembles \cite{MILC:2010pul, MILC:2012znn, PhysRevD.98.074512} details of which are provided in Table \ref{Lattice_details_ov_hisq}. We  use fat$7$ smeared gauge links with Lepage term in the fermion action \cite{Follana:2006rc,MILC:2009mpl}.  We also utilize the three-link (Naik) term in the HISQ fermion action for the valence quarks. The correction to the Naik term is tuned using dispersion relations of $\eta_c$ and $J/\psi$ mesons. The values of the lattice spacing for these ensembles are taken from Ref. \cite{HPQCD_2020}. 
As in Ref. \cite{HPQCD_2020}, on each of these ensembles, we tune the charm quark mass by equating lattice extracted $J/\psi$ meson mass with its experimental value.

To extract the ground state masses of $\eta_c$ and $J/\psi$ mesons, we use local $0-$link spin-taste naive quark bilinears \cite{Follana:2006rc, Golterman:1985dz}. The relevant operators utilized in this work are the following,
\begin{eqnarray}
    \mathcal{O}^{\eta} = \bar{c}\gamma_5\otimes\xi_5 c\; , \quad \mathcal{O}^{J/\psi}_i = \bar{c}\gamma_i\otimes\xi_i c\; ,
    \label{MesonOP}
\end{eqnarray}
where the $\gamma$ and $\xi$ matrices act on spinor and taste degrees of freedom of the $c$ quark field, respectively. For $\Omega_{ccc}$ baryons,   we utilize the following operator \cite{GoltermanBaryon},
\begin{eqnarray}
    \mathcal{O}_{\Omega_{ccc}}(t) =  \epsilon_{abc} D_1c^a(\mathbf{x},t)D_2c^b(\mathbf{x},t)D_3c^c(\mathbf{x},t) \; .\label{OmegaOp}
\end{eqnarray}
Here, $c^a(\mathbf{x},t)$ is the staggered charm quark field and $a$, $b$ and $c$ are the color indices and the action of $D_i$ operators on $c^a(\mathbf{x},t)$ amount to, $D_ic^a(\mathbf{x},t) = \frac{1}{2}\left(c^a(\mathbf{x}-\hat{i},t)+c^a(\mathbf{x}+\hat{i},t)\right) $. The flavor wavefunction of $\mathcal{O}_{\Omega_{ccc}}$ is fully symmetric, while the color wavefunction is totally antisymmetric. This operator transforms in the $8^{\prime}$ irreps of geometric time slice (GTS) group and lies in the $A_2^{-}$ irreps of the octahedral group $(\mathcal{O}_h)$. 
Details about this operator can be found in Refs. \cite{GoltermanBaryon, Bailey:2006zn}. With HISQ quarks this operator couples to $\Omega$ baryons with two different tastes, and they are degenerate only in the continuum limit. At a finite lattice spacing, this leads to a
difference in $\Omega_{ccc}$ baryons masses with a splitting of $\mathcal{O}(\alpha_s(1/a)(m_ca)^2)$. This systematic can be accounted for by incorporating this term in the continuum extrapolation.

\subsection{Extracting the Ground State Masses}
We extract the masses of 1S charmonia  $(\eta_c,\;J/\psi)$ and $\Omega_{ccc}$ baryons by studying the exponential decays of their respective two point correlation functions, 
\begin{eqnarray}
    C^{2\textit{-pt}}_{\mathcal{O}}(t) &=& \sum_{\mathbf{x}}\langle0\vert\mathcal{O}(\mathbf{x},t)\overline{\mathcal{O}}(\mathbf{x}_{src},t_{src})\vert0\rangle\;,\label{omega_2pt}
\end{eqnarray}
at large source-sink separation. Here, $\mathcal{O}$ represent their respective interpolating operators as discussed above. 
As in the overlap case, we use a wall-source point-sink setup.
The standard procedure of staggered quarks with relevant phase factors is followed.

By introducing a complete set of states in between two interpolating operators at the source and sink, one can show that the above two point correlation function simplifies to,
\begin{eqnarray}
    C^{2\textit{-pt}}(t) = \sum_{i=0}^{\infty}\vert \mathcal{Z}_i\vert^2(e^{-E_i\tau}+e^{-E_i(n_t-\tau)}) -  \sum_{i=0}^{\infty}(-)^{\tau}\vert \mathcal{Z}^{o}_i\vert^2(e^{-E^{o}_i\tau}+e^{-E^{o}_i(n_t-\tau)}).\label{specdecomp}
\end{eqnarray}
Here,  
$i$ runs over all possible finite volume eigenstates with increasing energy that can couple to their respective interpolating fields. The superscript ``$o$" in $E_i^o$ of the second term represents the oscillating contributions from the opposite parity states.
The time-oscillating terms 
in the above equation appear
as the staggered interpolating operators also couple to their respective opposite parity states with a phase oscillation in time \cite{Follana:2006rc, Golterman:1985dz}. 
At a large source-sink separation ($\tau = t - t_{src}$), the ground state ($i = 0$) dominates the correlation function with a functional form, 
\begin{eqnarray}
    C^{2\textit{-pt}}(t) \approx \vert \mathcal{Z}_0\vert^2(e^{-E_0\tau}+e^{-E_0(n_t-\tau)}) -  (-)^{\tau}\vert \mathcal{Z}^{o}_0\vert^2(e^{-E^{o}_0\tau}+e^{-E^{o}_0(n_t-\tau)}),\quad \textrm{for} \quad \textit{$t\gg t_{src}$}\;.\label{largeT}
\end{eqnarray}
As in Ref. \cite{DeTar:2014gla}, to determine the ground state mass of a single hadron, we employ the following strategy. To mitigate the oscillating phase, we first smooth out the two-point correlation function with appropriate time-shiftings, leading to the following asymptotic form, 
\begin{eqnarray}
        \overline{C^{2\textit{-pt}}}(t)^2 \approx \vert \mathcal{Z}_0 \vert^4 e^{-2E_0 \tau} \left(1 + (-)^{\tau} \mathcal{R} (1 - \cosh{\Delta})e^{-\Delta \tau} + \mathcal{O}\left(e^{-2\Delta \tau}\right)\right), \quad t \gg t_{src}. \label{check1}
\end{eqnarray}
Here, $\overline{C^{2\textit{-pt}}}(t)$ is the smoothed correlation function, \( \Delta = E_0^o - E_0 \), and \( \mathcal{R} = \frac{\vert \mathcal{Z}_0^o \vert^2}{\vert \mathcal{Z}_0 \vert^2} \). Since $\Delta > 0$ and large, the oscillating term in Eq. (\ref{check1}) gets highly suppressed beyond \( t \geq 15 \). 
For our fittings, we typically use  $\tau \geq 15$ for the ensembles S1 and S2, while those are $\geq 25$ for the ensembles S3, S4, and S5.

Next, we construct a correlation  matrix using these time-shifted smoothed correlation functions \cite{Aubin:2010jc},
    \begin{eqnarray}
        \mathbb{C}(t) = 
        \begin{pmatrix}
            \overline{C^{2\textit{-pt}}}(t) && \overline{C^{2\textit{-pt}}}(t+1) \\
            \overline{C^{2\textit{-pt}}}(t+1) && \overline{C^{2\textit{-pt}}}(t+2) 
        \end{pmatrix}\;.\label{TransferExple}
    \end{eqnarray}
    With such a correlation matrix, we then follow the standard GEVP procedure \cite{Michael:1985ne, Luscher:1990ck} to extract the lowest energy levels for the positive and negative parity $\Omega_{ccc}$ baryons.

\begin{figure*}[htb!]        \includegraphics[scale=0.28]
{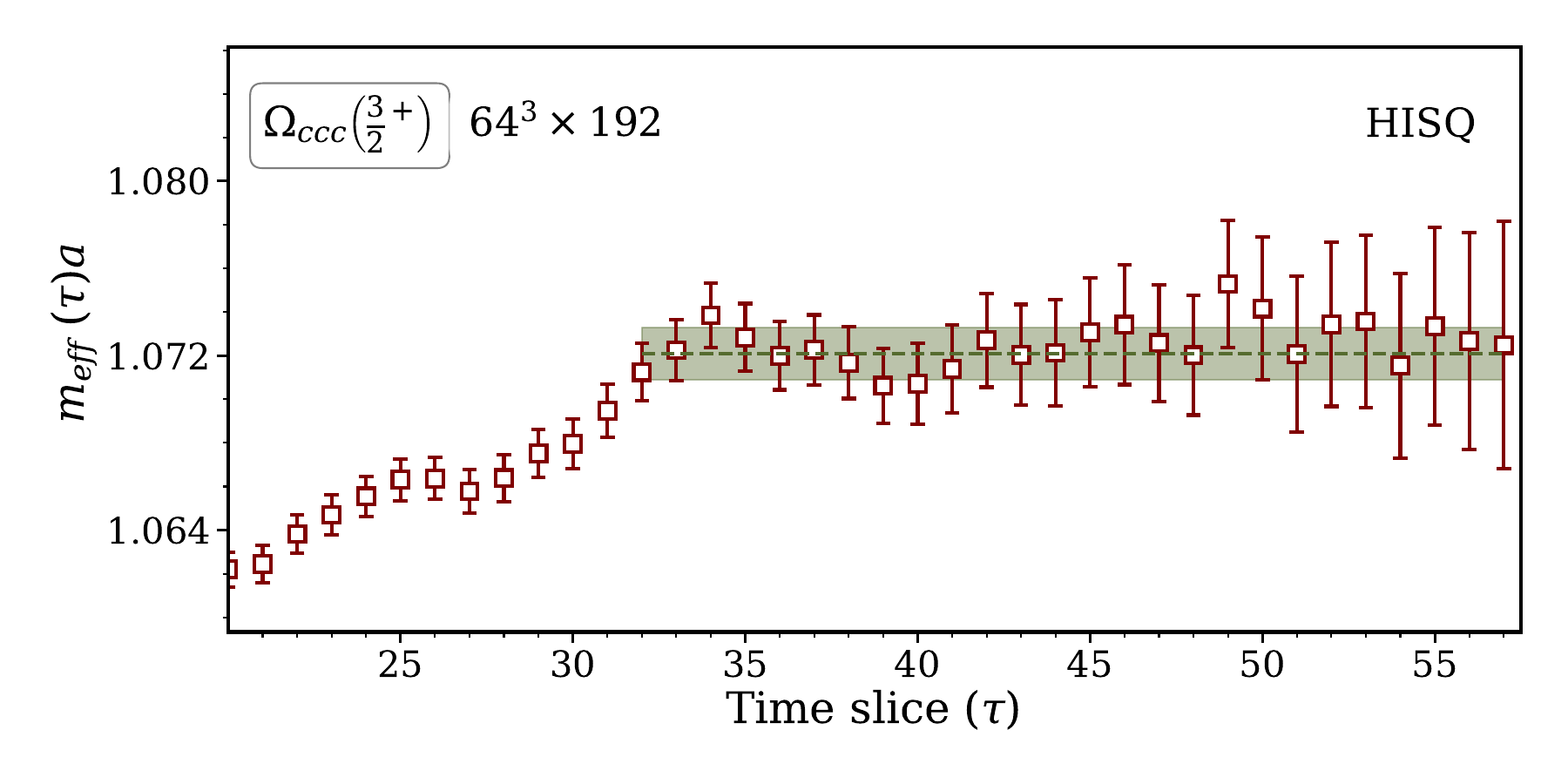} 
\includegraphics[scale=0.28]{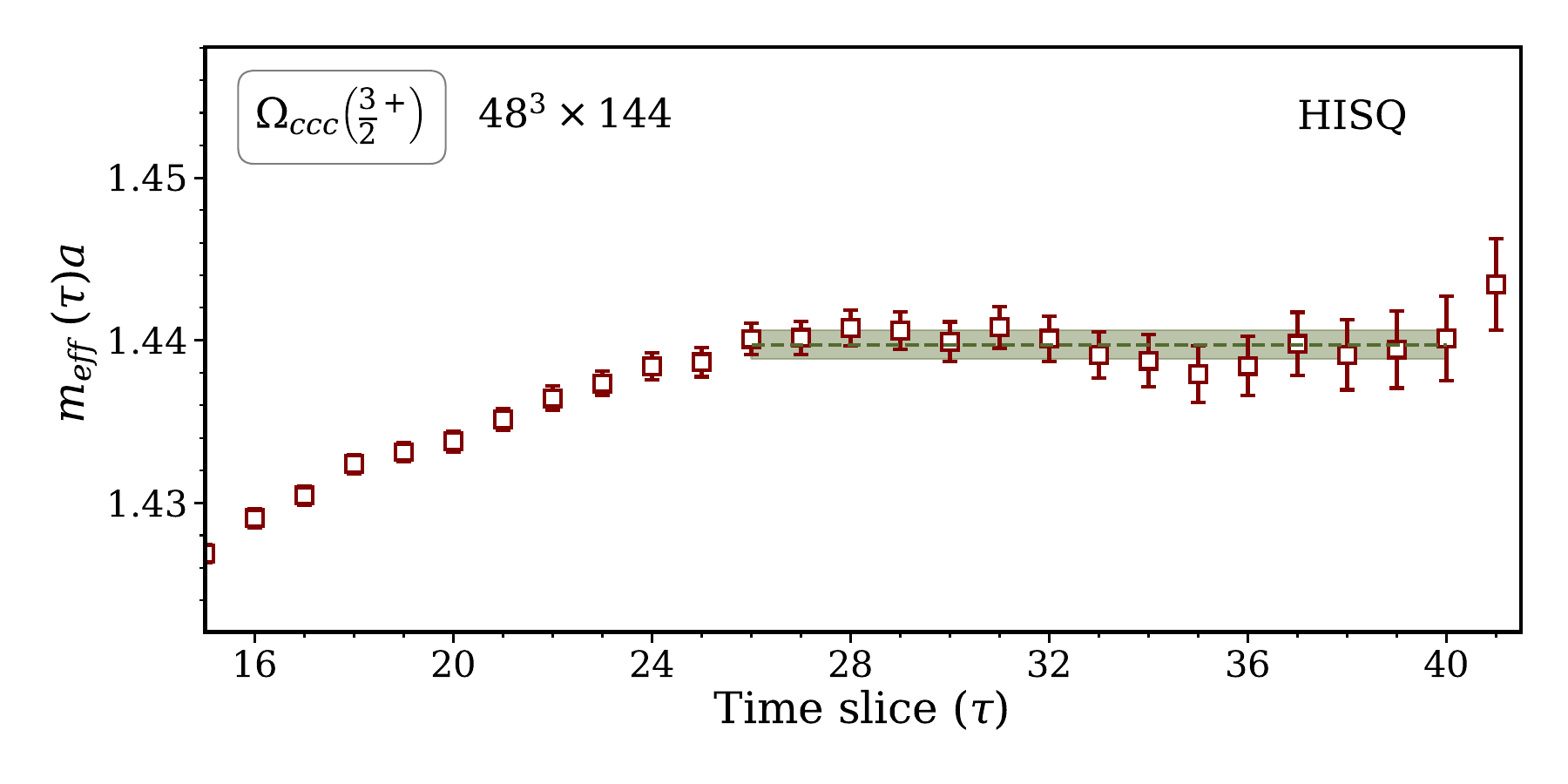}
\includegraphics[scale=0.28]{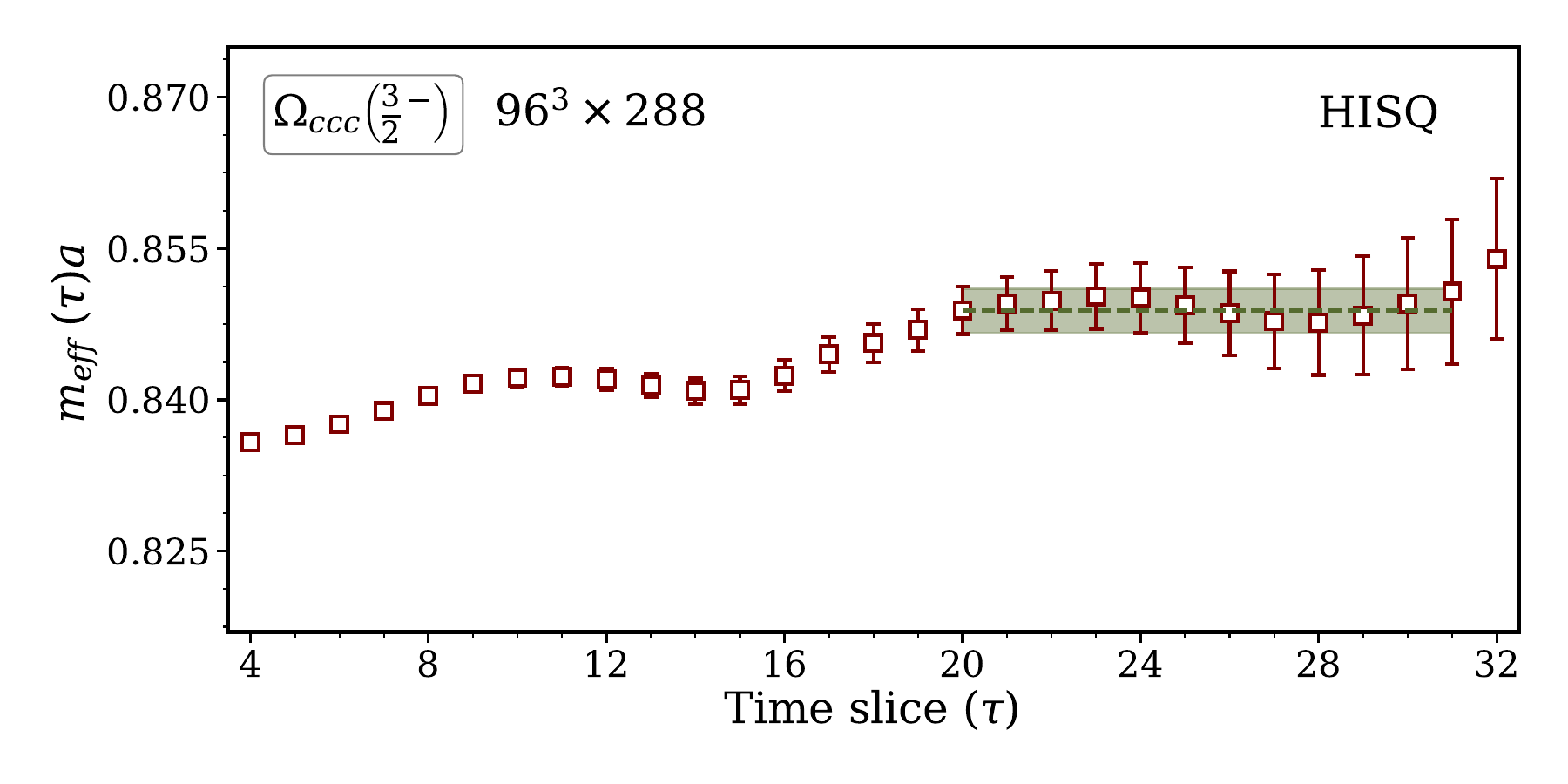} 
\includegraphics[scale=0.28]{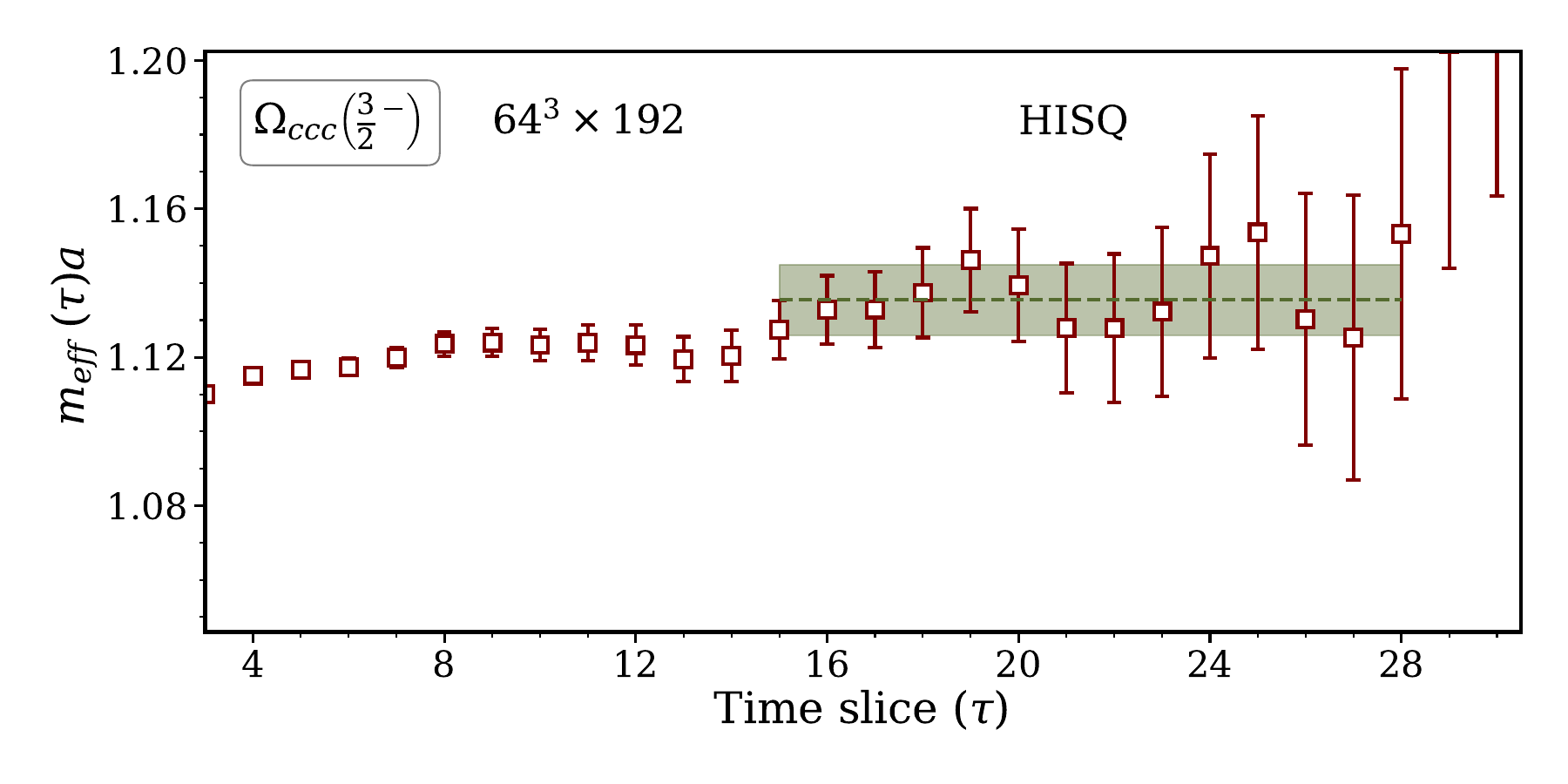} 
        \caption{Effective mass plots corresponding to the positive (top) and negative (bottom) parity lowest energy levels of $\Omega_{ccc}(3/2)$ baryons for different lattice ensembles. The dashed lines are the fit results with one exponential fits, and the bands show the fit ranges with the corresponding $1\sigma$ errors. }
        \label{fg:omega_eff_hisq}
\end{figure*}

In Fig. \ref{fg:omega_eff_hisq}, we show the representative effective mass plots on two different ensembles for the lowest energy levels of  $\Omega_{ccc}(3/2^+)$ (top pane) and $\Omega_{ccc}(3/2^-)$ (bottom pane) baryons. The fit ranges and $1\sigma$ fit errors are shown by the colored band. Following the methods mentioned above, we extract the ground state masses for $\eta_c$, $J/\psi$, $\Omega_{ccc}(3/2^+)$,  and $\Omega_{ccc}(3/2^-)$. We use both one and two exponential fits and observe no difference in fit values of the lowest state (much below percent level) at large $\tau$. After extracting the individual masses we calculate the 1S charmonia hyperfine splitting, $M_{J/\psi} - M_{\eta_c}$, and  {$\Omega_{ccc} - \frac{3}{2} M_{c\bar{c}} $}, with $c\bar{c} \equiv J/\psi$ and $\overline{1S}$ charmonia. The whole procedure is performed using the bootstrap method.

\begin{figure*}[htb!]        
\includegraphics[scale=0.28]{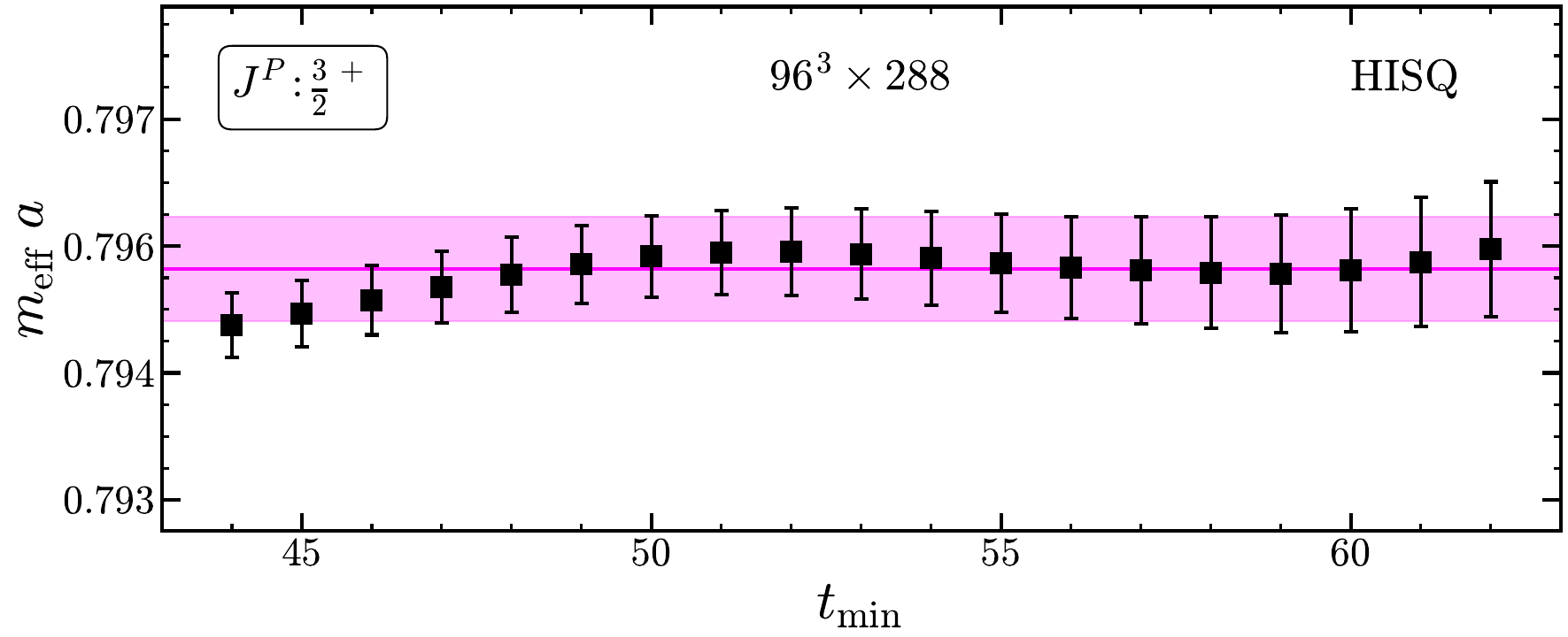} 
\includegraphics[scale=0.28]{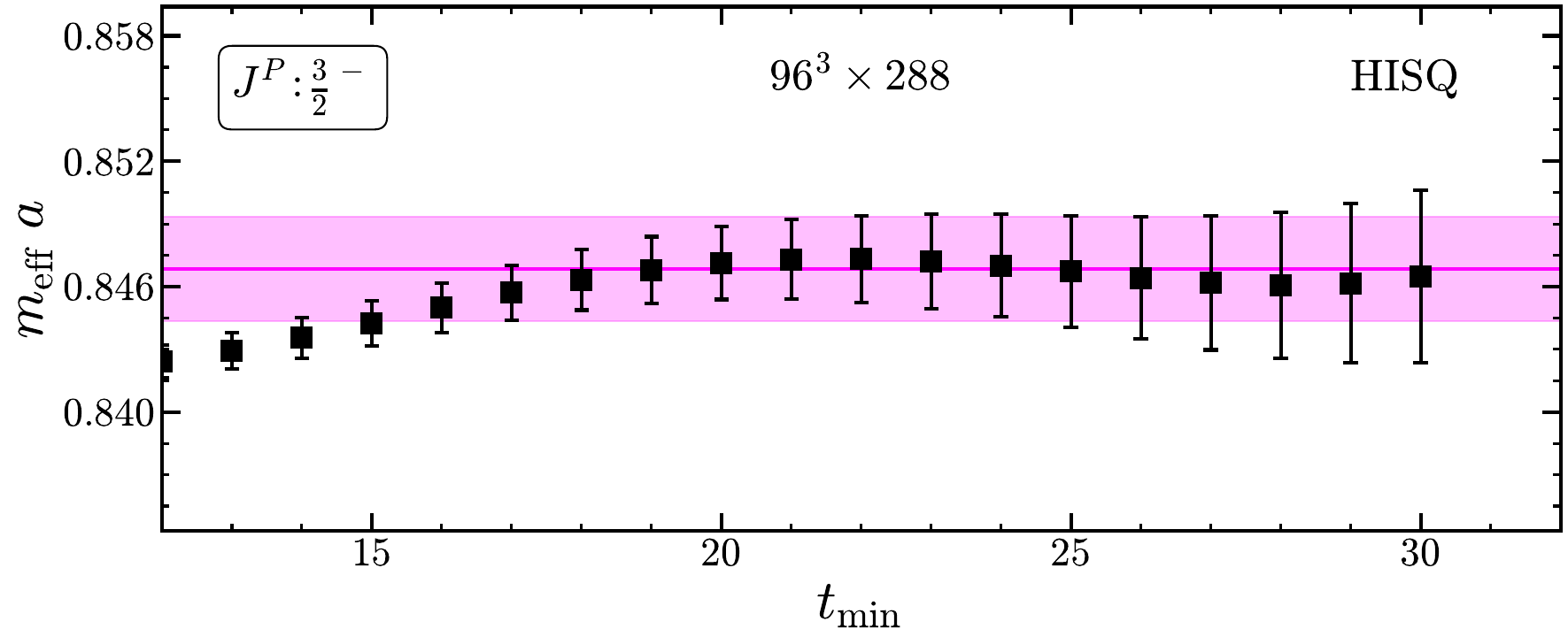}
\includegraphics[scale=0.28]{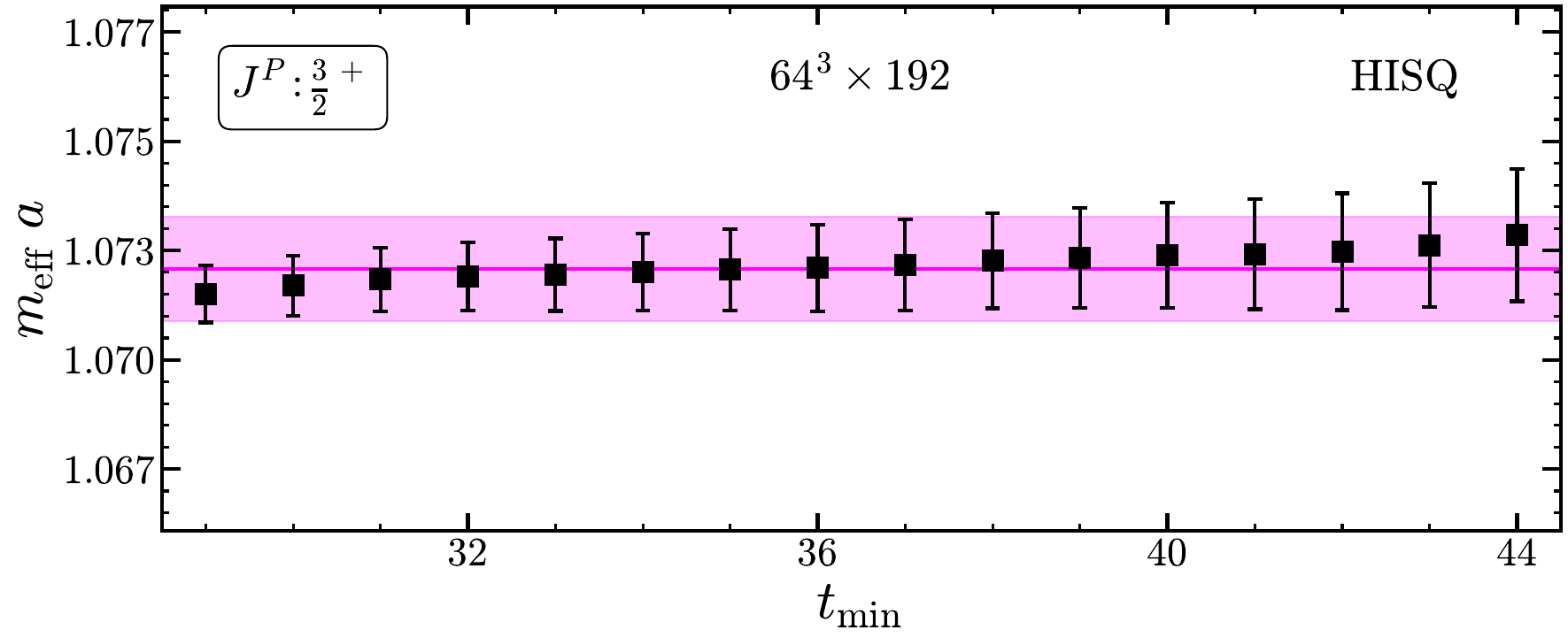} 
\includegraphics[scale=0.28]{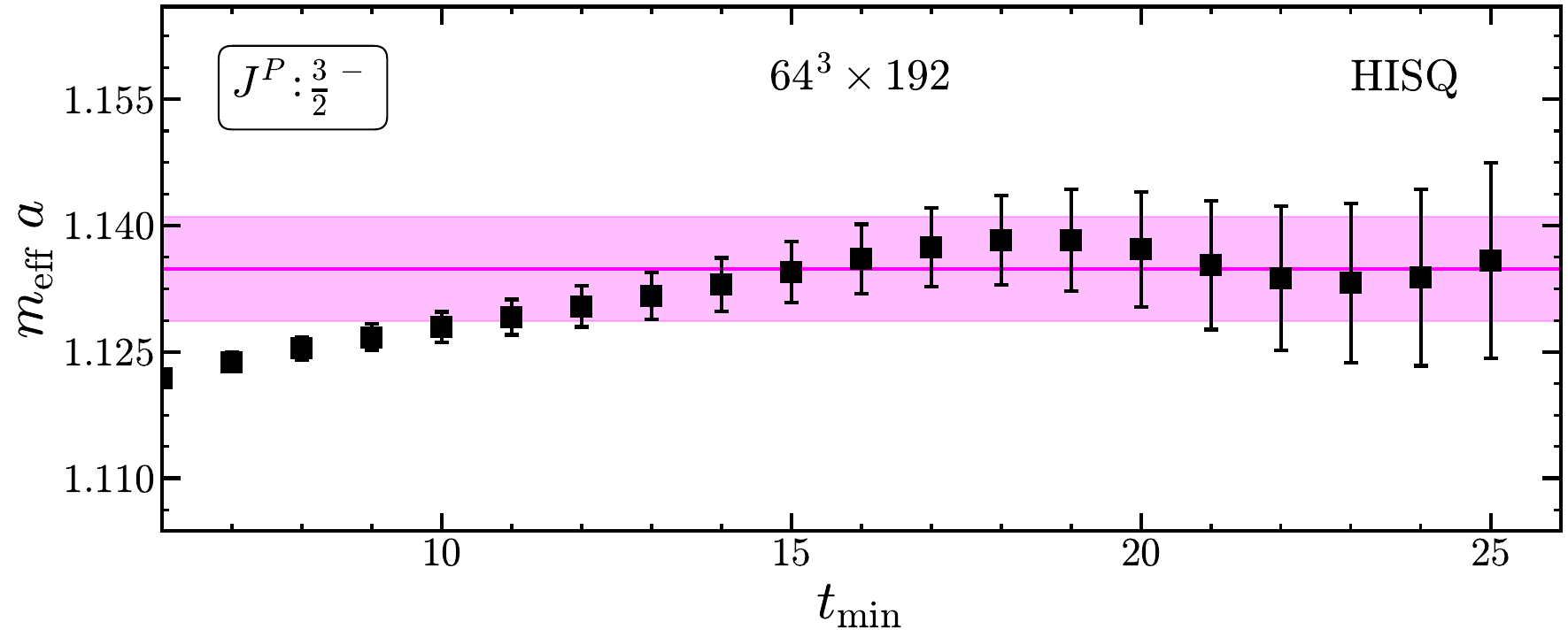} 

\caption{Fit results for the lowest energy levels of $\Omega_{ccc} (\frac{3}{2})$ baryons in both positive (left) and negative (right) parity channels, obtained using fit windows with varying minimum time $t_{\text{min}}$. The black points with error bars represent the central values and $1\sigma$ uncertainties of the fitted ground-state energy across different fit windows. The light magenta band represents the quoted fit result with $1\sigma$ uncertainty, incorporating the uncertainty from the fitting window selection.}

\label{fg:omega_eff_tmin_hisq}
\end{figure*}

\begin{figure*}[htb!]        
\includegraphics[scale=0.28]{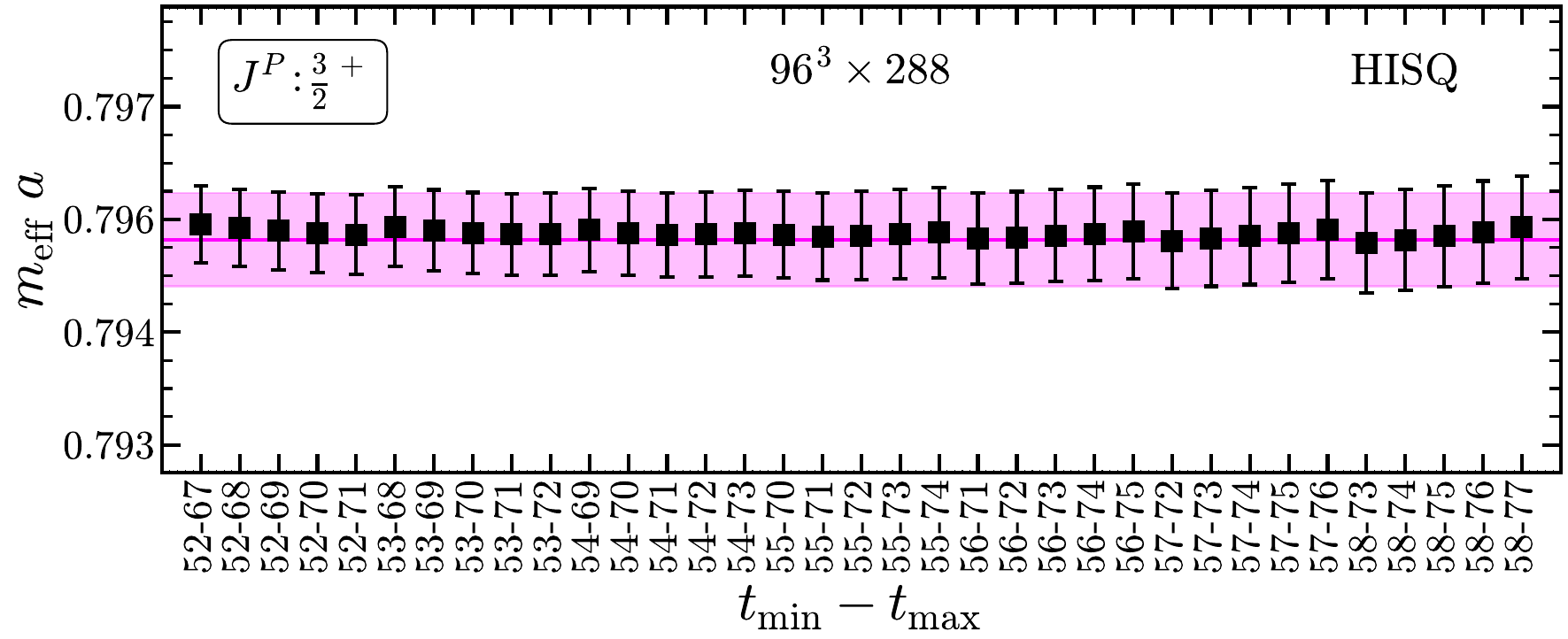} 
\includegraphics[scale=0.28]{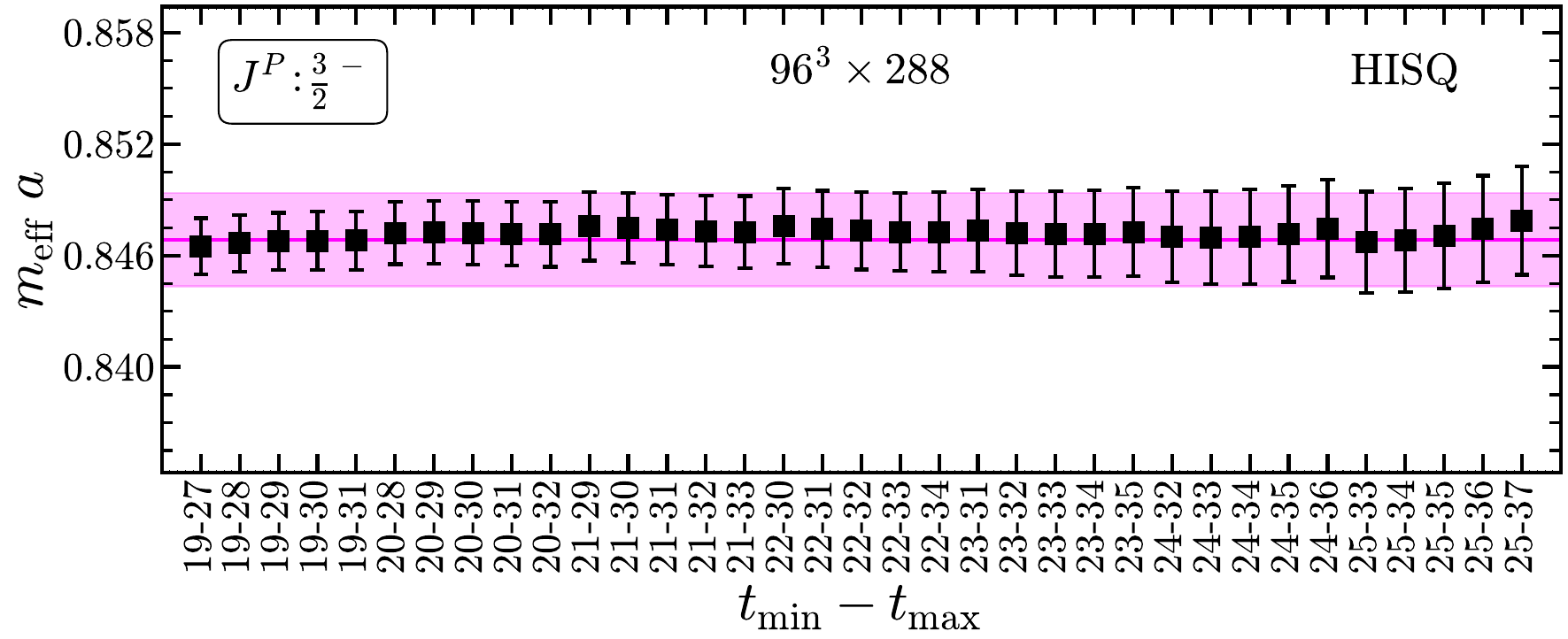} 
\includegraphics[scale=0.28]{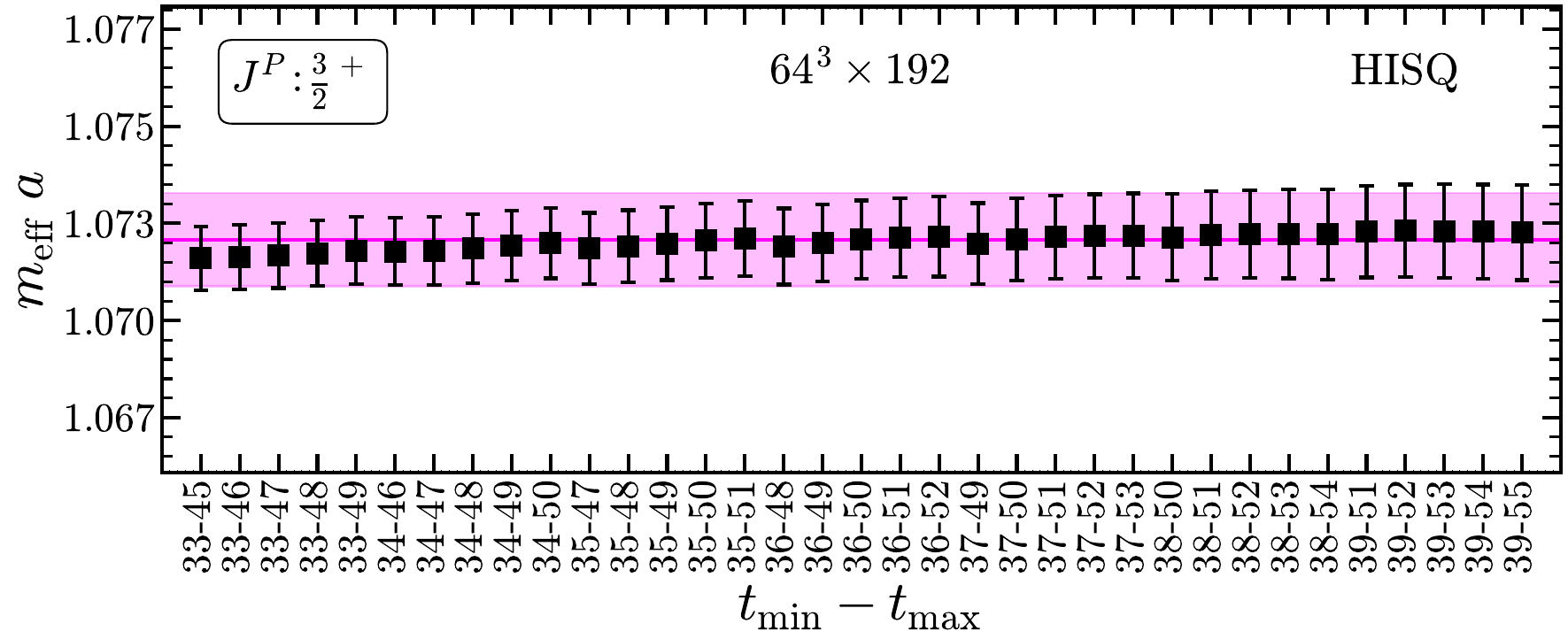} 
\includegraphics[scale=0.28]{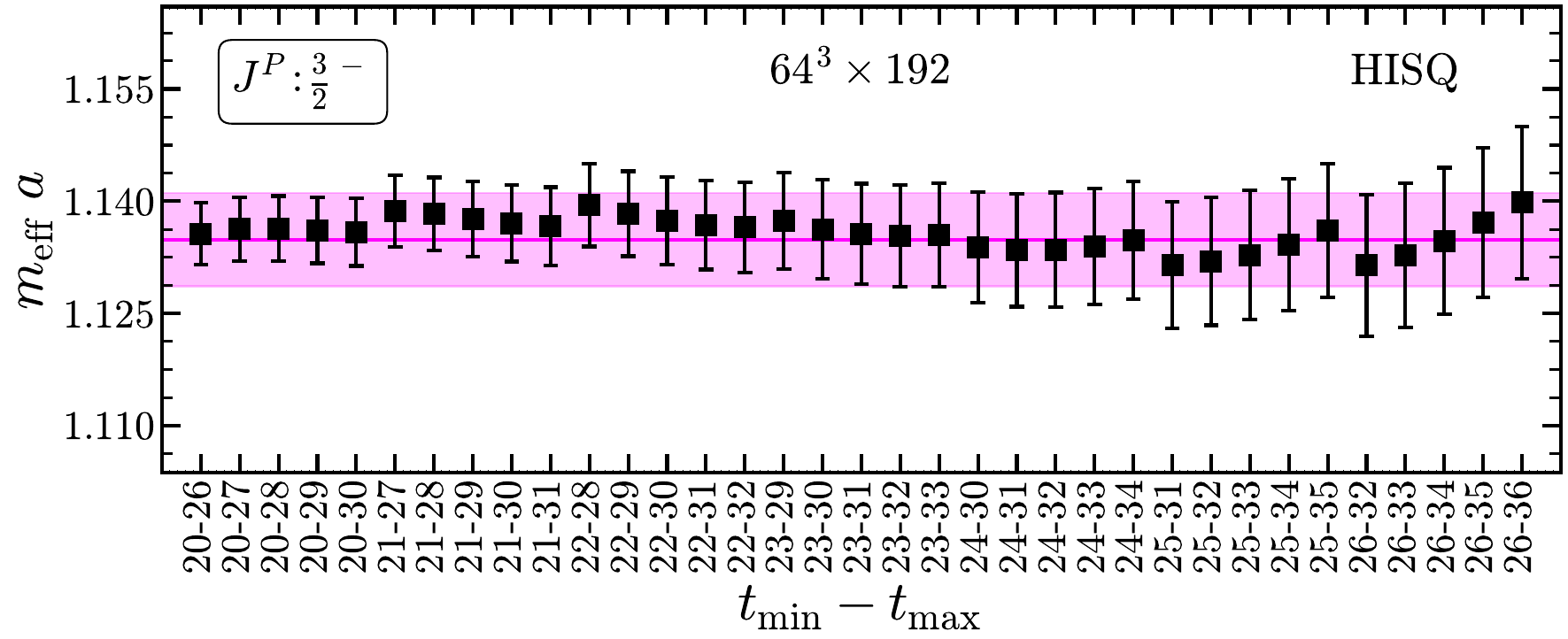} 
\caption{Fit results for the lowest energy levels of $\Omega_{ccc} (\frac{3}{2})$ baryons in both positive and negative parity channels, obtained using fit windows with varying minimum and maximum time i.e. $t_{\text{min}}$ and $t_{\text{max}}$. The black points with error bars represent the central values and $1\sigma$ uncertainties of the fitted ground-state energy across different fit windows. The light magenta band represents the quoted fit result with $1\sigma$ uncertainty, incorporating the uncertainty from the fitting window selection.}

\label{fg:omega_eff_tmin_tmax_hisq}
\end{figure*}

To quantify the uncertainties arising from the choice of the fitting window \((t_{\min}, t_{\max})\), we adopt the same procedure as discussed in the case of overlap fermions (section II) and showed in Figs. \ref{fg:omega_eff_tmin_overlap} and \ref{fg:omega_eff_tmin_tmax_overlap}. In that we first select \(t_{\max}\) to be as large as possible while maintaining a good signal-to-noise ratio. Then, we vary \(t_{\min}\) over a suitable range to assess the stability of the ground state energy estimate. The final choice of \(t_{\min}\) is made where a clear plateau is observed. To account for uncertainties in this selection, we take a conservative approach by computing a correlated average over neighboring \(t_{\min}\) values within the plateau. In Fig. (\ref{fg:omega_eff_tmin_hisq}), for the finest two lattice ensembles, we present the dependence of the fit results on \(t_{\min}\), along with the final estimate, incorporating the uncertainty from the chosen fitting window (indicated by the magenta bands). Additionally, we test the robustness of the fit by varying both \(t_{\min}\) and \(t_{\max}\). In Fig.~(\ref{fg:omega_eff_tmin_tmax_hisq}), we illustrate how the fit results depend on different choices of \(t_{\min}\) and \(t_{\max}\), along with the final estimate, considering the associated uncertainty from the fitting window (again shown by the magenta bands). This demonstrates the stability and reliability of our final fit result.   

\section{Results on mass splittings and continuum extrapolation}
In Table 
\ref{Tab:results} we show these mass splittings for each lattice ensemble. The errorbars in each fit include $1\sigma$ fit errors. 
Results using two different valence quark actions are presented in the separate columns for Overlap and HISQ.

\begin{table*}[htpb]
\centering
  \renewcommand\arraystretch{1.2}  
  \addtolength{\tabcolsep}{3 pt}   
  \begin{tabular}{|p{0.08\textwidth}|p{0.06\textwidth}|p{0.06\textwidth}|p{0.06\textwidth}|p{0.06\textwidth}|p{0.06\textwidth}|p{0.06\textwidth}|p{0.06\textwidth}|p{0.06\textwidth}|p{0.06\textwidth}|p{0.06\textwidth}|}
    \hline\hline
    Lattice Ensembles &  \multicolumn{2}{|c}{$M_{J/\psi}-M_{\eta_c}$} & \multicolumn{2}{|l}{$M_{\Omega_{ccc}\left(\frac{3}{2}\right)^+}-\frac{3}{2}M_{J/\psi}$}  & \multicolumn{2}{|l}{$M_{\Omega_{ccc}\left(\frac{3}{2}\right)^+}-\frac{3}{2}M_{\overline{1S}}$}& \multicolumn{2}{|l}{$M_{\Omega_{ccc}\left(\frac{3}{2}\right)^-}-\frac{3}{2}M_{J/\psi}$} & \multicolumn{2}{|l|}{$M_{\Omega_{ccc}\left(\frac{3}{2}\right)^-}-\frac{3}{2}M_{\overline{1S}}$}\\
    \cline{2-11}
           & Overlap & HISQ & Overlap & HISQ & Overlap & HISQ & Overlap & HISQ & Overlap & HISQ\\
    \hline
   $S_5$  & & 116.09 $\pm$ 1.15  &  & 149.77 $\pm$ 3.61 &  & 192.41 $\pm$ 3.72&  & 458.17 $\pm$ 15.11 & & 500.81 $\pm$ 15.21 \\
    \hline  
    $S_4$      & 113.0 $\pm$ 2.0 & 113.25 $\pm$ 1.16  & 147.14 $\pm$ 3.79 & 150.86 $\pm$ 5.61  & 189.29 $\pm$ 3.67 & 193.22 $\pm$ 5.64 & 448.89 $\pm$ 8.25 & 431.72 $\pm$ 23.15 & 491.04 $\pm$ 8.20 & 474.10 $\pm$ 23.07 \\
    \hline  
   $S_3$     & 119.0 $\pm$ 2.0 & 110.71 $\pm$ 0.83 & 147.68 $\pm$ 3.86 & 149.50 $\pm$ 3.20 & 192.11 $\pm$ 3.78 & 191.01 $\pm$ 3.24 & 439.15 $\pm$ 7.03 & 430.14 $\pm$ 11.89 & 483.59 $\pm$ 6.98 & 471.65 $\pm$ 11.96\\
    \hline 
   $S_2$   & 126.5 $\pm$ 2.0 & 105.79 $\pm$ 1.14  & 147.88 $\pm$ 5.69 & 148.85 $\pm$ 3.61 & 195.47 $\pm$ 5.61 & 188.52 $\pm$ 3.65 & 438.35 $\pm$ 14.03 & 427.96 $\pm$ 15.62 & 485.94 $\pm$ 14.00 & 467.62 $\pm$ 15.66\\
    \hline 
   $S_1$   & 147.0 $\pm$ 1.5 &  106.44 $\pm$ 0.33 & 135.80 $\pm$ 5.70 & 159.12 $\pm$ 1.97 & 190.64 $\pm$ 5.63 & 199.04 $\pm$ 1.98 & 392.06 $\pm$ 10.80 & 467.54 $\pm$ 17.88 & 446.89 $\pm$ 10.76& 507.45 $\pm$ 17.89\\
    \hline
    $L_1$     & 146.0 $\pm$ 1.5 & 106.94 $\pm$ 0.42 & 129.91 $\pm$ 3.03 & 158.50 $\pm$ 2.09 & 184.49 $\pm$ 2.96 & 198.61 $\pm$ 2.11 & 401.42 $\pm$ 13.05 & 442.32 $\pm$ 37.34 & 455.99 $\pm$ 13.04& 482.42 $\pm$ 37.35\\
    \hline\hline
  \end{tabular}
  \caption{Extracted mass differences (in units of MeV) obtained by fitting the individual correlators and then taking the difference of fitted masses by bootstrapping. The subtracted masses are obtained using both 
  $c\bar{c} \equiv J/\psi$ and $\overline{1S}$ in Eq. (2) of the main paper. We also show here the hyperfine splitting in 1S-charmonia. Results for both HISQ and overlap quarks are shown.}
  \label{Tab:results}
\end{table*}

\subsection{Hyperfine splitting in 1S charmonia}
\begin{figure*}[htb!]
 \includegraphics [width=0.32\textwidth, height=0.23\textwidth]{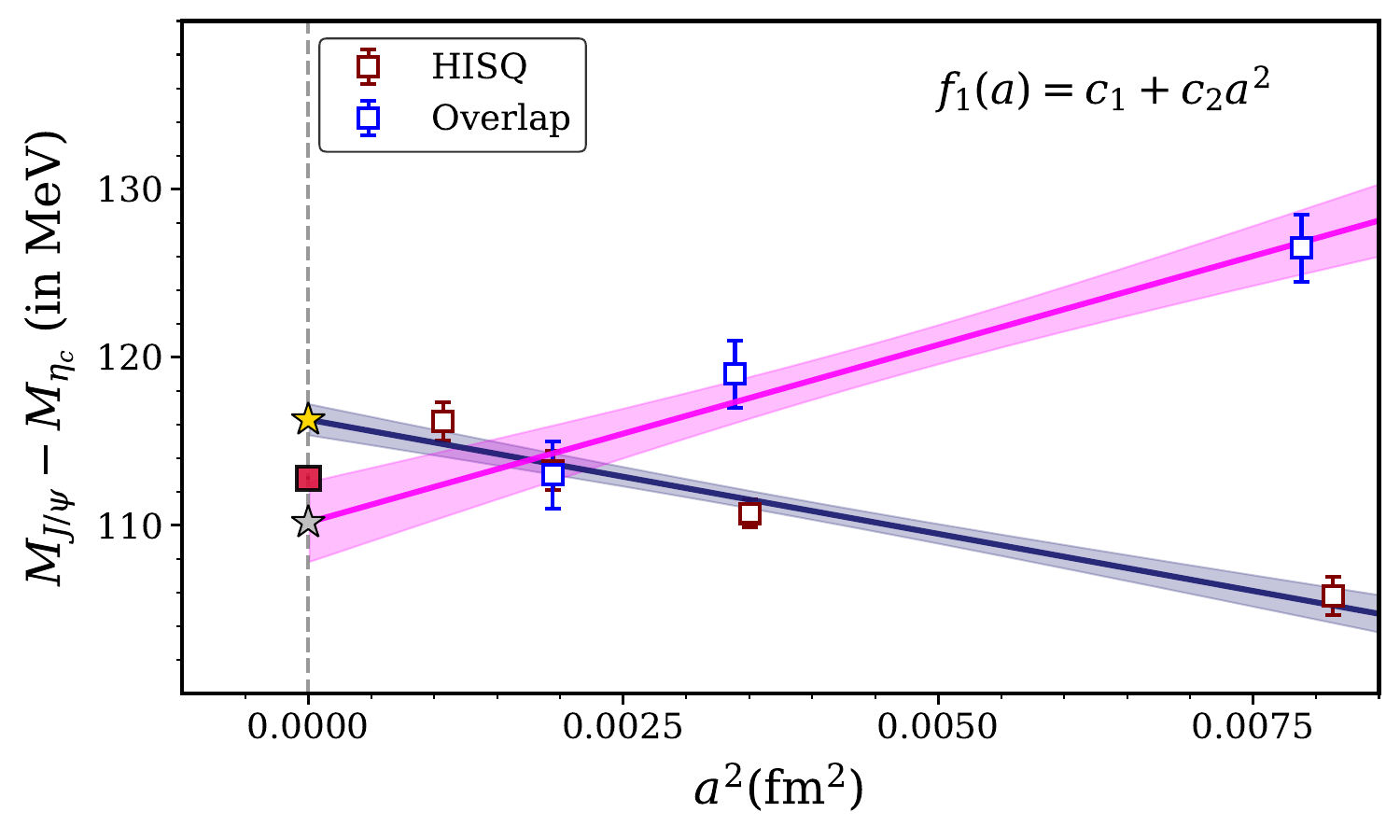}
 \includegraphics [width=0.32\textwidth, height=0.23\textwidth]{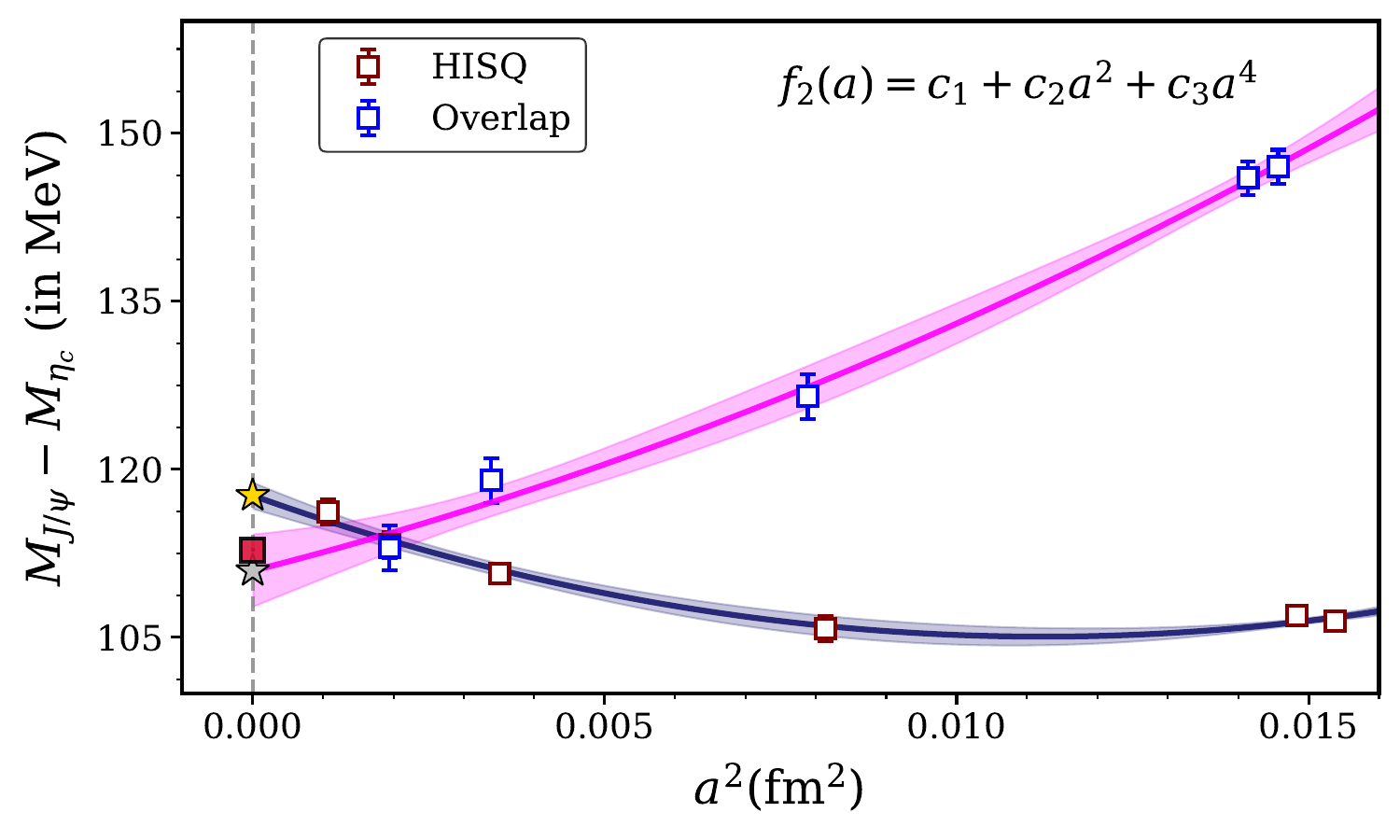}
 \includegraphics [width=0.32\textwidth, height=0.23\textwidth]{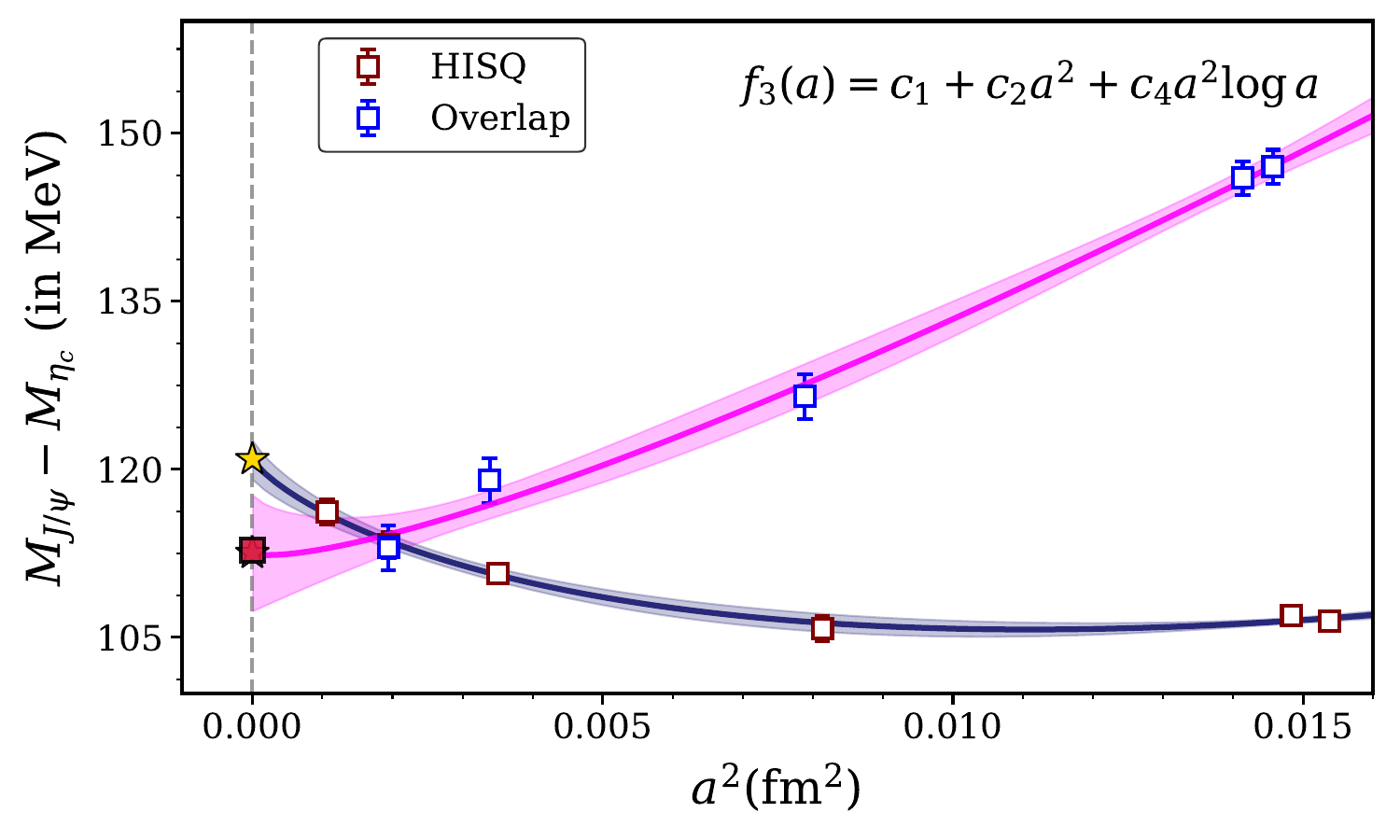}
        \caption{Continuum extrapolation of the hyperfine splitting ($M_{J/\psi} - M_{\eta_c}$) for both overlap and HISQ valence quarks. Left: Extrapolation using fitting model $f_1(a)$ from Table \ref{tab:hf_split2}; Middle: with fitting model $f_2(a)$; Right: with fitting model $f_1(a)$. 
        Considering the possibility of higher order discretization, we do not include the coarsest ensembles $S_1$ and $L_1$ with the fit form $f_1(a)$. The $1\sigma$ bands are estimated using the bootstrap resampling method, while the magenta square on the $a^2 = 0$ line represents the PDG average of ($M_{J/\psi} - M_{\eta_c}$) = $112.8 \pm 0.4$ MeV \cite{PhysRevD.110.030001}. 
        }
        \label{fg:hf_split_plot2}
\end{figure*}

The 1S charmonia splittings extracted on different ensembles at finite lattice spacings are presented in the second column of Table \ref{Tab:results}. Following this, we proceed with the continuum extrapolation of these splittings. Since both the overlap and HISQ actions are free from $\mathcal{O}(ma)$ errors, the leading term in the extrapolation is expected to be of $\mathcal{O}((ma)^2)$.
We use three different fit ansatzs; (i) $f_1(a) = c_1 + c_2 a^2$, (ii) $f_2(a) = c_1 + c_2 a^2 + c_3 a^4$, and (iii)
$f_3(a) = c_1 + c_2 a^2 + c_4 a^2 \,{\mathrm{log}} a$. The last form is motivated from the expectation that for an asymptotically free theory the leading asymptotic lattice spacing dependence may have $a^2 {\mathrm{log}} (a^{n=1})$ term, rather than a simple integer power-law term  \cite{BALOG2009188, Husung:2022kvi, Husung:2024cgc}.
The fit results are presented in Table \ref{tab:hf_split2} and shown in Fig \ref{fg:hf_split_plot2} with colored bands representing the $1\sigma$ fit errors. The stars at the continuum limit are the extrapolated values while solid red square represents the physical value of the hyperfine splitting. Notably, the approach to the continuum limit differs between the two versions of the discretized lattice actions: the overlap results approach from above, while the HISQ results approach from below. The results obtained with overlap quarks are below the physical value but remain consistent with it, whereas the results with HISQ quarks are just above the physical value. HISQ results also agree with that obtained in Ref. \cite{HPQCD_2020}. As noted in Ref. \cite{HPQCD_2020}, this discrepancy could be attributed to the exclusion of the charm self-annihilation diagrams while calculating $\eta_c$ mass. 
This contribution to $J/\psi$ correlation function appears at order $\mathcal{O}(\alpha_s^3)$ based on power counting of strong coupling $\alpha_s$ whereas this contributes to $\eta_c$ correlation function at order $\mathcal{O}(\alpha_s^2)$. 
A lattice calculation also found the $\eta_c$ mass increases by $3-4$ MeV, whereas the effects on $J/\psi$ mass are negligible at the level of per-mille precision \cite{ZhangRenQiang:2021gnn}.
This effect survives at the continuum limit and hence results in a discrepancy between lattice and physical results. For the overlap valence quarks, because of two different actions in sea and valence sectors, an 
extra systematic appears which is proportional to $\Delta_{mix}$ in the mixed action formalism \cite{Bar:2005tu,Chen:2007ug,PhysRevD.86.014501}. This term needs to be accounted in the continuum extrapolation and can amount to a systematic of a few MeV \cite{Basak:2014kma}.

\begin{table}[htpb]
    \centering
     \renewcommand\arraystretch{1.2}  
     \addtolength{\tabcolsep}{3 pt}   
    \begin{tabular}{|c|c|c|c|c|} 
        \hline\hline 
	\multirow{2}{*}{Fitting Model}&\multicolumn{2}{|c|}{Splitting(MeV)} & \multicolumn{2} 
        {|c|}{$\chi^2$/d.o.f.}\\\cline{2-5}
        & Overlap & HISQ & Overlap & HISQ\\\hline
       $f_1(a) = c_1 + c_2a^2 $ & $110.0 \pm 2.2$ & $116.3\pm0.9$ &  $1.14$ & $1.35$\\\hline
        $f_2(a) = c_1 + c_2a^2 + c_3a^4$ &$110.9\pm3.2$ & $117.7 \pm 1.2$ &$0.79$ &$1.21$ \\\hline
        $f_3(a) = c_1 + c_2a^2 + c_4a^2\log a$ & $112.5\pm5.2$&$120.9\pm1.8$ &$0.92$ & $0.73$ \\\hline\hline
        \end{tabular}
    \caption{Continuum extrapolation results for the hyperfine splitting ($M_{J/\psi} - M_{\eta_c}$). The first column lists the fitting ansatz, the second column provides the continuum extrapolated values in MeV, and the third column reports the respective $\chi^2$/d.o.f. 
    }
    \label{tab:hf_split2}
\end{table}

\subsection{Continuum extrapolation  and estimation of $\Omega_{ccc}$ masses}
In Table \ref{Tab:results}, we present the subtracted masses (as in Eq. (2) of the main draft) both for the overlap and HISQ quarks. Results are obtained at various lattice spacings within a range $ \sim 0.12-0.032$ fm. Next, we perform the continuum extrapolation of these mass splittings. As in the case of the continuum limit of hyperfine splitting, we use the following fit ansatzes: (i) $f_1(a) = c_1 + c_2 a^2$ and (ii) $f_2(a) = c_1 + c_2 a^2 + c_3 a^4$. Along with that, for HISQ quark we also include
 a fitting form similar to Eq. (5) of Ref. \cite{HPQCD_2020} (iiia)  $f_3(a)= c_1 + c_2a^2 + c_5\alpha_s(1/a)(m_ca)^2$, and $f_4(a)= c_1 + c_4a^4 + c_5\alpha_s(1/a)(m_ca)^2$, where $\alpha_s$  is the strong-coupling constant and $m_ca$ is the bare charm quark mass. The fitted results are presented in Tables  \ref{tab:omega_jpsi_split} and \ref{tab:omega_charmonia_split}. 
\begin{table}[htpb]
    \centering
     \renewcommand\arraystretch{1.2}  
     \addtolength{\tabcolsep}{3 pt}   
    \begin{tabular}{|c|c|c|c|c|} 
        \hline
	\multirow{2}{*}{Fitting Model}&\multicolumn{2}{|c|}{$M_{\Omega_{ccc}}(3/2^+) - \frac{3}{2}M_{J/\psi}$ (MeV)} & \multicolumn{2} 
        {|c|}{$\chi^2$/d.o.f.}\\\cline{2-5}
        & Overlap & HISQ & Overlap & HISQ\\\hline
        $f_1(a) = c_1 + c_2a^2 $ & $147.1\pm4.3$ & $149.7\pm3.2$  & $0.1$  & $0.16$ \\\hline 
        $f_2(a) = c_1 + c_2a^2 + c_3a^4$ & $144.3\pm 6.2$ & $151.3\pm 4.3$  & $1.08$ & $0.12$\\\hline
        $f_3(a) = c_1 + c_2a^2 + c_5\alpha_s(1/a)(m_ca)^2$ & $-$ & $148.5 \pm 7.9$ & $-$ & $0.15$\\\hline
        $f_4(a) = c_1 + c_4a^4 + c_5\alpha_s(1/a)(m_ca)^2$ & $-$ & $150.6 \pm 3.7$ & $-$ & $0.1$\\\hline
        \end{tabular}
    \caption{Continuum extrapolation results for the subtracted mass $\left[M_{\Omega_{ccc}}(3/2^+)- \frac{3}{2}M_{J/\psi}\right]$, using various possible fit ansatzs. Results are shown both for overlap and HISQ valence quarks.
    }
    \label{tab:omega_jpsi_split}
\end{table}

\begin{table}[htpb]
    \centering
     \renewcommand\arraystretch{1.2}  
     \addtolength{\tabcolsep}{3 pt}   
    \begin{tabular}{|c|c|c|c|c|} 
        \hline
	\multirow{2}{*}{Fitting Model}&\multicolumn{2}{|c|}{$M_{\Omega_{ccc}}(3/2^+) - \frac{3}{2}M_{\overline{1S}}$ (MeV)} & \multicolumn{2} 
        {|c|}{$\chi^2$/d.o.f.}\\\cline{2-5}
        & Overlap & HISQ & Overlap & HISQ\\\hline
  {$f_1(a) = c_1 + c_2a^2 $} & $187.9\pm4.2$ & $193.0\pm3.2$ & $0.1$ & $0.26$ \\\hline     
        $f_2(a) = c_1 + c_2a^2 + c_3a^4$ & $185.2\pm 6.4$ & $194.8\pm 4.3$  & $1.03$ & $0.31$\\\hline
        $f_4(a) = c_1 + c_4a^4 + c_5\alpha_s(1/a)(m_ca)^2$ & $-$ & $194.3 \pm 3.8$ & $-$ & $0.78$\\\hline
        \end{tabular}
    \caption{Similar to Table \ref{tab:omega_jpsi_split}, but for energy splittings $\left[M_{\Omega_{ccc}}(3/2^+)- \frac{3}{2}M_{\overline{1S}}\right]$. }
    \label{tab:omega_charmonia_split}
\end{table}

\begin{figure*}[htb!]
        \includegraphics[scale=0.3]  {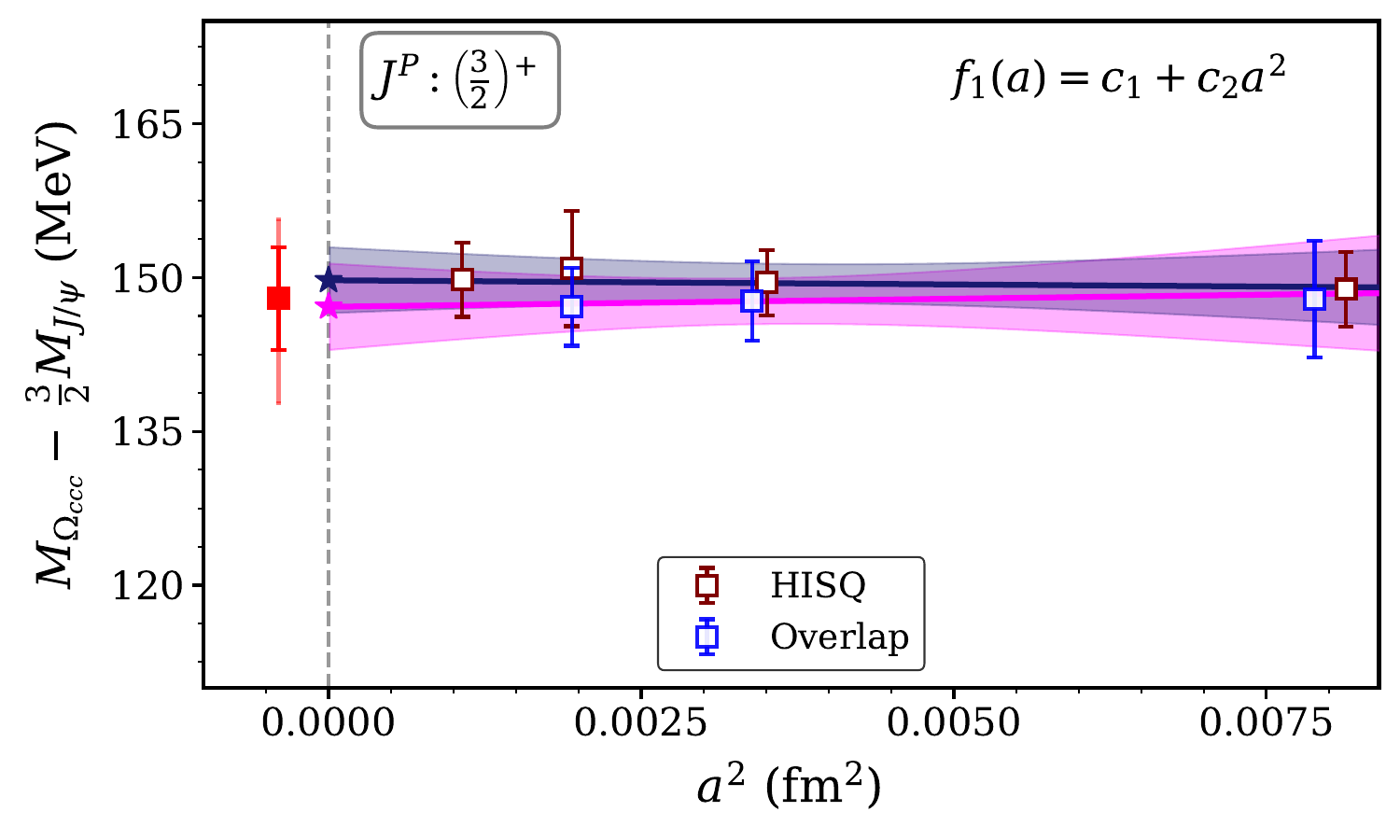}     
        \includegraphics[scale=0.3]  {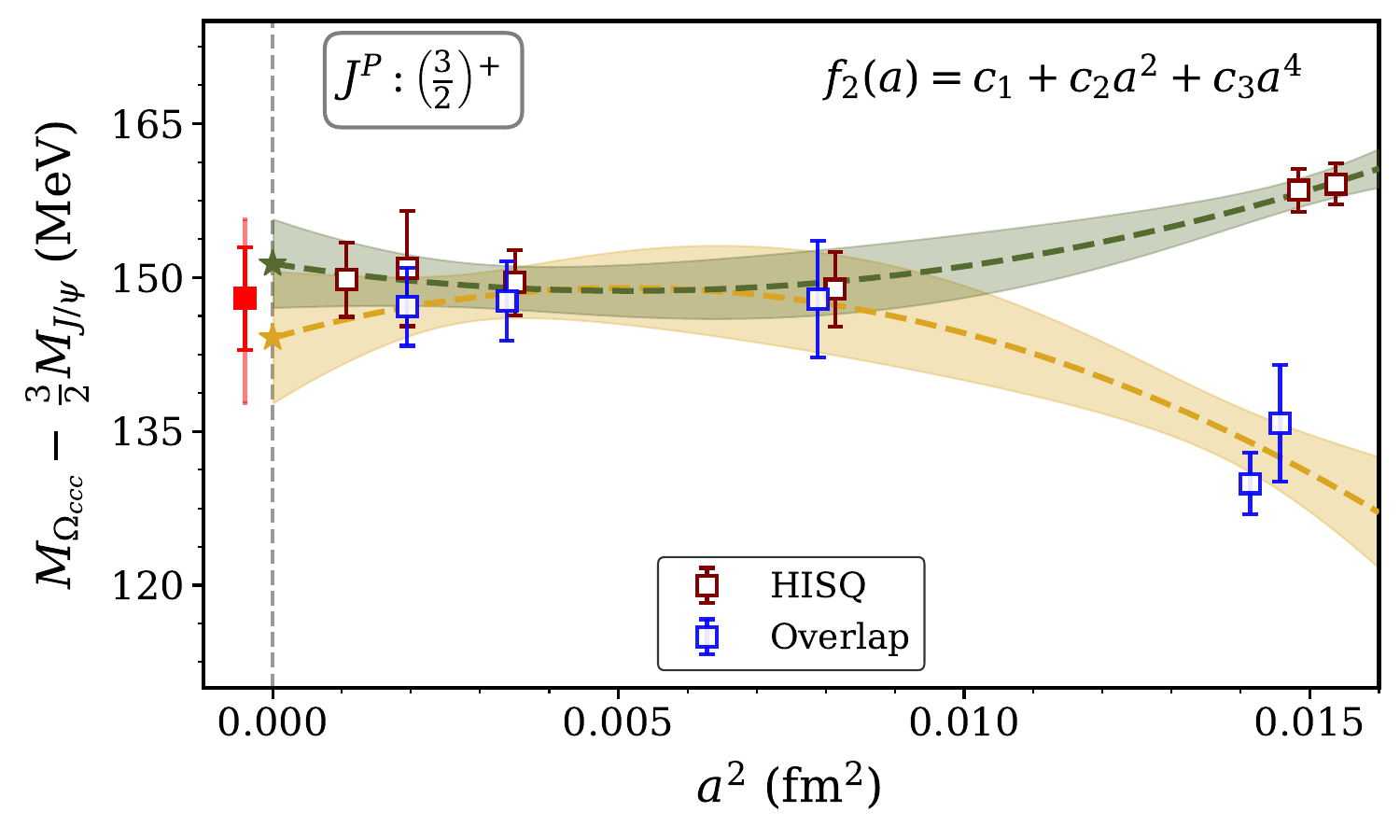} 
        \caption{Continuum extrapolation of the subtracted mass splitting, $M_{\Omega_{ccc}}(3/2^+) - \frac{3}{2}M_{J/\psi}$, for both the Overlap and HISQ actions. Left: Extrapolation using the quadratic fitting ansatz, $f_1(a)$, as in Table \ref{tab:omega_jpsi_split}. Here the coarser lattices, $S_1$ and $L_1$ ensembles, are not included since they may have higher order discretization. 
        Right: same with the ansatz of quadratic and quartic terms in lattice spacings,  $f_2(a)$, fitted with all data points. The extrapolated results at the continuum limits are shown by the star symbols. The $1\sigma$ confidence bands are estimated using the bootstrap resampling method. The solid square is a symmetrized average of two results obtained with the fit form $f_1(a)$, with the errorbar covering both the Overlap and HISQ fit errors. The thin line includes $1\sigma$ errors from the fit ansatzs  $f_1(a)$ and $f_2(a)$. }
        \label{fg:omega_jpsi_split_plot}
\end{figure*}

\begin{figure*}[htbp]
        \includegraphics[scale=0.3]{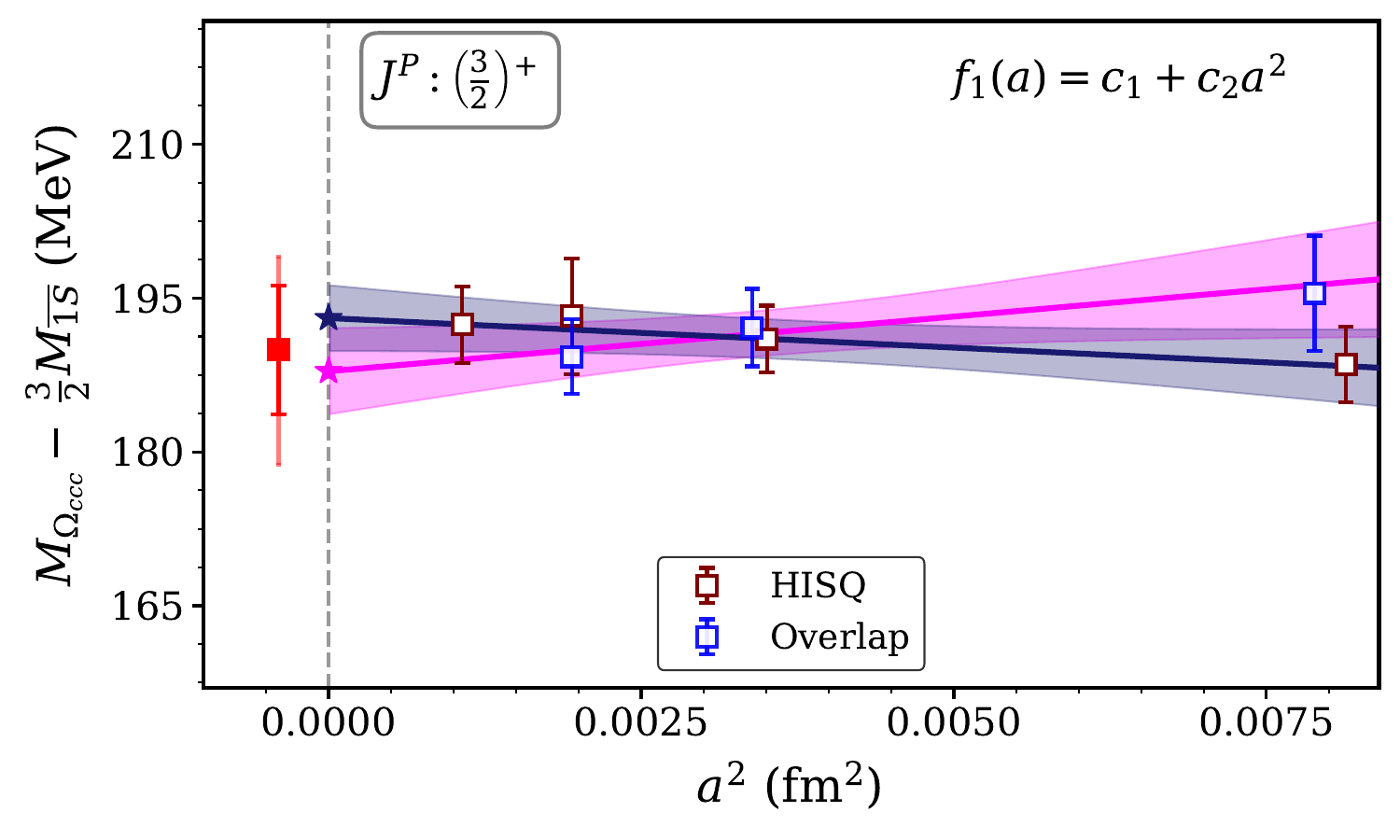}
        \includegraphics[scale=0.3] {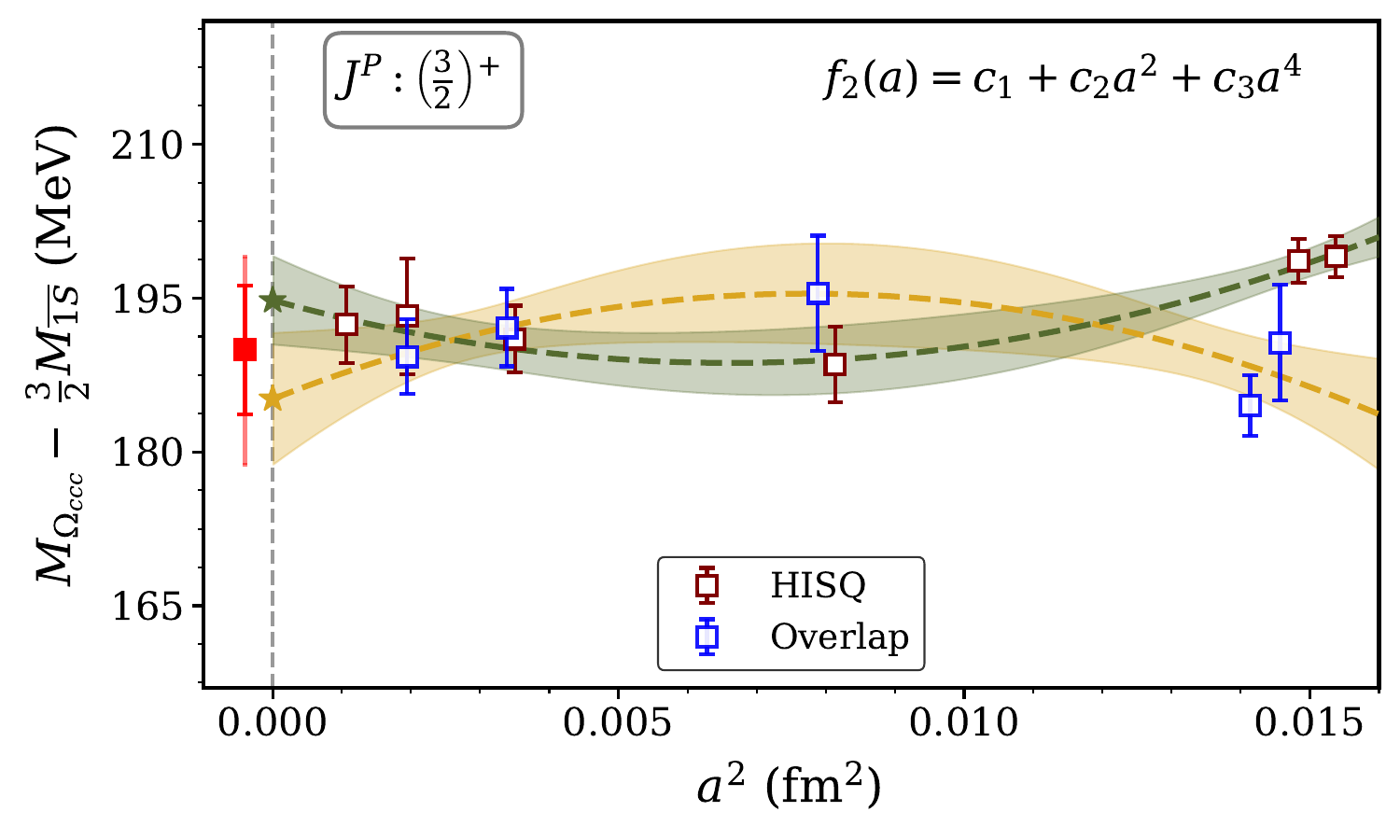} 
        \caption{Continuum extrapolation of the subtracted mass splitting, $M_{\Omega_{ccc}}(3/2^+) - \frac{3}{2}M_{\overline{1S}}$, for both the Overlap and HISQ actions. All other details are as in Fig. \ref{fg:omega_jpsi_split_plot}.}
        \label{fg:omega_1Sbar_split_plot}
\end{figure*}

In Fig. \ref{fg:omega_jpsi_split_plot} and \ref{fg:omega_1Sbar_split_plot}, we depict the above fitted results, for $\left[M_{\Omega_{ccc}}(3/2^+)- \frac{3}{2}M_{J/\psi}\right]$ and $\left[M_{\Omega_{ccc}}(3/2^+)- \frac{3}{2}M_{\overline{1S}}\right]$, respectively. In the left panes, on these figures, we show the results with the quadratic fit forms $f_1(a)$. Note we do not use the coarser lattices, $S_1$ and $L_1$ ensembles, in this case to avoid higher order discretization at largest lattice spacing. The extrapolated results at the continuum limit are shown by the star symbols. The solid square is a symmetrized average of two results obtained with the fit form $f_1(a)$, with the errorbar covering both the Overlap and HISQ fit errors.
In the right panes, we show results by including both quadratic and quartic terms in lattice spacings,  $f_2(a)$.    The thin line includes $1\sigma$ errors covering both the fit ansatzs,  $f_1(a)$ and $f_2(a)$.

Note that, unlike in the charmonia, there is no charm annihilation contribution in the lattice mass estimation of $\Omega_{ccc}$ baryons. Since  $J/\psi$ meson mass has a much lesser contribution from such effect, we choose our final values from the subtracted mass, $M_{\Omega_{ccc}} - M_{J/\psi}$. 
Given that our estimate for $\Omega_{ccc}$ baryon is arrived at following independent approaches to continuum  of the energy splittings from the spin average of the 1S charmonia as well as the $J/\psi$ meson mass, these effects due to charm annihilation are well under 1 MeV errors. Consistency between the $\Omega_{ccc}$ baryon mass estimates from either of these approaches gives further confidence in our numbers.

\begin{figure*}[htpb!]
        \includegraphics[scale=0.3]  {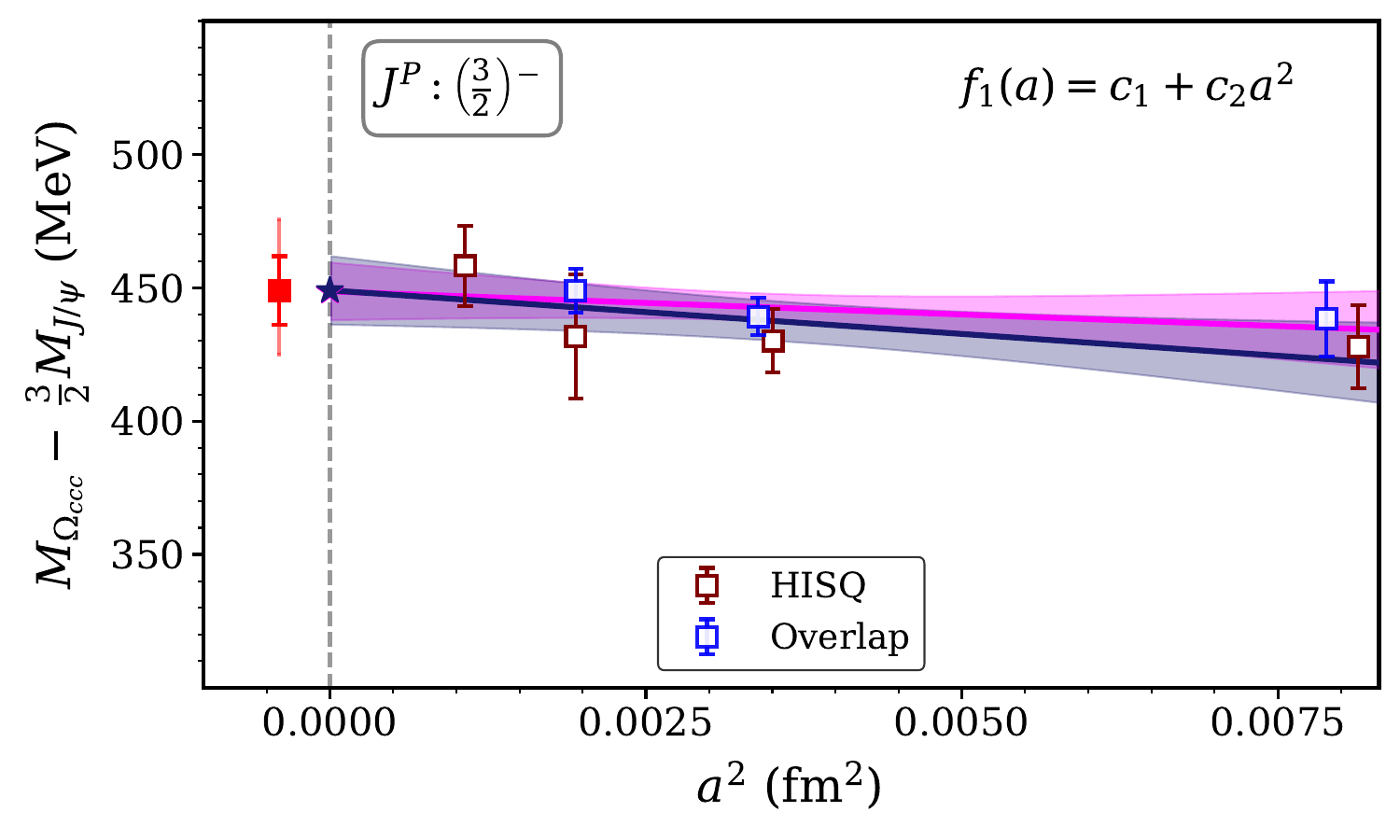}     
        \includegraphics[scale=0.3]  {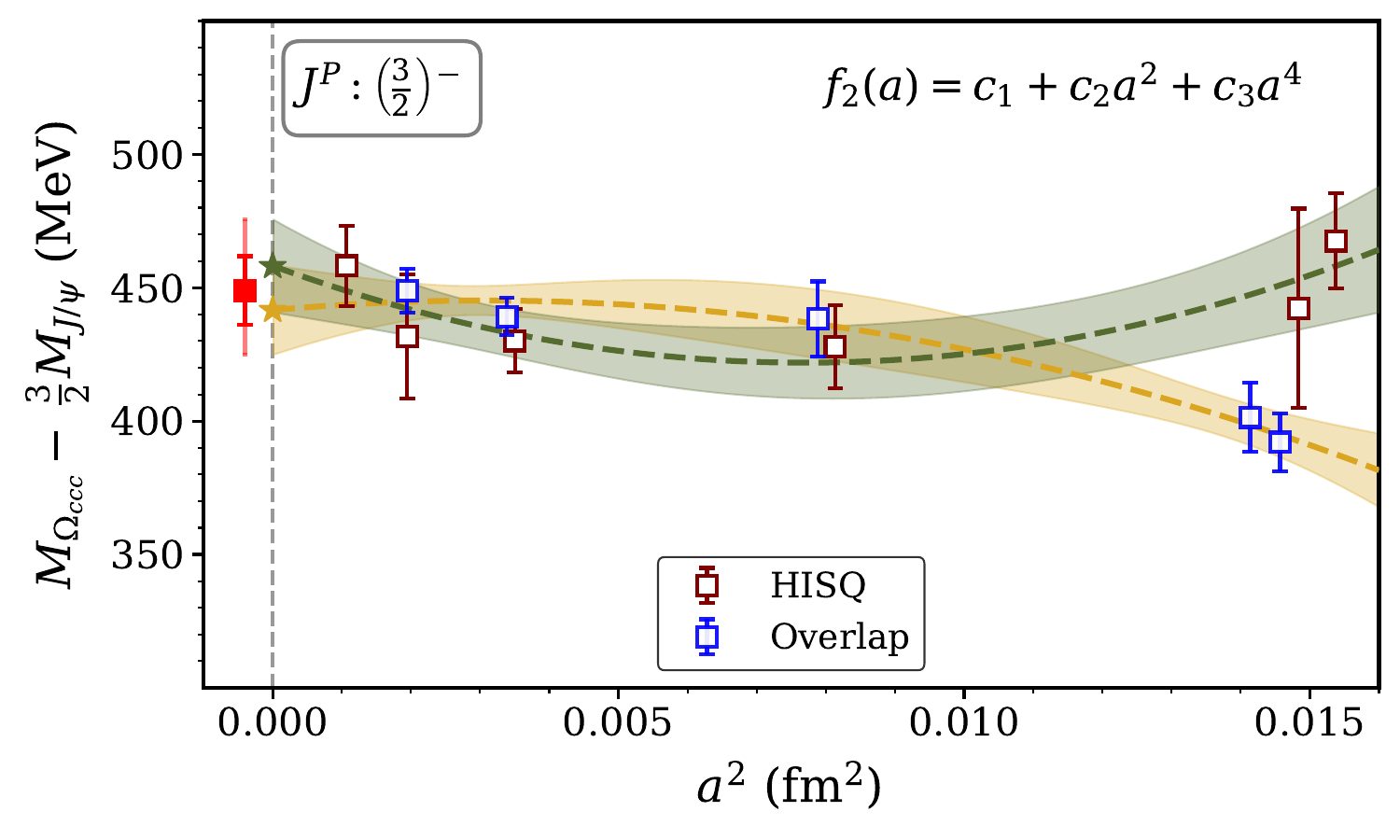}
        \caption{Continuum extrapolation of the subtracted mass splitting, $\left[M_{\Omega_{ccc}}(3/2^-) - \frac{3}{2}M_{J/\psi}\right]$, for both the Overlap and HISQ actions. The fit results are presented in Table \ref{tab:omega_neg_jpsi_split}. All other details are as in Fig. \ref{fg:omega_jpsi_split_plot}.
        }
        \label{fg:omega_neg_jpsi_split_plot}
\end{figure*}

\begin{figure*}[htb!]
        \includegraphics[scale=0.3]  {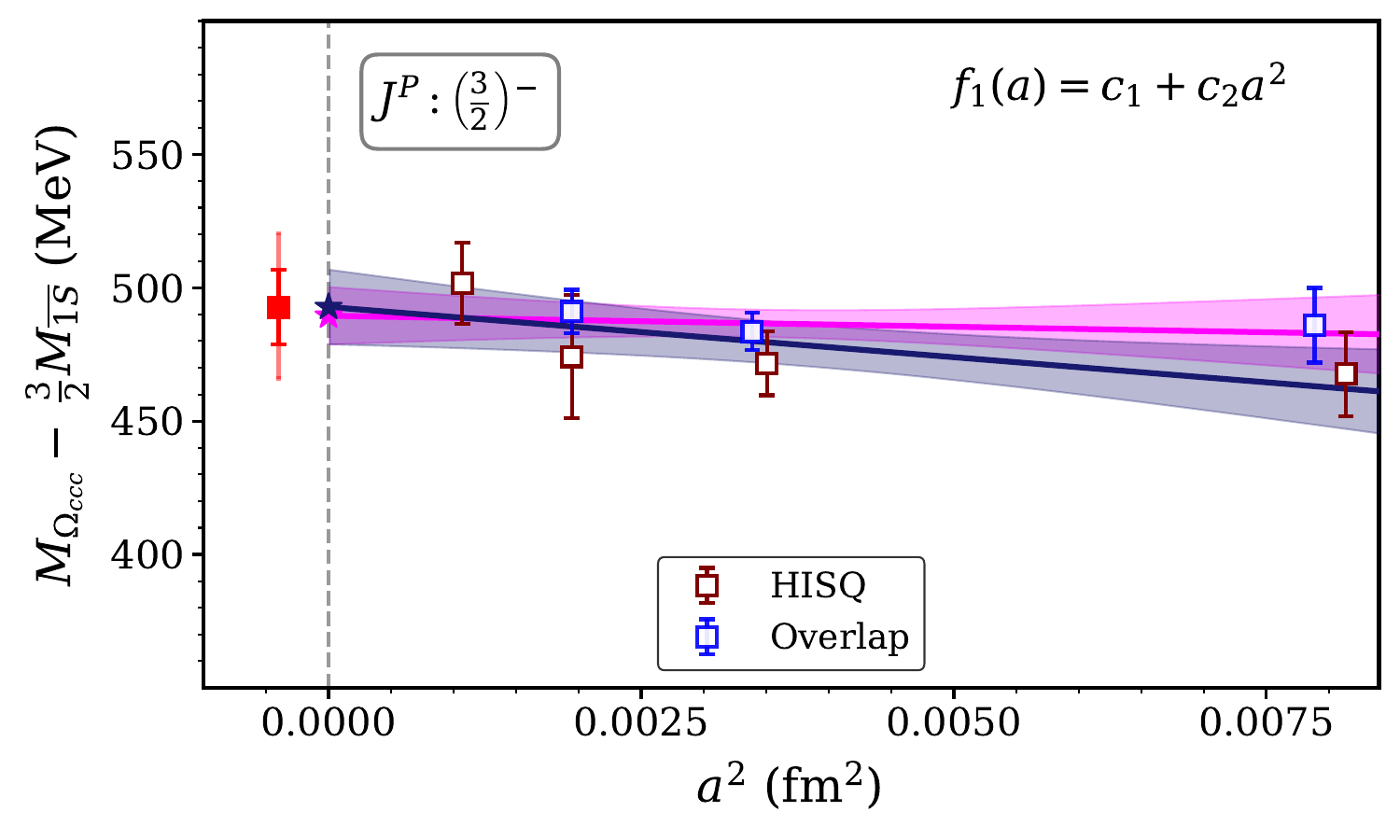}     
        \includegraphics[scale=0.3]  {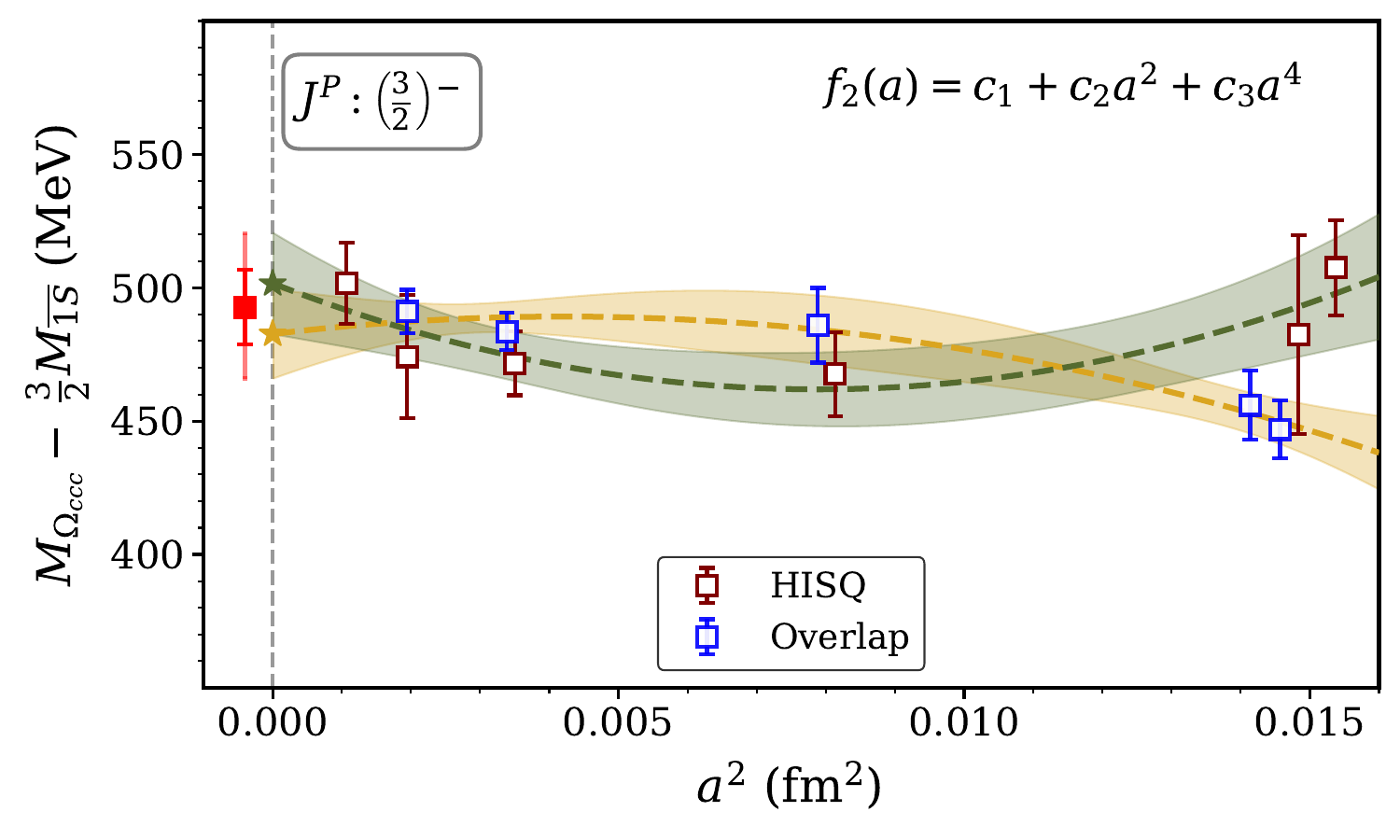}   
        \caption{Continuum extrapolation of the subtracted mass splitting, $\left[M_{\Omega_{ccc}}(3/2^-) - \frac{3}{2}M_{\overline{1S}}\right]$, for both the Overlap and HISQ actions.  The fit results are presented in Table \ref{tab:omega_neg_1Sbar_split}. All other details are as in Fig. \ref{fg:omega_jpsi_split_plot}. 
        }
        \label{fg:omega_neg_1Sbar_split_plot}
\end{figure*}

The corresponding continuum extrapolated results for the negative parity state, $\left[M_{\Omega_{ccc}}(3/2^-) - \frac{3}{2}M_{c\bar{c}}\right]$, with $c\bar{c} \, \equiv \, J/\psi$ and ${\overline{1S}}$ charmonia, are tabulated in Tables \ref{tab:omega_neg_jpsi_split} and 
\ref{tab:omega_neg_1Sbar_split} respectively. Here again, different possible fit forms are employed to find the extent of discretization effects. The fit results are also presented in Figs.  
\ref{fg:omega_neg_jpsi_split_plot} and 
\ref{fg:omega_neg_1Sbar_split_plot}, respectively, with the similar symbols and color-coding as in Figs. \ref{fg:omega_jpsi_split_plot} and \ref {fg:omega_1Sbar_split_plot}.

\begin{table}[htpb]
    \centering
     \renewcommand\arraystretch{1.2}  
     \addtolength{\tabcolsep}{3 pt}   
    \begin{tabular}{|c|c|c|c|c|} 
        \hline
	\multirow{2}{*}{Fitting Model}&\multicolumn{2}{|c|}{$M_{\Omega_{ccc}}(3/2^-) - \frac{3}{2}M_{J/\psi}$ (MeV)} & \multicolumn{2} 
        {|c|}{$\chi^2$/d.o.f.}\\\cline{2-5}
        & Overlap & HISQ & Overlap & HISQ\\\hline
        $f_1(a) = c_1 + c_2a^2 $ & $448.6\pm 10.8$ & $449.0\pm12.8$  & $0.52$  & $0.72$ \\\hline   
        $f_2(a) = c_1 + c_2a^2 + c_3a^4$ & $441.6\pm 16.8$ & $458.1\pm 17.4$  & $0.56$ & $0.38$\\\hline
        \end{tabular}
    \caption{Continuum extrapolation results for the splitting $\left[M_{\Omega_{ccc}}(3/2^-)  - \frac{3}{2}M_{J/\psi}\right]$ with various fit ansatzs for the Overlap and HISQ actions. 
    }
    \label{tab:omega_neg_jpsi_split}
\end{table}
\begin{table}[htpb]
    \centering
     \renewcommand\arraystretch{1.2}  
     \addtolength{\tabcolsep}{3 pt}   
    \begin{tabular}{|c|c|c|c|c|} 
        \hline
	\multirow{2}{*}{Fitting Model}&\multicolumn{2}{|c|}{$M_{\Omega_{ccc}}(3/2^-) - \frac{3}{2}M_{\overline{1S}}$ (MeV)} & \multicolumn{2} 
        {|c|}{$\chi^2$/d.o.f.}\\\cline{2-5}
        & Overlap & HISQ & Overlap & HISQ\\\hline
        $f_1(a) = c_1 + c_2a^2 $ & $489.5\pm 10.7$ & $492.7\pm13.9$  & $0.39$  & $0.76$ \\\hline   
        $f_2(a) = c_1 + c_2a^2 + c_3a^4$ & $480.7\pm 16.7$ & $501.5\pm 18.9$  & $0.47$ & $0.39$\\\hline
        \end{tabular}
    \caption{Continuum extrapolation results for the energy splitting $\left[M_{\Omega_{ccc}}(3/2^-)  - \frac{3}{2}M_{\overline{1S}}\right]$ with various fit ansatzs for the Overlap and HISQ actions.  }
    \label{tab:omega_neg_1Sbar_split}
\end{table}

\begin{figure*}[h!]
\includegraphics[width=0.5\textwidth]
{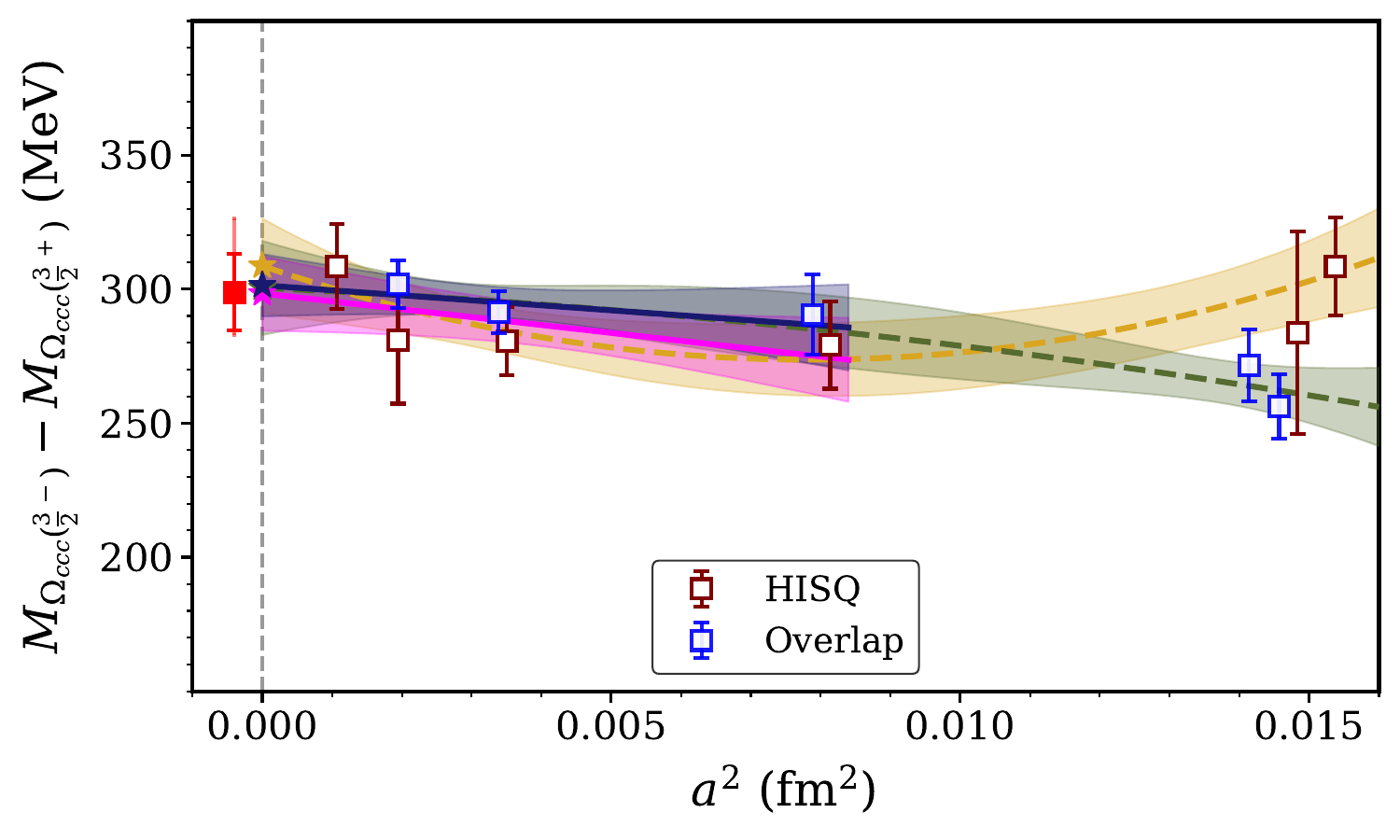}      \caption{Continuum extrapolation of the energy splitting between the ground state masses of $\Omega_{ccc}(\frac{3}{2}^-)$ and $\Omega_{ccc}(\frac{3}{2}^-)$ baryons. Color scheme: magenta and blue solid line represents the fit results using the fit form i) $\Delta M(a) = c_1 + c_2 a^2$, for the HISQ and overlap valence actions, respectively. The green and yellow dashed line show the fit results using the fit form ii) $\Delta M(a) = c_1 + c_2 a^2 + c_3 a^4$, for the HISQ and overlap valence action, respectively. The stars indicate the continuum extrapolated values, whereas the red square is a symmetrized average of results from either procedures, with thick errorbars representing the errors from quadratic extrapolation. The thin errorbar account the uncertainties arising out of difference in results between quadratic and quartic extrapolations.
        }
        \label{fg:omega_pos_neg}
\end{figure*}

Next we present the mass splitting between 
$\Omega_{ccc}(3/2)^{-}$ and $\Omega_{ccc}(3/2)^{+}$ baryons, in  Table \ref{Tab:results_2}. For this mass splitting we do not use the ratio method as the fit ranges for the ground states of  $\Omega_{ccc}(3/2)^{-}$ and $\Omega_{ccc}(3/2)^{+}$ are different. Instead, we use the difference in their fit values within the bootstrap procedure. In 
Fig. \ref{fg:omega_pos_neg} we show these fit results. The continuum extrapolated results are shown by the star symbol, and the solid red square is our final value with the thick bar as the $1\sigma$ error. The thin bar at the continuum point shows the extent of discretization considering the fit forms, $f_1$ and $f_2$. Our final result for this mass splitting is: $\Omega_{ccc}(3/2)^{-}$ and $\Omega_{ccc}(3/2)^{+} = 301 (13)(14)$ MeV, with statistical and systematic errors respectively.

\begin{table*}[htpb]
\centering
  \renewcommand\arraystretch{1.2}  
  \addtolength{\tabcolsep}{3 pt}   
  \begin{tabular}{|c|c|c|}
  \hline\hline
    Lattice Ensembles  &  \multicolumn{2}{|l|}{$M_{\Omega_{ccc}\left(\frac{3}{2}\right)^-}-M_{\Omega_{ccc}\left(\frac{3}{2}\right)^+}$ (MeV)}\\
    \cline{2-3}
           & Overlap & HISQ \\
    \hline
    $S_5$   & &  308.41 $\pm$ 15.78    \\
    \hline  
    $S_4$      & 301.76 $\pm$ 8.84  & 280.86 $\pm$ 23.59  \\
    \hline  
    $S_3$     & 291.47 $\pm$ 7.82 & 280.64 $\pm$ 12.77 \\
    \hline 
    $S_2$       & 290.47 $\pm$ 15.00 & 279.11 $\pm$ 16.24   \\
    \hline 
    $S_1$       & 256.26 $\pm$ 12.06 & 308.42 $\pm$ 18.23\\
    \hline
    $L_1$       & 271.51 $\pm$ 13.33 & 283.82 $\pm$ 37.81 \\
    \hline\hline
  \end{tabular}
  \caption{The mass splitting between the lowest energy levels of positive and negative parity $\Omega_{ccc}$ baryons.}
  \label{Tab:results_2}
\end{table*}

{}

\vspace*{0.2in}

In Table \ref{tab:lat_reslt_sum} we tabulate the results obtained by various lattice QCD collaborations for the ground states of $\Omega_{ccc}(3/2^+)$ and $\Omega_{ccc}(3/2^-)$ baryons. The references are shown in the first column. Note that, in addition to other details, we have also highlighted whether continuum extrapolations were included in these calculations. Results from this work are listed in the last row.
\begin{table*}[htp]
\setlength{\tabcolsep}{0.5mm}{
\centering
\renewcommand\arraystretch{1.2} 
\begin{tabular}{|c|c|c|c|c|c|c|c|c|}
\hline 
\hline 
Ref. (Year)      &$N_{f}$           &$a(fm)$         &$m_{\pi}$ (MeV)    &$S_{q}^{sea}$       &$S_{c}^{val}$     & Continuum &$M_{\Omega_{ccc}}(\frac{3}{2}^{+})$  &$M_{\Omega_{ccc}}(\frac{3}{2}^{-})$ 
\\
 &          &        &    & && Extrapolation & (MeV) &(MeV)
\\

\hline
\cite{Chiu:2005zc} (2005)     &quench   &0.0882      &-    & Wilson    &DW    & $\textcolor{blue!50!black!100}{\textbf{No}}$&4681(28)      &5066(48)   \\
\cite{Alexandrou:2012xk} (2012)     & 2,2+1 &   0.056-0.089   &  260-470  &  TM  & OS  & $\textcolor{orange!80!yellow!100}{\textbf{Yes}}$ (3) &4677(5)(3)    & -  \\
\cite{Briceno:2012wt} (2012)    &2+1+1       &0.06-0.12     &220-310    &HISQ           &RHQA &  $\textcolor{orange!80!yellow!100}{\textbf{Yes}}$ (3)   &4761(52)(21)(6)    &-   \\
\cite{Basak:2012py} (2012) &  2+1+1     &  0.06-0.09   &  316-329  &     HISQ      &  Overlap   &  $\textcolor{blue!50!black!100}{\textbf{No}}$ &4765(10)  &-   \\
\cite{Durr:2012dw} (2012) & 2     &  0.0728 & 280 & Clover  &   Brillouin        & $\textcolor{blue!50!black!100}{\textbf{No}}$  &4774(24)    & -\\
\cite{PACS-CS:2013vie} (2013) &  2+1    & 0.0899  & 135  & Clover &    RHQA       & $\textcolor{blue!50!black!100}{\textbf{No}}$ & 4789(22)    & - \\
\cite{Padmanath:2013zfa} (2013) &  2+1    & 0.0351  & 390 & Clover  &    Clover       & $\textcolor{blue!50!black!100}{\textbf{No}}$ & 4763(7)     & 5124(14)\\
\cite{Alexandrou:2014sha} (2014) &  2+1+1    &  0.065-0.094 & 210-430 & TM &  OS         &  $\textcolor{orange!80!yellow!100}{\textbf{Yes}}$ (3)&4734(12)(11)(9)     & -\\
\cite{Brown:2014ena} (2014) &  2+1    &  0.085-0.11  & 227-419 & DW &   RHQA        & $\textcolor{orange!80!yellow!100}{\textbf{Yes}}$ (2)  & 4796(8)(18)    & -\\
\cite{Can:2015exa} (2015) &  2+1    &  0.0907  & 156 & Wilson &  Clover         &  $\textcolor{blue!50!black!100}{\textbf{No}}$  & 4769(6)   & - \\
\cite{Alexandrou:2017xwd} (2017) & 2     & 0.0938  & 130 & TM Clover &  OS         & $\textcolor{blue!50!black!100}{\textbf{No}}$ &4746(4)(32)     & -\\
\cite{Chen:2017kxr} (2017) &  2+1+1    &  0.063 &  280 & DW &   DW    &  $\textcolor{blue!50!black!100}{\textbf{No}}$  & 4766(5)(11)      & 5168(37)(51)\\ 
\cite{Bahtiyar:2020uuj} (2020) &  2+1    &  0.0907 & 156 & Clover & Clover      &  $\textcolor{blue!50!black!100}{\textbf{No}}$  &  4817(12)     & 5083(67)\\
\cite{Lyu:2021qsh} (2021) &   2+1   & 0.0846  & 146 & Clover &  RHQA     &  $\textcolor{blue!50!black!100}{\textbf{No}}$  &   4796(1)    & -\\
\cite{Li:2022vbc} (2022) &  2+1     & 0.0711-0.0828  & 278-300 & DW &  Overlap     &  $\textcolor{blue!50!black!100}{\textbf{No}}$  &  4793(21)     & 5071(27)\\
\cite{Alexandrou:2023dlu} (2023) & 2+1+1     & 0.057-0.080  & 137-141 & TM Clover &  OS     &   $\textcolor{orange!80!yellow!100}{\textbf{Yes}}$ (3) &    4785(71)   & -\\
This work &  2+1+1    & 0.0327-0.1207  & 216-329 \cite{PhysRevD.98.074512} & HISQ & HISQ, Overlap      &  $\textcolor{red}{\textbf{Yes}}$ (5)  &   4793(5)($^{+11}_{-8}$)    & 5094(12)($^{+19}_{-17}$)\\
\hline \hline
\end{tabular}}
\caption{A compilation of all lattice results for masses of $\Omega_{ccc}(3/2^+)$ and $\Omega_{ccc}(3/2^-)$, along with various relevant technical details involved. The result from this work is presented in the bottom most row. Several relevant lattice specific details listed above include: the number of dynamical quark flavors ($N_{f}$), lattice spacing ($a$), pion mass ($m_{\pi}$), as well as the actions employed for the sea ($S^{sea}{q}$) and valence charm quarks ($S^{val}{c}$) in each work. Additionally, it is indicated whether the calculation includes a continuum extrapolation, and if so, the number of distinct lattice spacings used in the extrapolation is indicated inside brackets. Abbreviations: HISQ (highly-improved staggered quark action), DW (Domain-wall), TW (Twisted mass), OS (Osterwalder-Seiler), and RHQA (relativistic heavy-quark action). Quoted values are rounded to the nearest integer. For calculations on multiple lattice spacings without continuum extrapolation, results from the finest lattice are listed.}\label{tab:lat_reslt_sum}
\end{table*}

{\vspace*{0.4in}}

\section{Sea-Quark Mass dependence}
\begin{figure}[h!]
    \centering
     \includegraphics[width=0.6\linewidth]{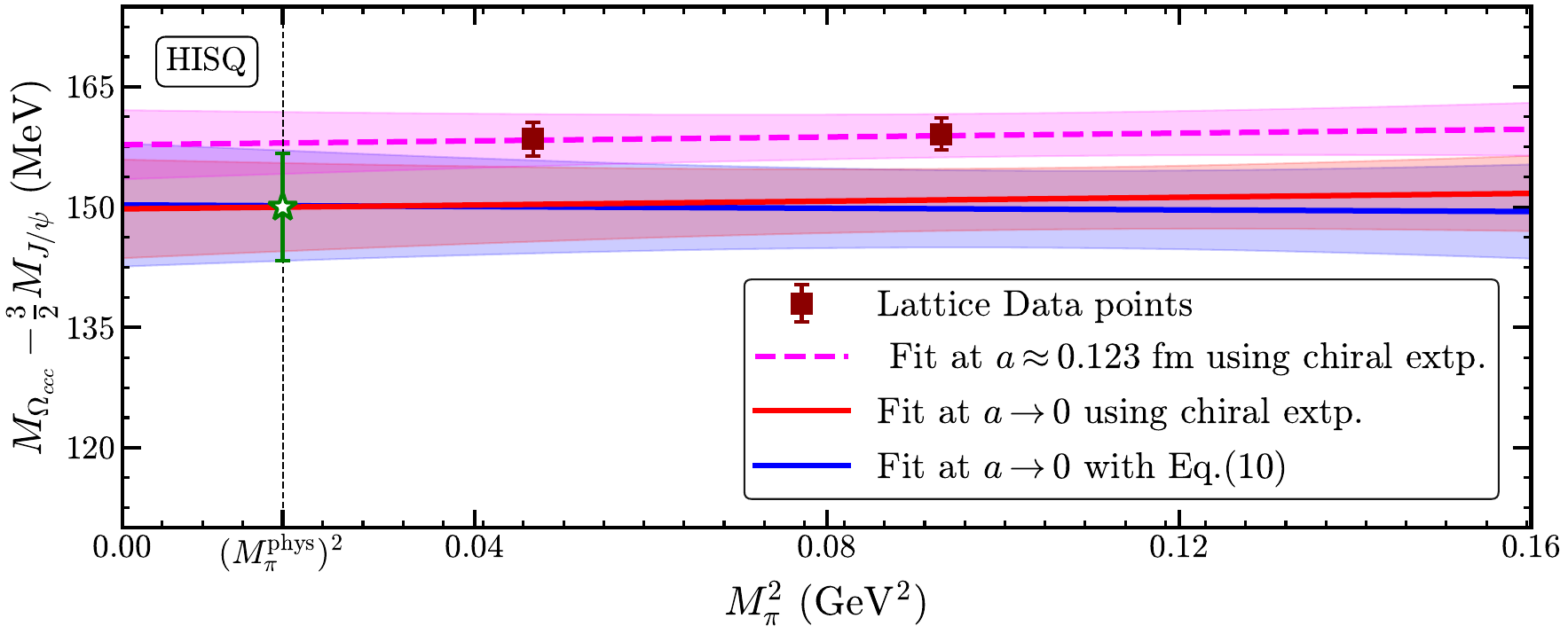}
    \caption{Chiral extrapolation of the subtracted mass splitting, 
$M_{\Omega_{ccc}}(3/2^+) - \frac{3}{2} M_{J/\psi}$ for the HISQ action. 
The deep blue line represents the central value of the chiral and continuum extrapolation ($a \to 0$) (obtained using Eq. \ref{chiral_continuum}), with the light blue band as the corresponding $1\sigma$ uncertainty. 
The dark red data points with errorbars correspond to the $S_1$ and $L_1$ ensembles, while the dashed magenta line represents the chiral extrapolation at $a = 0.12$ fm. The solid red line is obtained with continuum extrapolation of the data obtained using the fitted chiral-slope at different lattice spacings.}

    \label{fig:chiral_extrap}
\end{figure}

{Since 
the $\Omega_{ccc}$ baryons have only charm quark in their valence sector, 
it is naively expected to have small effects from unphysical light sea quarks \cite{McNeile:2012qf, Dowdall:2012ab, Chakraborty:2014aca}. However, for a precise estimation of 
$M_{\Omega_{ccc}}$ it is necessary to estimate an uncertainty due to the presence of unphysical light sea quarks. 
For the case of HISQ fermions, with six points for extrapolation, as described in Ref. \cite{Brown:2014ena}, one can investigate the sea quark mass dependence by performing a simultaneous chiral and continuum extrapolation of the energy splitting
$\Delta M_{\Omega_{ccc}} = M_{\Omega_{ccc}} - \frac{3}{2} M_{J/\psi}$, using a functional form,
\beq
\Delta M_{\Omega_{ccc}}(a, M_{\pi}) = A + B a^2 + C a^4 + D M_{\pi}^2.
\eeq{chiral_continuum}
Here $M_{\pi}$ represents the sea pion mass. With the above fit form we find the chiral and continuum extrapolated number for this energy splitting to be $\Delta M_{\Omega_{ccc}} = 150.3 \pm 6.5$ MeV.  

In Fig. \ref{fig:chiral_extrap}, we present the chiral approach of the estimates for $\Delta M_{\Omega_{ccc}}$. The two points in the figure correspond to the two lattice ensembles around the lattice spacing $0.12$ fm, but differ in the light sea quark mass (see Table \ref{Lattice_details_ov_hisq}), from which one could gauge its effects. As can be seen, there is hardly any variation between the estimates from either ensembles, indicating the sea quark mass independence. One can naively estimate this dependence with a fit form $A+B M_{\pi}^2$ and utilize the slope $B$ for other lattice spacings, and finally arrive at a continuum value. The two bands represent the sea quark mass dependence for lattice spacing $a\sim0.12$ fm [magenta] and $a\rightarrow0$ [blue]. It can be observed that there is barely any significant sea quark mass dependence considering the statistical uncertainties. Moreover these results are also consistent with the continuum extrapolated ones that does not explicitly incorporate sea pion mass effects, albeit with slightly larger errorbars due to the inclusion of chiral corrections. The increased uncertainty reflects the impact of including the pion mass dependence in our fit, ensuring a more thorough treatment of the systematic effects. Based on the above analysis an error of 5 MeV can be assigned to $M_{\Omega_{ccc}}$, accounting the uncertainty for the light sea-quark mass dependence (with and without pion mass term in \eqn{chiral_continuum}).} 

\section{Electromagnetic correction}
The presence of two units of electric charge is expected to affect the mass of the $\Omega_{ccc}^{++}$ baryons. Moreover, since we estimates its mass utilizing the subtracted method with $J/\psi$, the electromagnetic corrections to the masses of $\Omega_{ccc}^{++}$ and $J/\psi$ need to be estimated.
Since our lattice QCD simulations do not include electromagnetic interactions, we estimate these corrections, for i)   $\Omega_{ccc}^{++}$: with a leading-order perturbative electromagnetic (EM) corrections, and for ii)$J/\psi$: using a potential model based calculation with and without the coulomb term.

A naive perturbative estimate for the leading order electromagnetic correction to the mass of $\Omega_{ccc}^{++}$ baryon should be proportional to $\alpha_{em}\hbar c/r_c$, where $\alpha_{em}$ is the electromagnetic fine structure coefficient, whereas $\hbar c/r_c$ is the natural energy scale corresponding to $r_c$, the root-mean-square charge radius of the $\Omega_{ccc}^{++}$ baryon. Using $\alpha_{em} = 1/137$ and the lattice-determined estimate for $r_c = 0.410(6)$ fm from Ref. \cite{Can:2015exa}, one can find that this correction is approximately 3.5 MeV. 

To elaborate this assessment, one could apply the standard time independent perturbation theory, by assuming the Coulombic interaction potential as a perturbative part and a trial wave function for the $\Omega_{ccc}^{++}$ baryon ground state that could provide an upper bound for this electromagnetic correction. To this end, we consider two trial wave functions that respects the symmetries of the $\Omega_{ccc}^{++}$ baryon ground state and has the same root-mean-square charge radius $r_c = \sqrt{\langle \psi \vert r^2 \vert \psi\rangle }$. The first choice is the $1$S radial solution of Hydrogen atom which has been utilized also in Ref. \cite{Lyu:2021qsh}, and the second one is a Gaussian wavefunction in the radial direction, as given below,
\begin{eqnarray}
\psi_1(\mathbf{r}) &=& \left( \frac{8}{\pi a^3} \right)^{1/2} \exp \left( -\frac{2 \vert \mathbf{r} \vert}{a} \right),\\
\psi_2(\mathbf{r}) &=& \left( \frac{2}{\pi a^2} \right)^{3/4} \exp \left( -\frac{\vert \mathbf{r} \vert^2}{a^2} \right).
\label{Bohr1Swavefunc}
\end{eqnarray}
The parameter $a$ is chosen to be as $a = 2r_c/\sqrt{3}$, such that the average charge radius is $r_c$ for both choices of wavefunctions. 
With these two ansatzs, the leading order corrections are given by, $\frac{5}{2\sqrt{3}}~\alpha_{em}\hbar c/r_c$ and $\frac{4}{\sqrt{3\pi}}~\alpha_{em}\hbar c/r_c$, respectively. It is evident from the expressions that the leading order electromagnetic corrections are of the same order ($\alpha_{em}\hbar c/r_c$) and is independent of the trial wave function, modulo factors $\mathcal{O}(1)$.  This amounts to the corrections $5.1$ MeV and $4.6$ MeV, respectively, which are in good agreement with the previous estimate. Based on these estimates along with the previous perturbative estimate we take a mass correction to be about 4 MeV.

The $1S$ charmonium mass is subject to electromagnetic corrections, which could also potentially impact our estimates for $\Delta M_{\Omega_{ccc}}$. In fact, considering the opposite charges involved in the charm-anticharm system, the electromagnetic effects on $J/\psi$ meson could decrease its mass from a purely strongly noninteracting scenario. We utilize a nonrelativisitc quark model \cite{BGS_model} to estimate the corrections associated with $J/\psi$ meson. The quark-antiquark potential in charmonia can be taken as 
$V^{c\bar{c}}_0(r) = V_{st}(r) + V_{em}(r)$, where the strong-interaction potential is given by,
\begin{equation}
    V_{st}(r) = -\frac{4}{3} \frac{\alpha_s}{r} + br + \frac{32\pi\alpha_s}{9m_c^2} \tilde{\delta}_{\sigma}(r) \mathbf{S}_c \cdot \mathbf{S}_{\bar{c}},
    \label{BGS_model}
\end{equation}
with  $\tilde{\delta}_{\sigma}(r) = \left(\frac{\sigma}{\sqrt{\pi}}\right)^3 \exp(-\sigma^2 r^2)$. The electromagnetic potential is given by  $V_{em}(r) = -\frac{4}{9} \frac{\alpha_{em}}{r}$, where the prefactor $\frac{4}{9}$ arises from the squared of the electric-charge of the charm quark. There are four parameters in this model $\alpha_s$, $b$, $m_c$, and $\sigma$ and we adopt the values from Ref. \cite{BGS_model} that are tuned to reproduce the low lying charmonium spectrum. 
The corrections to the $1^1S_0$ and $1^3S_1$$-$corresponding to $\eta_c$ and $J/\psi$ mesons$-$are 
determined by comparing the mass estimates from the model with and without the inclusion of the Coulombic term, $V_{em}(r)$. The results are summarized in Table \ref{tab:QED_correction}.
\begin{table}[htpb]
    \centering
    \renewcommand\arraystretch{1.2}  
    \addtolength{\tabcolsep}{3 pt}   
    \begin{tabular}{|c|c|c|} 
        \hline
        State & \(V_{st}\) (MeV)  & \(V_{st} + V_{em}\) (MeV) \\
        \hline
        \(1^3S_1 (J/\psi)\) & 3091.2 & 3088.8 \\
        \(1^1S_0 (\eta_c)\) & 2983.9 & 2981.1 \\ \hline 
    \end{tabular}
    \caption{The ground-state masses of 1S charmonium with and without electromagnetic corrections based on a quark model {\it \'a la} Ref. \cite{BGS_model}.}
    \label{tab:QED_correction}
\end{table}

{}
From these estimates, it can be immediately inferred that the electromagnetic correction for the $J/\psi$ meson
is $-2.4$ MeV. In addition to this potential-model-based electromagnetic correction, one could also observe a leading perturbative estimate on the QED systematic to be $\sim$ $-2.5$ MeV, following the similar arguments presented above for the $\Omega_{ccc}$ baryon. The total corrections due to electromagnetic effects can then be estimated by adding the respective errors as, $4 + 3/2\times2.5\sim 7.8$ MeV. Note that we present this estimate as an asymmetric error of $\left(^{+7.8}_{+0.0}\right)$ MeV, as these effects can effectively increase the masses of $\Omega_{ccc}$ baryons. These electromagnetic correction estimates are subject to the model dependence and quantification of any higher order contributions is beyond the scope of this work.

\end{document}